\documentclass[usenatbib]{mn2e}
\include{epsf}
\usepackage{graphicx}
\usepackage{epstopdf}

\usepackage{color}

\def \farcs{\hbox{$.\!\!^{\prime\prime}$}}

\def \msun{\hbox{$\rm{M}_\odot$}}

\title[Updated CCCP weak lensing masses]{The Canadian Cluster Comparison Project:
detailed study of systematics and updated weak lensing masses\thanks{Based on observations from the Canada-France-Hawaii
    Telescope, which is operated by the National Research Council of
    Canada, le Centre National de la Recherche Scientifique and the
    University of Hawaii.}}

\author[Hoekstra et al.]{Henk Hoekstra$^{1}$\thanks{E-mail:
    hoekstra@strw.leidenuniv.nl}, Ricardo Herbonnet$^{1}$, Adam
  Muzzin$^{1}$,  Arif Babul$^2$, \and Andi Mahdavi$^3$, Massimo
  Viola$^1$, Marcello Cacciato$^1$
  \vspace*{3mm}\\
  $^1$~Leiden Observatory, Leiden University, PO Box 9513, 2300 RA,
  Leiden, the  Netherlands\\
  $^2$~University of Victoria, Dept. of Physics \& Astronomy, 3800
  Finnerty Rd, Victoria, BC V8P 5C2, Canada\\
  $^3$~San Francisco State University, Dept. of Physics \& Astronomy,
  1600 Holloway Avenue, San Francisco, CA 94132}

\begin{document}

\date{Accepted. Received; in original form}

\maketitle

\begin{abstract}
  Masses of clusters of galaxies from weak gravitational lensing
  analyses of ever larger samples are increasingly used as the
  reference to which baryonic scaling relations are compared. In this
  paper we revisit the analysis of a sample of 50 clusters studied as
  part of the Canadian Cluster Comparison Project. We examine the key
  sources of systematic error in cluster masses. We quantify the
  robustness of our shape measurements and calibrate our algorithm
  empirically using extensive image simulations. The source redshift
  distribution is revised using the latest state-of-the-art
  photometric redshift catalogs that include new deep near-infrared
  observations. Nonetheless we find that the uncertainty in the
  determination of photometric redshifts is the largest source of
  systematic error for our mass estimates. We use our updated masses
  to determine $b$, the bias in the hydrostatic mass, for the clusters
  detected by {\it Planck}. Our results suggest $1-b=0.76\pm 0.05{\rm
    (stat)}\pm0.06{\rm (syst)}$, which does not resolve the tension
  with the measurements from the primary cosmic microwave background.
\end{abstract}

\begin{keywords}
cosmology: observations $-$ dark matter $-$ gravitational lensing $-$
galaxies: clusters
\end{keywords}

\section{Introduction}

The observed number density of clusters of galaxies as a function of
mass and redshift depends sensitively on the expansion history of the
Universe and the initial conditions of the density
fluctuations. Comparison with predictions from a model of structure
formation can thus constrain cosmological parameters, such as the mean
density $\Omega_m$ and the normalization of the matter power spectrum
$\sigma_8$ \citep[e.g.][]{Bahcall98,Henry00, Reiprich02, Henry09}, or
the dark energy equation-of-state $w$ \citep[e.g.][]{Vikhlinin09,
  Mantz10, Mantz15}.  For a recent review see \cite{Allen11}.

The fact that the observations do not provide actual cluster counts as
a function of mass, but rather the number density of objects with
certain observational properties, such as the number of red galaxies
or the X-ray flux within a given aperture, complicates a direct
comparison with predictions: the cosmological interpretation requires
knowledge of the selection function and the scaling relation between
the observable and the underlying mass. Furthermore, scaling relations
typically have intrinsic scatter that also needs to be measured, or at
least accounted for.

One way forward is to simulate the observable properties of clusters,
but the complex non-linear physics involved limits the fidelity of
such approaches, at least for the moment. Therefore direct estimates
of the clusters masses are needed. This can be achieved through
dynamical analyses, such as the measurement of the motion of cluster
members, or by measuring the temperature of the hot intra-cluster
medium (ICM). However, in both cases the cluster is assumed to be in
equilibrium, which is generally not a valid assumption. For instance,
simulations suggest that hydrostatic X-ray masses are biased low
\citep[e.g.][]{Rasia06, Nagai07,Lau09}.

A more direct probe of the (dark) matter distribution would be
preferable, which is provided by the gravitational lensing distortion
of background galaxies: the gravitational potential of the cluster
perturbs the paths of photons emitted by these distant galaxies,
resulting in a slight, but measurable, coherent distortion. This in
turn provides a direct measurement of the gravitational tidal field,
which can be used to directly infer the projected mass distribution.
Note, however, that the comparison to baryonic tracers does typically
depend on the assumed geometry of the cluster.  For a recent review of
the use of gravitational lensing to study cluster masses and density
profiles we refer the reader to \cite{Hoekstra13}.

The sizes of cluster samples are increasing rapidly thanks to
wide-angle surveys at various wavelengths, especially at millimeter
wavelengths thanks to the inverse-Compton scattering of cosmic
microwave background (CMB) photons off hot electrons in the
intracluster medium, the Sunyaev-Zel'dovich Effect
\citep[SZE;][]{Sunyaev72}. In fact, the lack of calibrated scaling
relations is currently the dominant limitation of blind searches that
exploit this effect, such as those carried out using the South Pole
Telescope \citep[SPT;][]{Reichardt13} or the Atacama Cosmology
Telescope \citep[ACT;][]{Hasselfield13}. The importance of accurate
mass calibration is furthermore highlighted by the tension between the
cosmological parameters determined from the primary CMB measured by
{\it Planck} \citep{PlanckXVI} and those inferred from the cluster
counts \citep{PlanckXX}.

Fortunately it is not necessary to determine masses for all clusters,
but instead it is sufficient to calibrate the appropriate scaling
relation and its scatter. However, doing so still requires substantial
samples of clusters for which weak lensing masses need to be
determined. Even for the most massive clusters the uncertainty in the
projected mass is $\sim 10\%$. The triaxial nature of cluster halos,
however, leads to an additional intrinsic scatter of $\sim 15-20\%$
\citep[e.g.][]{Corless07,Meneghetti10, Becker11}. Hence to calibrate the
normalization of a scaling relation to a few percent requires a sample
of 50 or more clusters.

To examine the relation between the baryonic properties of clusters
and the underlying matter distribution the Canadian Cluster Comparison
Project (CCCP) started with the study of archival observations of 20
clusters of galaxies, described in \cite{Hoekstra07} and
\cite{Mahdavi08}. This sample was augmented by observations of an
additional 30 clusters with $0.15<z<0.55$ with the
Canada-France-Hawaii Telescope (CFHT). A detailed description of the
sample can be found in \cite{Hoekstra12} (H12 hereafter) and
\cite{Mahdavi13}. The comparison to the X-ray properties, presented in
\cite{Mahdavi08} and \cite{Mahdavi13} confirmed the predictions from
numerical simulations that the hydrostatic mass estimates are biased
low.

Other groups have carried out similar studies. The Local Cluster
Substructure Survey (LoCuSS) used the Subaru telescope to carry out a
weak lensing study of 50 clusters, with the most recent results
presented in \cite{Okabe13}. A thorough analysis of a sample of 51
clusters was presented by Weighing the Giants
\citep[WtG;][]{vonderLinden14,Applegate14}. For a large fraction of
the clusters the latter study also obtained photometric redshifts for
the sources \citep{Kelly14}. Most recently \cite{Umetsu14} presented
results for a sample of 20 massive clusters, 17 of which were observed
by WtG. In general there is significant overlap as these studies all
target massive well-known clusters of galaxies. This was exploited by
\cite{Applegate14} who compared the masses from the various
studies. Although they find an excellent correlation with the results
from H12, the CCCP masses are on average $\sim 20\%$
lower than their estimates. This is much larger than the statistical
uncertainties and warrants further investigation. This is the aim of
this paper.

A correct interpretation of the inferred weak lensing signal relies on
accurate shape measurements and knowledge of the redshifts of the
sources used in the analysis. The former has been examined quite
extensively over the past decade, for instance in several blind
studies using simulated images \citep{STEP1,STEP2,GREAT08,GREAT10,GREAT3}.
The results of such simulations have been used to quantify the biases
in shape measurements, but the sensitivity of the calibration to the
input of the simulations has not been investigated in much detail.
However, thanks to an improved understanding of the sources of bias,
and how they propagate \citep[e.g.][]{Massey13,
  Semboloni13,Miller13,Viola14}, it has become evident that a correct
interpretation of these simulations depends critically on how well
they match the specific observations under consideration. In \S2 we
examine the importance of the fidelity of the image simulations. To
calibrate our method, we create an extensive set of images, varying a
number of input parameters.

Another important source of uncertainty is the redshift distribution
of the sources. In \S3 we present our photometric redshift estimates
based on measurements in 29 bands in the COSMOS field
\citep{Scoville07,Capak07} including new deep observations in five NIR
bands from UltraVISTA \citep{McCracken12}. We also revisit the issue
of contamination by cluster members. We present new weak lensing mass
estimates in \S4 and use these in \S5 to calibrate the hydrostatic
masses used by \cite{PlanckXX} to infer cosmological parameters.
Throughout the paper we assume a cosmology with $\Omega_m=0.3$,
$\Omega_\Lambda=0.7$ and $H_0=70 h_{70}$~km/s/Mpc.

\section{Calibration of Shape Measurements}\label{sec:shapes}

The measurement of the shapes of small, faint galaxies is one of two
critical steps in order to derive accurate cluster masses from weak
gravitational lensing, the other step involving knowledge of the
source redshift distribution.  We discuss the latter in
\S\ref{sec:redshift} and focus first on the algorithms used to measure
galaxy shapes. Most studies to date have focussed on the correction
for the blurring by the PSF, which leads to rounder images (due to the
size of the PSF) and preferred orientations (if the PSF is
anisotropic). An incomplete correction for the former leads to a
multiplicative bias $\mu$ and a residual in the latter to an additive
bias $c$; the observed shear and true shear are thus related by
\citep[e.g.][]{STEP1}:

\begin{equation}
\gamma_i^{\rm obs}=(1+\mu)\gamma_i^{\rm true}+c,
\end{equation}

\noindent where we implicitly assumed that the biases are the same for
both shear components. For cosmic shear studies the additive bias is a
major source of concern because the (residual) PSF introduces power on
relevant scales \citep[e.g.][]{Hoekstra04}. For cluster lensing the
additive bias is less important because the measurement of cluster
masses involves the azimuthally averaged tangential shear and PSF
patterns largely average out for our data. We study the residual
additive bias in Appendix~\ref{sec:psfan} and find that we can indeed
ignore the residuals arising from PSF anisotropy in our analysis.

One approach to recover the true galaxy shape is to assume a suitable
model for the galaxy light distribution, which is subsequently
sheared, convolved with the PSF and pixellated. The model parameters
are varied until a best fit to the data is obtained. This has the
advantage that the detrimental effects of the PSF (and other
instrumental biases) can be incorporated into a Bayesian framework
\citep[e.g.][]{Miller13, Bernstein14}.  The challenge, however, is to
use a model that provides a good description of the galaxies, while
having a limited number of parameters in order to avoid
over-fitting. A model that is too rigid will lead to model bias
\citep[e.g.][]{Bernstein10}, whereas a model that is too flexible
tends to fit noise in the images
\citep[e.g.][]{Kacprzak12}. Furthermore, accurate priors for the size
and ellipticity distributions (and any other parameter entering the
model) are required to obtain an unbiased estimate for the shear.

An alternative approach, which we use here, involves measuring the
moments of the galaxy images, which are subsequently corrected for the
PSF. The shapes can be quantified by the polarization:

\begin{equation}
e_1=\frac{I_{11}-I_{22}}{I_{11}+I_{22}},~{\rm and~}
e_2=\frac{2I_{12}}{I_{11}+I_{22}},
\end{equation}

\noindent where the quadrupole moments $I_{ij}$ are given by

\begin{equation}
I_{ij}=\frac{1}{I_0}\int{\rm d}^2{\bf x}x_i x_j W({\bf x})f({\bf x}),
\end{equation}

\noindent where $f({\bf x})$ is the observed galaxy image, $W({\bf
  x})$ a suitable weight function to suppress the noise and $I_0$ the
weighted monopole moment. In the case of unweighted moments, $I_0$
corresponds to the flux and the correction for the PSF is
straightforward as the PSF corrected moments are given by\footnote{We
  assume that the measurement is centered on the location where the
  dipoles vanish.}

\begin{equation}
I^{\rm true}_{ij}=I^{\rm obs}_{ij}-I^{\rm PSF}_{ij},
\end{equation}

\noindent i.e., one only needs to subtract the moments of the PSF from
the observed moments. The result provides an unbiased estimate of the
polarization. However, the change in polarization $\delta e_i$ due to
a shear $\delta\gamma_i$ depends on the unsheared shape $e_{\rm int}$:
it is more difficult to change the shape of an object that is already
elongated. This response is quantified by the polarizability
$P^\gamma$, defined such that $\delta e_i=P^\gamma \delta\gamma_i$.
As the shear is obtained from an ensemble of galaxies, an unbiased
estimate thus requires knowledge of the intrinsic ellipticity
distribution \citep[e.g.][]{Viola14}.

Unfortunately real data contain noise and thus unweighted moments are
not practical. To suppress the effects of noise a weight function
needs to be chosen, ideally matched to the size and shape of the
galaxy image. However, as discussed in e.g. \cite{Massey13} and
\cite{Semboloni13} this complicates matters as the correction for the
PSF now involves higher order moments, which themselves are affected
by noise. Limiting the expansion in moments is similar to the model
bias in fitting methods.

Recent studies using simulated data have shown that multiplicative
biases depend strongly on the signal-to-noise ratio
\citep[SNR;][]{GREAT08,GREAT10,Miller13} with some hints already present in the
dependence of the bias on magnitude in \cite{STEP2}. As the origin of
this bias is now better understood, it has also become clear that the
performance of a particular algorithm will depend on the data it is
applied to. Hence the performance evaluation, such as the
determination of the bias that one wishes to correct for, depends on
the input of the simulations: if the input does not match the actual
data, the inferred bias may be different from the actual value.
Although SNR is the most critical parameter, the bias may also depend
on the galaxy profile, or the size and ellipticity distributions
\citep[e.g.][]{Miller13, Melchior12, Kacprzak12,Viola14}.  Unless the fidelity
of the simulation can be somehow guaranteed, the sensitivity of a
method to the input parameters needs to be quantified and the
uncertainties propagated.

In this paper we focus on the commonly used KSB method developed by
\cite{KSB95} and \cite{LK97} with corrections provided in
\cite{Hoekstra98} and \cite{Hoekstra00}. It was used to determine
masses for the CCCP sample in \cite{Hoekstra07} and H12; we refer the
interested reader to these papers for more details. The object
detection is done using the hierarchical peak finder described in
\cite{KSB95}, which gives an estimate for $r_g$, the Gaussian scale
radius of the object. This value is used to compute the weighted
moments, which are corrected following \cite{Hoekstra98}. In addition
we also compute $\sigma_e$, the uncertainty in the polarization, which
is approximately $\propto 1/\nu$, where $\nu$ is the signal-to-noise
ratio of the detection \citep{Hoekstra00}. This allows us to
downweight the noisy galaxies and we therefore estimate the average
shear for an ensemble of galaxies as

\begin{equation}
  \langle\gamma_i\rangle=\frac{\sum w_i e_i/{\tilde P}^\gamma}{\sum w_i},
  ~{\rm with}~w_i=\frac{1}{\langle \epsilon^2\rangle+
    \left( \frac{\sigma_e}{{\tilde P}^\gamma}\right)^2},
\label{eq:avshear}
\end{equation}

\noindent where $\langle\epsilon^2\rangle$ is the intrinsic variance
of the galaxy ellipticity components. This is the dominant source of
uncertainty for the shear for bright objects, and we adopt a value of
$\langle\epsilon^2\rangle^{1/2}=0.25$ \citep{Hoekstra00}. In our image
simations we vary the input ellipticity distribution (see
\S\ref{sec:input}), which in principle would require adjusting the
value for $\langle\epsilon^2\rangle^{1/2}$ accordingly to optimally
weight objects. However, for simplicity we keep it fixed when we
quantify the multiplicative bias. 

\begin{figure}
\centering
\includegraphics[width=8.5cm]{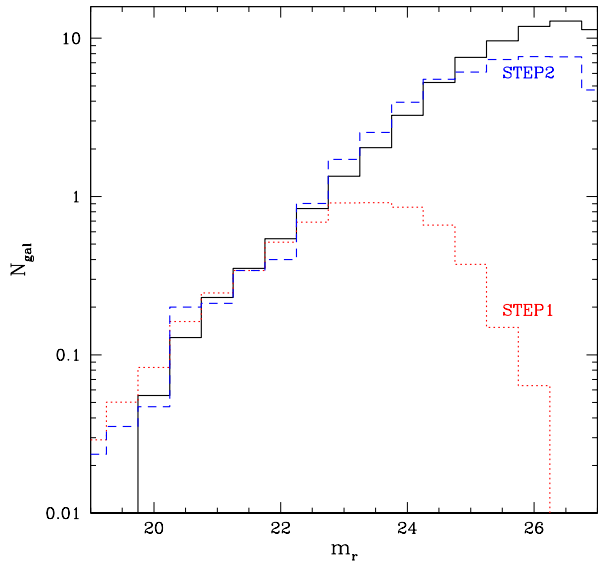}
\caption{Number counts as a function of $r$-band magnitude for our image
simulations (solid black histogram), STEP1 (dotted red histogram) and STEP2 
(dashed blue histogram). The counts were normalized in the range $20<m_r<22$ 
and we adjusted the magnitudes of the STEP simulations for the use of
different filters.
\label{fig:input_counts}}
\end{figure}

\subsection{Input galaxy properties}
\label{sec:input}

To populate our image simulations we use a sample of galaxies for which
morphological parameters were measured from resolved $F606W$ images
from the GEMS survey \citep{Rix04}. These galaxies were modeled as
single Sersic models with {\tt galfit} \citep{Peng02} and for our
study we use the half-light radius, magnitude and Sersic index $n$. We
only consider galaxies fainter than $m_r=20$ because bright objects
might cause unrealistic features in the simulated images. Excluding
these does not impact our results as we do not use them in our source
sample anyway: to measure the lensing signal we use galaxies with
$22<m_r<25$.

The resulting number density of galaxies as a function of apparent
magnitude $m_r$ is presented in Figure~\ref{fig:input_counts} (solid
black histogram). The results suggest a power-law for the counts,
where the flattening for $m_r>25.5$ is caused by incompleteness of
the input catalog. In principle, faint unresolved galaxies can affect
shape measurements of brighter galaxies through modulation of the
background noise and blending. In \S\ref{sec:sim_input} we therefore
examine the need to include fainter galaxies in the simulations.

H12 based their assessment of the accuracy of the shape measurements
on the results from the Shear TEsting Programmes
\citep[STEP;][]{STEP1, STEP2}. These were blind challenges with the
aim to benchmark the performance of shape measurement algorithms,
especially for cosmic shear studies. In both cases the implementation
used by H12 performed well, with an average multiplicative bias of
$\sim 2\%$. As a consequence, H12 ignored the multiplicative bias in
their mass estimates.

STEP1 \citep{STEP1} simulated CFHT observations in the $I$-band with
an integration time of 3600s, which should be quite comparable to our
data (a total integration time of 1 hour in the $r'$ band using CFHT).
The red dotted histogram in Figure~\ref{fig:input_counts} shows the
galaxy number counts that were used as input for STEP1 (converted to
$r$-band assuming a mean galaxy color of $r-i=1$). The counts are
normalized such that the sum is the same for all three examples in the
range $20<m_r<22$. Comparison to the GEMS catalog shows that STEP1
lacks the faint galaxies that are present in real observations, even
if they are unresolved. As we will show in \S\ref{sec:sim_input}, this
leads to a significant underestimate of the multiplicative bias for
the actual CCCP data.

STEP2 \citep{STEP2} simulated images that would be obtained with an
exposure time of 40 minutes in good conditions with SuprimeCam on
Subaru. Given the larger aperture and throughput of Subaru compared to
CFHT this corresponds to a total exposure time that is $\sim 4$ times
longer than the CCCP data. The input galaxy images were based on a
{\tt shapelet} decomposition of resolved galaxy images from the {\it
  Hubble Space Telescope} COSMOS survey \citep{Scoville07}, as
described in \cite{Massey04}. As a result the simulations should
better capture the complex morphologies of real galaxies. The number
counts, shown by the blue dashed histogram in
Figure~\ref{fig:input_counts}, match the GEMS input counts much better
than STEP1, although incompleteness occurs at $m_r\sim 24.5$.

\begin{figure}
\centering
\includegraphics[width=8.5cm]{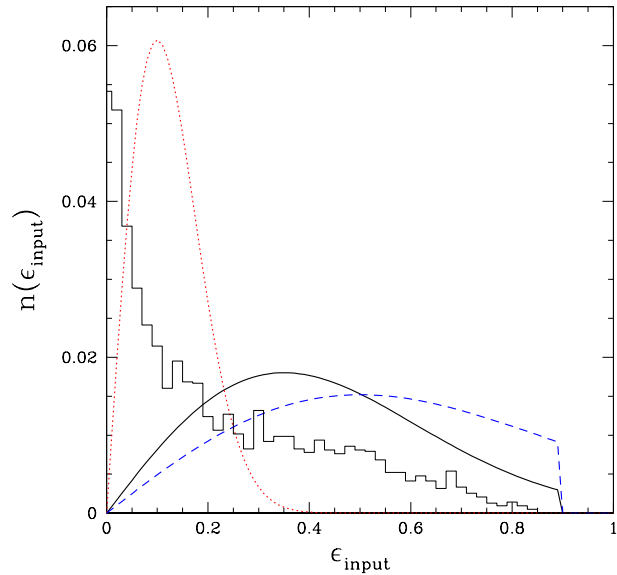}
\caption{Input ellipticity distributions, described by a Rayleigh
  distribution truncated at $\epsilon=0.9$, for $\epsilon_0=0.35$
  (thick black curve), $\epsilon_0=0.1$ (red dotted curve) and
  $\epsilon_0=0.5$ (blue dashed curve). For reference we also show a
  histogram of ellipticities from one of the STEP1 simulations.
\label{fig:input_e}}
\end{figure}

In addition to the magnitudes, the GEMS catalog provides values for
the Sersic index of the galaxies, as well as their half-light radii
and ellipticities. The use of Sersic profiles to describe the galaxies
may limit the fidelity of the simulations \citep[see][for a study of
the biases that may arise]{Kacprzak14}. We examine the bias as a
function Sersic index in \S\ref{sec:sim_input} and find that our shape
measurement algorithm is not particularly sensitive to the profile,
especially when compared to other sources of bias. We therefore expect
that the difference with using realistic galaxy morphologies is
small. One of the aims of the third gravitational lensing accuracy
testing challenge\footnote{http://www.great3challenge.info}
\citep[GREAT3;][]{Mandelbaum14} is to compare the results of shape
measurement methods on postage stamps of actual (PSF corrected) HST
observations and the corresponding Sersic fits.

The different parameters describing the galaxies are jointly sampled
from the GEMS catalog to account for their intrinsic correlations
(e.g., brighter galaxies are on average larger). However, we do not
use the ellipticities provided by \cite{Rix04}, because of concerns
that these do not match our data, as discussed in
\S\ref{sec:description}. Instead we use a parametric description,
which allows us to investigate the role of the ellipticity
distribution. We assign ellipticities\footnote{The ellipticity is
  defined as $(a-b)/(a+b)$, with $a$ and $b$ the major and minor axes,
  resp. The polarization for such a galaxy would be $\sim
  (a^2-b^2)/(a^2+b^2)$} $\epsilon$ that are drawn from a Rayleigh
distribution given by

\begin{equation}
P(\epsilon,\epsilon_0)=\frac{\epsilon}{\epsilon_0^2}e^{-\epsilon^2/
2\epsilon_0^2},
\end{equation}

\noindent where the value of $\epsilon_0$ determines the width of the
distribution, as well as the average
$\langle\epsilon\rangle=\epsilon_0 \sqrt{\pi/2}$. We need to truncate
the distribution because the ellipticity cannot exceed unity, but also
because galaxy disks have a finite thickness. We therefore set
$P(\epsilon,\epsilon_0)=0$ if $\epsilon>0.9$. We assume that the
ellipticity distribution is independent of other galaxy properties,
whereas e.g., \cite{Uitert11} did observe different distributions for
early and late type galaxies. Given the accuracy we require here, we
find that this assumption does not impact our results. The ellipticity
distribution of the GEMS catalog matches that of $\epsilon_0=0.35$ for
moderate ellipticities \citep{Melchior12} which is indicated by the
black curve in Figure~\ref{fig:input_e}. We also show input
ellipticity distributions for $\epsilon_0=0.1$ (red dotted curve), and
$\epsilon_0=0.5$ (blue dashed curve). For comparison we also show the
input ellipticity distribution used by STEP1 \citep{STEP1}, which
peaks at very low ellipticities.

\subsection{Description of the simulations}
\label{sec:description}

\begin{figure}
\centering
\includegraphics[width=8.5cm]{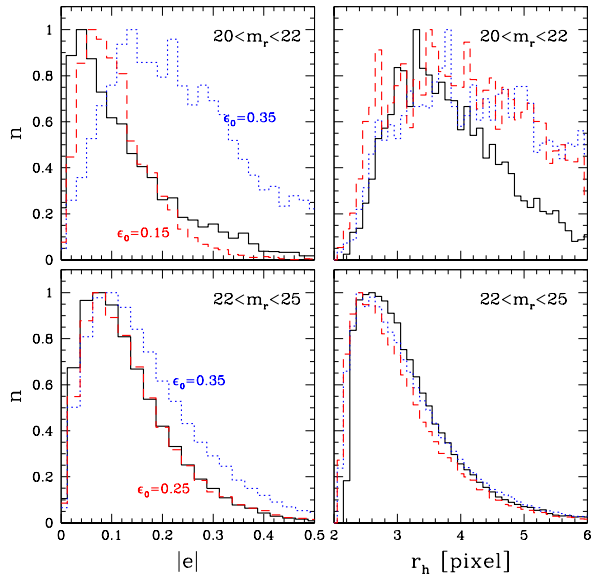}
\caption{Comparison of the simulations and the actual CCCP data for 
Abell~1835 (black lines) for bright and faint sources. The left panels
show the observed polarizations, i.e. uncorrected for the PSF. We find
that the ellipticity distribution for bright galaxies $(20<m_r<22)$ is
best approximated if we take $\epsilon_0=0.15$, whereas the ellipticity
distribution for fainter galaxies  $(22<m_r<25)$ requires a higher
value of $\epsilon_0=0.25$. For reference we show $\epsilon_0=0.35$
which is ruled out by the data. As shown in the right panels, the 
simulations match the observed distribution of half-light radii 
fairly well.
\label{fig:compare}}
\end{figure}

To create the images we use {\tt galsim} \citep{Rowe14}, a publicly
available code that was developed for GREAT3 \citep{Mandelbaum14}. The
main input is a list of galaxies with a position, flux, half-light
radius, Sercic index and ellipticity, from which sheared images are
computed. We limit the sample to objects with $0.5<n<4.2$ because of
limitations of the version of {\tt galsim} we used. To create the
simulated images we draw objects from the GEMS catalog. Given the
limited number of galaxies observed by GEMS, objects typically appear
multiple times in the simulation, but with a different ellipticity and
orientation.

The weak lensing analysis of CCCP Megacam data is done on stacks with
a total integration time of 1 hour each. For each cluster two such
stacks are observed, which are merged at the catalog stage (see H12
for details). Our image simulations therefore assume the same noise
level as observed in these data. To simulate the observed data we also
need to provide a realistic PSF, for which we use a circular Moffat
profile with a FWHM=$0\farcs67$ and $\beta=3.5$. This resembles our
observations of Abell~1835, which are typical for the CCCP sample, and
we also adopt the noise level observed in these data. We include a low
number of stars to measure the PSF in the images. In
Appendix~\ref{sec:star_contam} we quantify the impact of realistic
star densities. We find that the observed star densities in the CCCP
data are sufficiently low that they do not impact the results.

We create pairs of images where the galaxies are rotated by 90 degrees
in the second image to reduce the noise due to the intrinsic
ellipticity distribution \citep[see e.g.][]{STEP2}: by construction
the mean intrinsic ellipticity when both are combined is zero.  We
analyse the images separately and thus, due to noise in the images,
this is no longer exactly true, especially for faint galaxies. The
input shears typically range from -0.06 to 0.06 in steps of 0.01 (for
both components), yielding 169 image pairs for each ellipticity
distribution. Each image has a size of 10,000 by 10,000 pixels, with a
pixel scale of $0\farcs 185$, the same as our MegaCam data. This
results in a sample of $\sim 10^7$ galaxies with $20<m_r<25$ for each
value of $\epsilon_0$. To examine the dependence on seeing and PSF
anisotropy we create somewhat smaller sets, consisting of 49 pairs of
images.

We analyse these images in the same way as the CCCP
data. Figure~\ref{fig:compare} shows the distribution of observed
polarizations and half-light radii from the actual data (solid
histograms) and simulated data (dotted and dashed histograms). We
reproduce the magnitude distribution (not shown) and the size
distribution for galaxies fainter than $m_r=22$.  As shown in the top
right panel of Figure~\ref{fig:compare}, there are many simulated
bright galaxies that have half-light radii that are large compared to
our CCCP data. For these galaxies the polarizations are significantly
smaller than the distribution with $\epsilon_0=0.35$ found by
\cite{Melchior12}. Although the input catalog is based on HST data, we
suspect that the use of {\tt galfit} may give too much weight to the
outer regions of the galaxies, which are down-weighted in moment-based
methods.  This highlights the difficulty in establishing the input
ellipticity distribution, which remains rather uncertain. For the main
sample of sources, with $22<m_r<25$, we find a good match if we adopt
$\epsilon_0=0.25$. We take this value as our reference in the
remainder of this paper. We will conservatively assume that
$0.15<\epsilon_0<0.3$ when we estimate systematic uncertainties in the
empirical bias correction.

\subsection{Multiplicative bias as a function input parameters}
\label{sec:sim_input}

\begin{figure}
\centering
\leavevmode \hbox{%
\includegraphics[width=8.5cm]{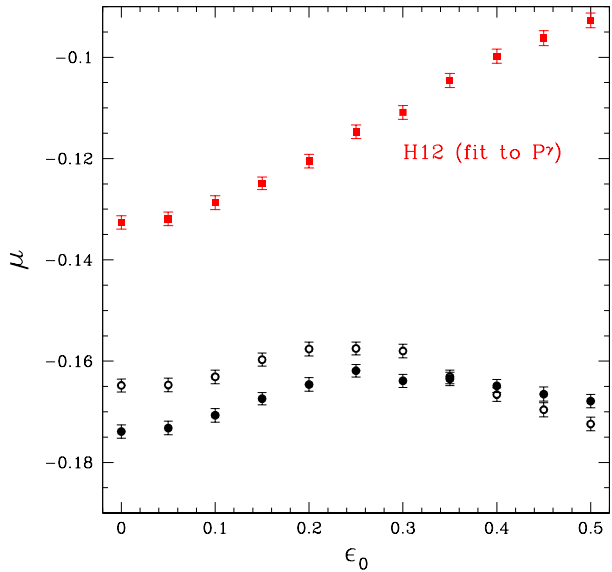}}
\caption{Multiplicative bias for sources with $20<m_r<25$ as a
  function of $\epsilon_0$ (see text). The red squares show the results for the
  implementation of KSB used in H12. The black points show
  the results when we use the observed estimate for $P^\gamma$ for
  individual galaxies. Although the bias is larger in the latter case,
  it depends only weakly on $\epsilon_0$, especially in the relevant
  range of 0.15-0.3. The open circles indicate the bias if {\tt SExtractor}
is used instead of {\tt hfindpeaks}.
\label{fig:bias_e}}
\end{figure}

As the underlying ellipticity distribution remains uncertain, we start
by examining the average bias of KSB as a function of $\epsilon_0$ for
galaxies with $20<m_r<25$, i.e. the range in magnitude of the sources
used in the CCCP analysis by H12. To detect objects we use the
hierarchical peak finder described in \cite{KSB95}, which is the
default algorithm in our analysis. The main difference between the
various implementations of the KSB algorithm is the way the shear
polarizability $P^\gamma$ is estimated. As the observed values are
noisy, H12 used a parametric fit to average values as a function of
size for different magnitude bins \citep[also see][for a concise
description]{STEP1}. The red squares in Figure~\ref{fig:bias_e} show
the bias as a function of $\epsilon_0$ for this implementation of KSB.
The bias changes by 0.04, which corresponds to a relative change
  of $\sim 40\%$, over the rather extreme range in ellipticity
distribution. For $\epsilon_0=0.25$ we find a bias of $\mu\sim
-0.115$, which is much larger than the value reported in \cite{STEP1}
and \cite{STEP2}. Consequently we cannot ignore the multiplicative
bias, as was done in H12.

The black points show the results if we use the measured value of
$P^\gamma$ for each galaxy. In this case the bias is larger
($\mu\sim-0.165$ for $\epsilon_0=0.25$), but also less sensitive to
the ellipticity distribution. For this reason, as well as simplicity,
we adopt this implementation as our reference.  We note that the full
chain of detection and shape analysis needs to be simulated. This is
highlighted by the open black points, which indicate the bias if we
use {\tt SExtractor} \citep{Bertin96} to detect objects and use the
value for {\tt FLUX\_RADIUS} to compute the corresponding value for
$r_g$: the observed bias is affected at the per cent level\footnote{In the
process of making this comparison we discovered that {\tt SExtractor}
(we used version 2.5.0) incorrectly assigns objects {\tt FLAG=16} if
they are elongated horizontally. This problem can be avoided by
adopting a value for {\tt MEMORY\_BUFSIZE} larger than the image
dimensions. We note that \cite{Gruen14} discovered the same problem and
reported on this.}.

\begin{figure}
\centering
\leavevmode \hbox{%
\includegraphics[width=8.5cm]{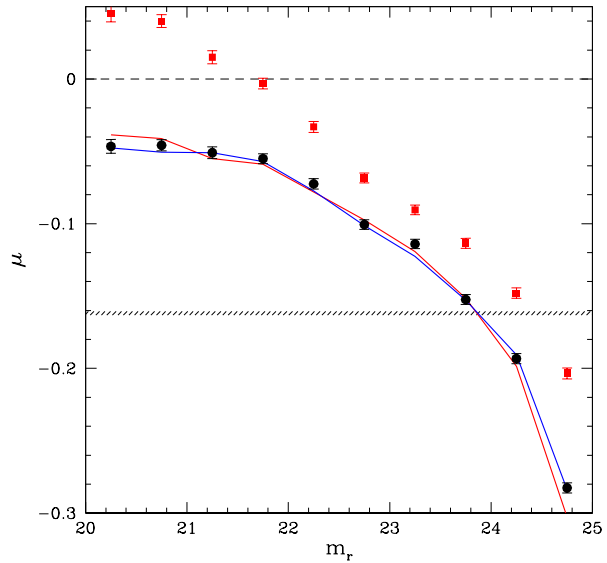}}
\caption{Multiplicative bias as a function of apparent magnitude for
  the simulated CCCP data. The measurements are averages for bins with
  a width of 0.5 magnitude. The black points correspond to an
  elliptictity distribution with $\epsilon_0=0.25$. The red and
  blue lines are for $\epsilon_0=0.15$ and $\epsilon_0=0.3$,
  respectively. The red squares indicate the bias if we
  follow the procedure to evaluate $P^\gamma$ used in H12. For comparison, the 
  hatched region indicates the 68\% confidence interval for the average 
  bias for galaxies with $20<m_r<25$ (for $\epsilon_0=0.25$).
  \label{fig:bias_mag}}
\end{figure}

Figure~\ref{fig:bias_mag} shows that the bias increases quickly for
fainter galaxies, irrespective of the ellipticity distribution.  This
is also the case when we consider the implementation used by H12 (red
squares). A strong dependence of the bias on the SNR was already
observed in \cite{GREAT08} and \cite{GREAT10}. The lack of faint
galaxies in STEP1 is the main reason that a small bias was observed in
\cite{STEP1}. When we restrict the analysis to the magnitude range
simulated by STEP1 we reproduce the small bias for the implementation
used for that paper. Note that STEP2 simulated data that are deeper
than our CCCP data. The implementation used by H12 gives smaller
biases when considering the full range in magnitude, but overcorrects
bright galaxies (i.e. $\mu>0$). It appears that the choice of the
fitting function partly compensated for the bias due to noise.

\begin{figure}
\centering
\leavevmode \hbox{%
\includegraphics[width=8.5cm]{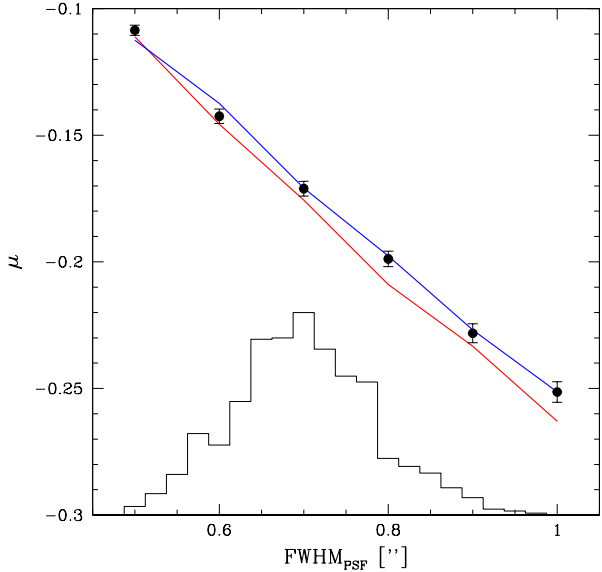}}
\caption{Multiplicative bias as a function of seeing for galaxies with
  $20<m_r<25$ for an elliptictity distribution with $\epsilon_0=0$
  (red line), the reference case with $\epsilon_0=0.25$ (black
  points) and an extreme case with $\epsilon_0=0.5$ (blue
  line). The histogram shows the distribution of PSF sizes of the
  CCCP data measured for each chip. The image quality is typically
  best in the inner regions of the field-of-view, which are most
  relevant for the mass estimates.\label{fig:bias_seeing}}
\end{figure}

The SNR is also affected by the PSF size: the larger the PSF, the
lower the SNR as the flux is spread over more pixels. The seeing also
determines how well galaxies are resolved, which impacts the bias as
well (see Appendix~\ref{app:correction}).
Figure~\ref{fig:bias_seeing} shows the value of $\mu$ for galaxies
with $20<m_r<25$ as a function of seeing for $\epsilon_0=0$ (red
line), $\epsilon_0=0.25$ (black points) and an extreme case with
$\epsilon_0=0.5$ (blue line). Note that we keep the range in apparent
magnitude the same. The results demonstrate the importance of good
image quality: the bias more than doubles from -0.11 to -0.25 as the
seeing deteriorates from $0\farcs5$ to 1$''$.

\begin{figure}
\centering
\leavevmode \hbox{%
\includegraphics[width=8.5cm]{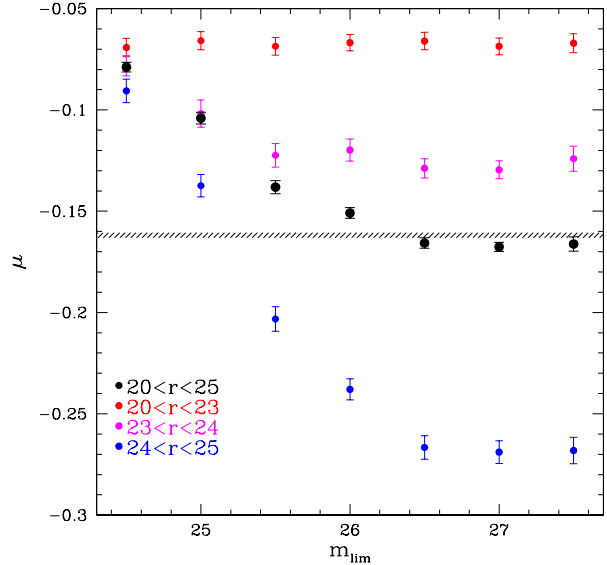}}
\caption{Multiplicative bias for an ellipticity distribution with
  $\epsilon_0=0.25$ where only galaxies with magnitudes brighter than
  $m_{\rm lim}$ are included in the simulation. The black points show the 
  bias for $20<m_r<25$. For comparison, the hatched region indicates the 
  68\% confidence interval for the average bias for galaxies using the GEMS input catalog. Irrespective of 
  the magnitude range, the bias converges when $m_{\rm lim}$ is 1.5 mag 
  fainter than the magnitude limit of the source sample. 
  \label{fig:bias_lim}}
\end{figure}

The number of faint galaxies increases rapidly
(cf. Fig.~\ref{fig:input_counts}), which results in source galaxies
blending with fainter ones. Even if a faint galaxy is not detectable,
it will impact the noise level, effectively introducing correlated
noise that affects the local background determination. Both of
these will modify the multiplicative bias in a way that that can only
be quantified through simulations. In \cite{GREAT08}, \cite{GREAT10}
and \cite{GREAT3} only postage stamps of isolated galaxies were
analyzed, and thus the effects of blending and faint galaxies were not
included. Figure~\ref{fig:bias_lim} shows that this is an important
effect, and cannot be neglected. To obtain these results we create
images where we include galaxies down to a limiting magnitude $m_{\rm
  lim}$. The input GEMS catalog is incomplete for $m_r>25.5$ (see
Figure~\ref{fig:input_counts}) and we augment the catalog by
duplicating the fainter galaxies such that the input counts follow the
power-law relation seen at brighter magnitudes. At the faintest
magnitudes these galaxies are unresolved in the simulated ground-based
data, and hence the details of their structural properties are not
critical.

The black points in Figure~\ref{fig:bias_lim} show that the bias for
galaxies with $20<m_r<25$ increases until $m_{\rm lim}>26.5$; in
general we find that the bias converges if we include sources that are
1.5 magnitude fainter than the magnitude limit of the sample of
sources used to measure the weak lensing signal. This also appears to
be true if we consider narrow bins in magnitude, such as the bin with
$24<m_r<25$ for which the bias is large, but converges for $m_r>26.5$.
The dominant contribution of these faint galaxies is to act as a
source of correlated noise, affecting the shape measurements of
brighter galaxies. These results demonstrate that it is important to
ensure that the input catalog used for image simulations contains a
sufficient number of galaxies fainter than the magnitude limit one is
interested in.

For comparison the hatched area in Figure~\ref{fig:bias_lim} indicates
the 68\% confidence region for the bias we obtain when we use the GEMS
input catalog, without introducing additional faint galaxies to
account for incompleteness (for $m_r>25.5$). Comparison with the black
points suggests that the input catalog is sufficient for the
interpretation of the CCCP data, and we use it to compute the
corrections in \S\ref{sec:correction}.

\begin{figure}
\centering
\leavevmode \hbox{%
\includegraphics[width=8.5cm]{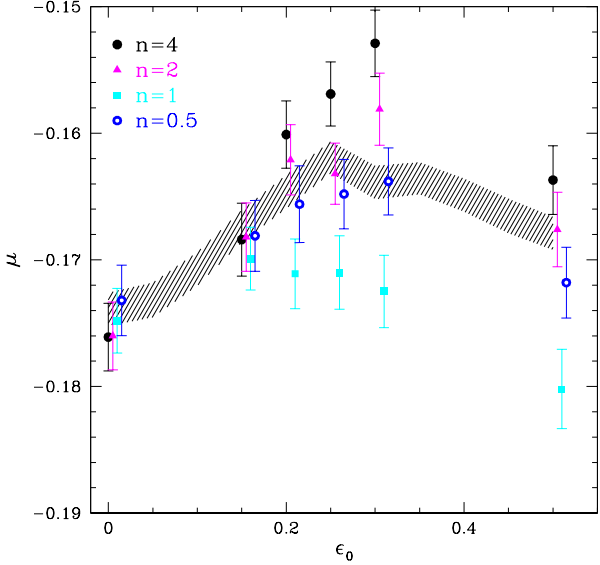}}
\caption{Multiplicative bias as a function of input ellipticity
  distribution for four different Sersic indices (as indicated) for
  galaxies with $20<m_r<25$.  The hatched region indicates the 68\%
  confidence region when the distribution of Sersic indices from GEMS
  is used. The bias depends on the value of the Sersic index, although
  we note that the results cannot be compared directly, as explained
  in the text. \label{fig:bias_profile}}
\end{figure}

Finally we examine whether the bias depends on the assumed
distribution of Sersic indices. To do so, we create images where all
galaxies have the same Sersic index $n$, while keeping the other
parameters the same. The results are presented in
Figure~\ref{fig:bias_profile} for different values of $\epsilon_0$.
The bias depends on the value of $n$, although the range is small
($\sim 0.02$ for $\epsilon_0=0.3$), with the results for $n=1$
(corresponding to exponential profiles) most discrepant. We note,
however, that the half-light radii were kept to the values listed in
the GEMS catalog, which can lead to small changes in the SNR,
complicating a direct comparison. Given the small variation in $\mu$,
and the fact that the observed distribution of Sersic indices is well
constrained, we assume that the uncertainty in this distribution can
be ignored. Hence, the dominant uncertainty in our bias estimate
arises from the uncertainty in the ellipticty distribution.

\subsection{Empirical correction}
\label{sec:correction}

If the simulated data resemble the actual observations sufficiently
well, the average bias for the source sample could be used to adjust
the cluster masses accordingly. Note that one would still have to
determine the bias as a function of seeing. An additional complication
arises, however, because we lack redshifts for the sources: the bias
depends on the fluxes and sizes of the sources. As a result the bias may
be redshift dependent, which is not captured by the image simulations
as the same shear is applied to all sources. In reality the amplitude
of the shear signal depends on the geometry of the lens-source
configuration, quantified by the critical surface density $\Sigma_{\rm
  crit}$ \citep[e.g.][]{Bartelmann01}

\begin{equation}
\Sigma_{\rm crit}=\frac{c^2}{4\pi G}\frac{D_s}{D_l D_{ls}},
\end{equation}

\noindent where $D_s$, $D_l$ and $D_{ls}$ are the angular diameter
distances between the observer and the source, the observer and the
lens, and the lens and source, respectively. The sensitivity to the
source redshift distribution is quantified by the ratio
$\beta=D_{ls}/D_s$. The average shear for an ensemble of galaxies is
proportional to $\langle (1+\mu) \beta\rangle$. If photometric
redshifts for the individual sources are available, the redshift
dependence of the lensing signal can be accounted for on an
object-by-object basis and an average correction for the
multiplicative bias is possible. An alternative route, which we
  take here, is to compute the multiplicative bias using the observed
  properties of individual galaxies. The correction, however, will
  still depend on the intrinsic ellipticity distribution of the
  sources.

We assume that the bias is only a function of signal-to-noise ratio
\citep[e.g.][]{Melchior12,Kacprzak12} and the size relative to that of
the PSF \citep[e.g.][]{Massey13}. We quantify the latter by the
parameter ${\cal R}$, defined as:

\begin{equation}
{\cal R}^2=\frac{r^2_{h,*}}{r^2_{h,{\rm gal}}-r^2_{h,*}},
\end{equation}

\noindent where $r_{h,*}$ denotes the half-light radius of the PSF and
$r_{h,{\rm gal}}$ that of the observed galaxy. Despite being a simple
prescription, we show in \S\ref{sec:test} that this captures the
dependence on PSF size quite well. As a proxy for the signal-to-noise
ratio we take $\nu=1/\sigma_e$, the reciprocal of the uncertainty in
the polarization \citep{Hoekstra00}. We refer the interested reader to
Appendix~\ref{app:correction} for more details about our empirical
correction, which is given by

\begin{equation}
\mu(\nu)=\frac{b(\nu)}{1+\alpha(\epsilon_0) {\cal R}}.\label{eq:correct}
\end{equation}

\noindent The dependence on the resolution parameter ${\cal R}$ is
described by a single parameter $\alpha$ that is a function of
$\epsilon_0$ only. We require three free parameters to describe
dependence of the bias on $\nu$: $b(\nu)=b_0+b_1/\sqrt(\nu)+b_2/\nu$,
with fit parameters that vary smoothly with $\epsilon_0$. The best fit
parameters as a function of $\epsilon_0$ are listed in
Table~\ref{tab:corpar}. Although this parametrization does not
describe the simulated data perfectly, and obvious improvements can be
suggested, we prefer our choice as it provides a sufficently accurate
correction, with a relatively small number of parameters. Including
more parameters did not improve the robustness of the correction.

We find that our parametrization of the bias does not perform well for
galaxies with large observed sizes ($r_h>5$ pixels; see
e.g. Figure~\ref{fig:alpha_par}). As discussed in more detail in
\S\ref{sec:update} we find that the recovered lensing signal for these
galaxies is biased low (see Figure~\ref{fig:re_bgsize}).  Closer
investigation of the simulated data shows that most of these galaxies
are intrinsically small and faint. In some of the cases the sizes are
increased by noise in the images, but a large fraction is blended with
other galaxies.  The large increase in galaxy density in clusters of
galaxies is expected to exacerbate this problem, which is not captured
by our simulations (which are representative of the field). To
minimize the impact of blending, we include only galaxies with $r_h<5$
pixels in the lensing analysis. This size cut is applied to the tests
presented below, as well as our actual measurements.

\begin{figure}
\centering
\leavevmode \hbox{%
\includegraphics[width=8.5cm]{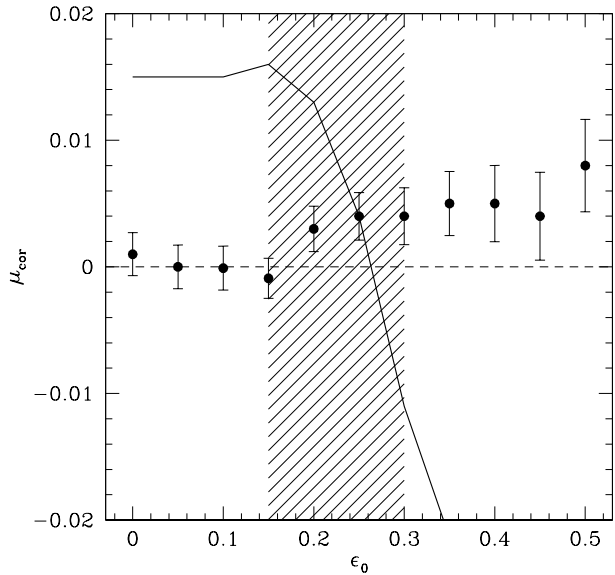}}
\caption{Residual multiplicative bias as a function of input
  ellipticity distribution for galaxies with $22<m_r<25$ and
$r_h<5$ pixels. The black
  points indicate the results when the correct value for $\epsilon_0$
  is used in the correction.  The line indicates the residual bias if
  we assume $\epsilon_0=0.25$ in the correction, instead of the correct
  value for the simulated distribution. Adopting a value $\epsilon_0=0.25$
  for the correction results
  in $|\mu_{\rm cor}|<0.015$ over the expected range in actual 
  $\epsilon_0$ values (indicated by the hatched region).
  \label{fig:bias_cor_e}}
\end{figure}

\subsection{Testing the empirical correction}
\label{sec:test}

To quantify how well the correction works when we apply it to the
simulated data, we first examine the residual bias $\mu_{\rm cor}$ as
a function of $\epsilon_0$. As explained in more detail in
\S\ref{sec:contam}, we restrict the source sample to galaxies with
$22<m_r<25$ to allow for a better correction for the contamination by
cluster members. In addition we apply a size cut, requiring that
  $r_h<5$ pixels. This is motivated by our image simulations where we
  found that the correction for large galaxies is biased, because they
  are blended or too faint to have their shapes measured reliably.
We therefore limit the discussion of the performance of the empirical
correction to this range in apparent magnitude and galaxy
  size. The results are presented in Figure~\ref{fig:bias_cor_e},
which shows that for the range of interest for $\epsilon_0$ (indicated
by the hatched region) $|\mu_{\rm cor}|<0.005$.

As the intrinsic ellipticity distribution remains uncertain, it is
useful to examine the bias that is introduced when an incorrect value
for $\epsilon_0$ is used for the empirical correction. If we use the
parameters corresponding to $\epsilon_0=0.25$ to correct the
measurements from other input distributions we find that $\mu_{\rm
  cor}$ is still small, as indicated by the black line in
Figure~\ref{fig:bias_cor_e}. Our empirical correction is
quite robust against the uncertainty in the input ellipticity
distribution (if we take $\epsilon_0=0.25$).  As discussed
in Appendix~\ref{app:correction} the parametrization for the size
dependence of the bias is not accurate for large galaxies, which are
typically bright. This is indeed reflected in the residual bias as a
function of apparent magnitude: we observe $\mu_{\rm cor}\sim 0.02$
for $m_r<22$, with a bias $\sim 0$ for galaxies with $m_r>22$.

\begin{figure}
\centering
\leavevmode \hbox{%
\includegraphics[width=8.5cm]{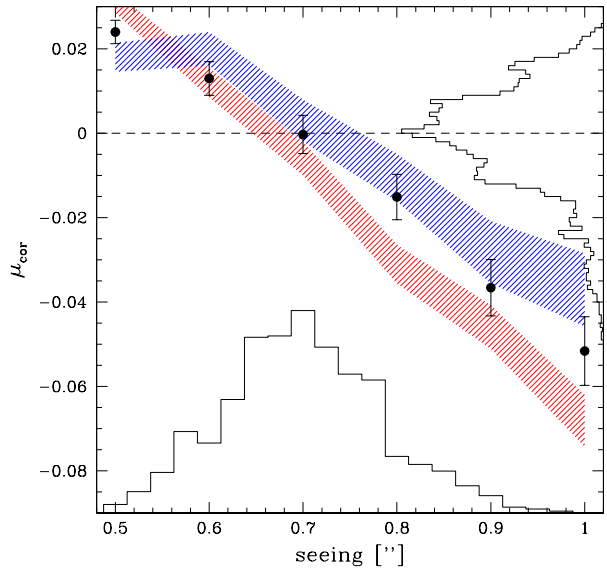}}
\caption{Residual multiplicative bias as a function
  of seeing for sources with $22<m_r<25$. The black points show the
  results for $\epsilon_0=0.25$. The red (blue) hatched regions
  indicate the 68\% confidence region for the bias if we use the parameters for $\epsilon_0=0.25$ to correct
  the simulations with  input distribution with $\epsilon_0=0.15$
 ($\epsilon_0=0.3$) instead. The bottom histogram shows the seeing distribution for each  chip in CCCP, the side histogram shows the corresponding
  distribution of residual bias, with
  $\langle\mu_{cor}\rangle=-0.001$.
  \label{fig:bias_cor_seeing}}
\end{figure}

The empirical correction was determined for a particular PSF and
integration time. Although it is in principle possible to create
simulated data sets for each set of observing conditions, a useful
correction scheme should be more generally applicable. As discussed in
Appendix~\ref{sec:rcs2} we also simulated data from the second
Red-sequence Cluster Survey (RCS2). These data are shallower, but the
results presented in Figure~\ref{fig:rcs2_cor} indicate that the
correction works fairly well for these shallower data. This suggests
that the modeling of the SNR-dependence is adequate.

More interesting is whether our approach to quantify how well galaxies
are resolved, i.e. the choice of ${\cal R}$, can be used for a range
of seeing values. To this end we correct the set of images used to
study the seeing dependence of the bias (see
Fig.~\ref{fig:bias_seeing}).  The results for galaxies with
$22<m_r<25$ are presented in Figure~\ref{fig:bias_cor_seeing}, which
shows $\mu_{\rm cor}$ as a function of the FHWM of the PSF. 
Even for a FWHM of $1''$ the bias is reduced significantly. Nonetheless
the residual bias can still be substantial.  However, as is shown by
the seeing histogram, the CCCP data span a relatively narrow range,
and the mean bias for the full sample is $\langle\mu_{\rm
  cor}\rangle=-0.001$.  Furthermore, the largest FWHM values occur on
chips far away from the cluster location.  We therefore ignore the
seeing dependence, as the residual bias is still much smaller than the
statistical uncertainties for individual clusters, and the ensemble
average bias very small.

We conclude that our empirical correction is adequate to determine
cluster masses for the CCCP sample. Based on the residuals we assign a
systematic uncertainty of 2\% in the cluster masses due to the
uncertainty in the input ellipticity distribution, the limited
exploration of the role of morphology, and the variation in image
quality.

\begin{figure*}
\centering
\leavevmode \hbox{%
\includegraphics[width=8.5cm]{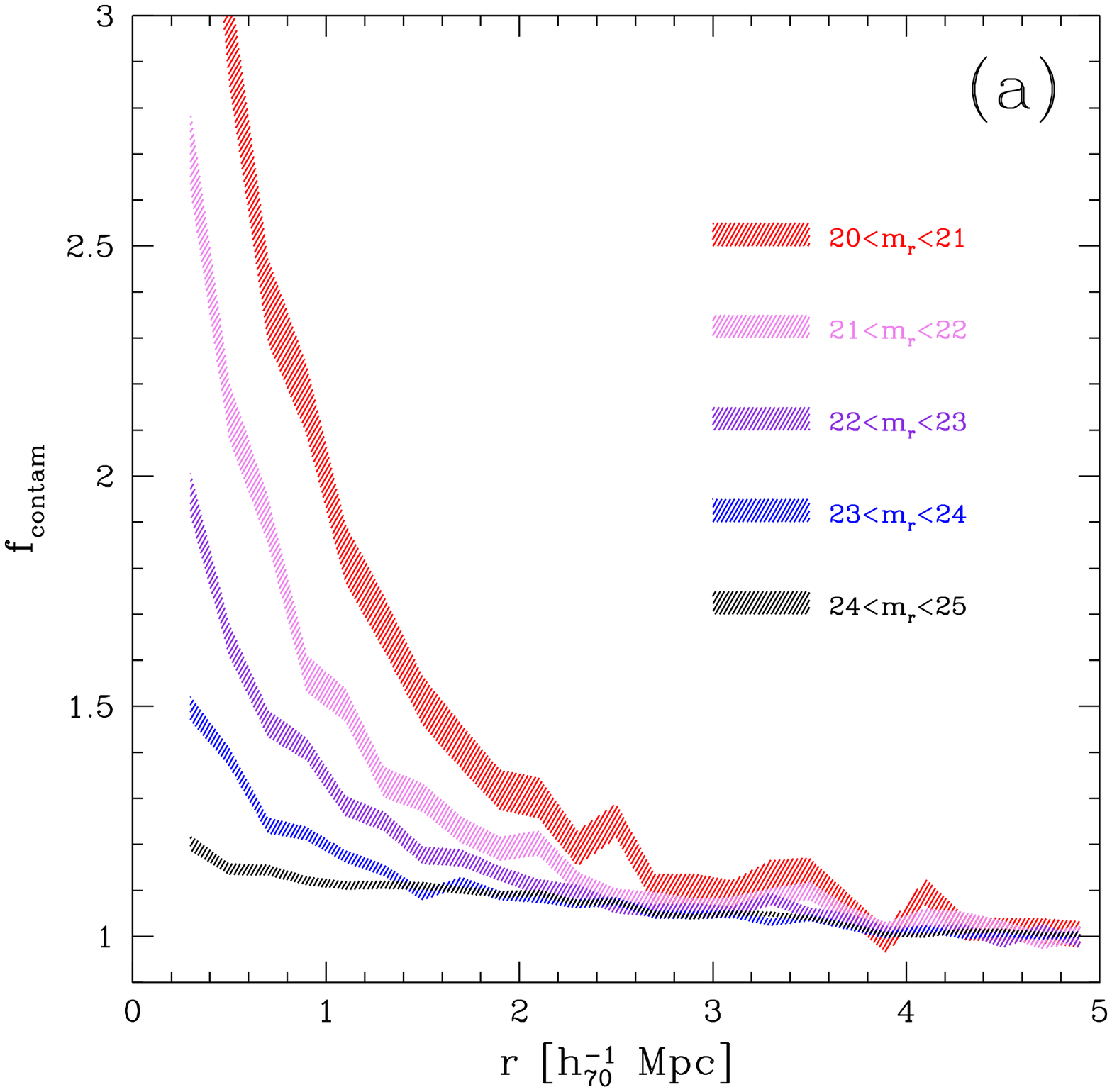}
\includegraphics[width=8.5cm]{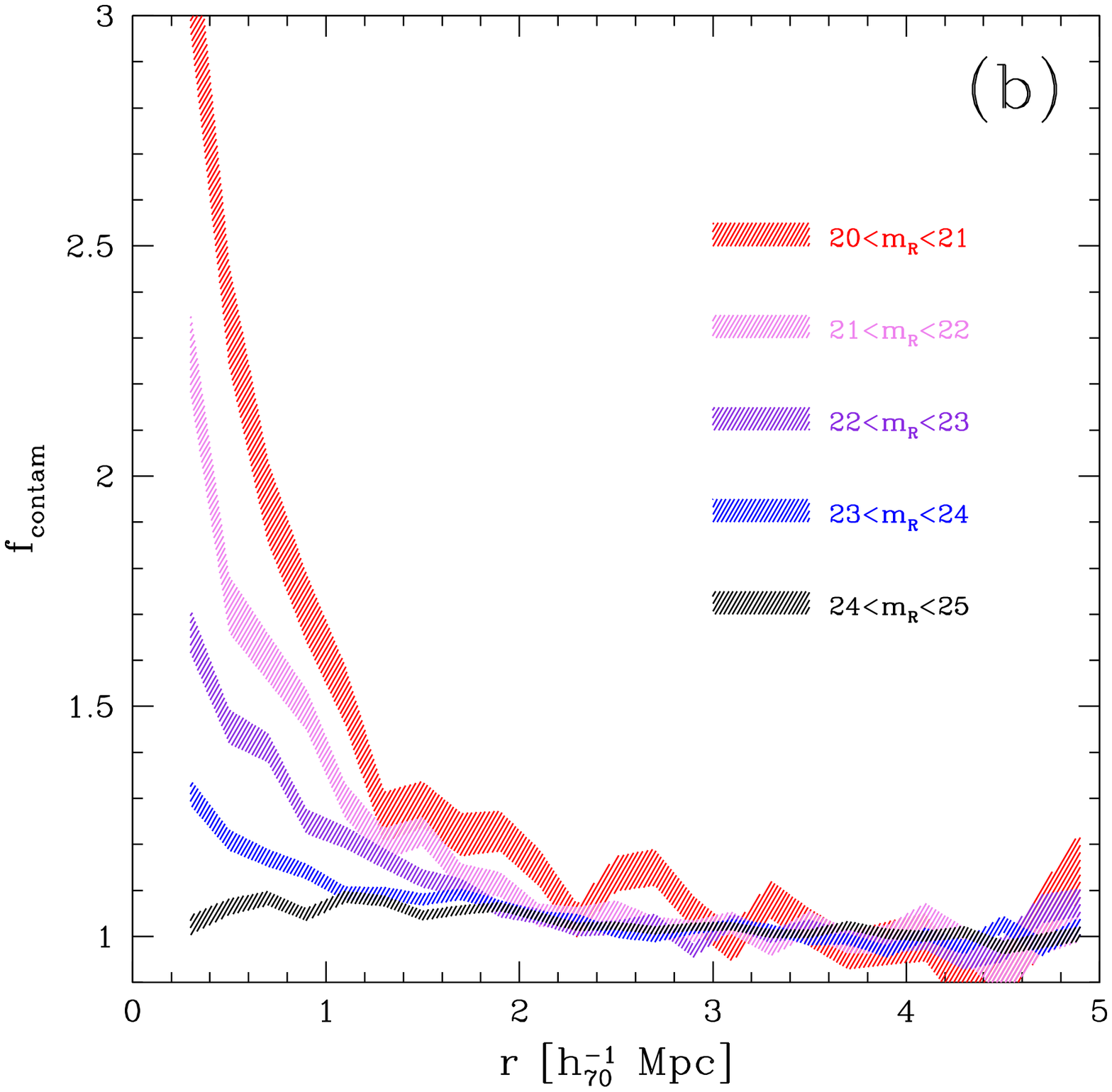}}
\caption{The ensemble averaged correction factor for contamination by
  cluster members as a function of distance for different bins of
  apparent magnitude. Panel~(a) shows the results for clusters
  observed with Megacam, whereas panel~(b) is for the CFH12k
  observations.}  \label{fig:contam_mag}
\end{figure*}
 
\section{Source redshift distribution}\label{sec:redshift}

We lack the color information to derive photometric redshifts for the
individual sources, as opposed to e.g. \cite{Applegate14} and
\cite{Umetsu14}.  Fortunately it is sufficient to know the average
source redshift distribution, which we discuss in more detail in
\S\ref{sec:photoz}. We note, however, that photometric redshifts
enable an optimal weighting of the sources. In particular photometric
redshifts allow for a separation of source galaxies and cluster
members. The latter are unlensed, as are foreground galaxies, and thus
dillute the observed lensing signal by a factor $f_{\rm
    contam}(r)=1+n_{\rm cl}(r)/n_{\rm fld}$, where $n_{\rm cl}(r)$ is
  the number density of cluster galaxies and $n_{\rm fld}$ the number
  density of field galaxies. This correction is especially important
  at small distances from the cluster center. As we describe in
  \S\ref{sec:contam} we can correct for the reduction in signal by
  quantifying the level of contamination. This assumes that the
  orientations of the cluster members are random, which is supported
  by observations \citep{Sifon14}.

  In doing so, we assume that the change in counts is solely caused by
  contamination by cluster members. However, gravitational lensing not
  only changes the galaxy shapes, but also magnifies sources. As a
  result the background sources appear brighter, leading to an
  increase in the observed counts. On the other hand, the actual
  volume is reduced, because the observed solid angle corresponds to a
  smaller solid angle behind the cluster. Consequently, the net change
  depends on the number density of background galaxies as a function
  of apparent magnitude \citep[see e.g. \S3.4 in][]{Mellier99}. In our
  case we observe a slope ${\rm d}\log N_{\rm gal}/{\rm d}M\sim
  0.38-0.4$ for galaxies with $22<m_r<24$. This is somewhat
    steeper than the slope of $\sim 0.33$ observed by \cite{Hogg97} in
    the $R-$band. In either case the net effect is minimal: even for
    $\kappa=0.1$ the change in observed counts is $1-3\%$. Hence, it
  is safe to assume that the excess counts are solely caused by
  contamination by cluster members. Note, however, that the source
  redshift distribution is somewhat changed, as we do see
  intrinsically fainter galaxies.  We verified that the resulting
  change in mean redshift can be safely neglected.

\subsection{Contamination by cluster members}\label{sec:contam}

To reduce contamination by cluster members, H12 used their limited
color information to identify and remove galaxies on the
red-sequence. As shown in \cite{Hoekstra07}, this does lower the
contamination, but only by $\sim 30\%$ as many faint cluster members
are blue. Furthermore H12 assumed that the excess number density of
galaxies can be described as $f_{\rm contam}\propto r^{-1}$, with the
amplitude determined for each cluster. The analysis presented here
differs from H12 in several ways. Rather than applying a color cut, we
restrict the magnitude range of the sources. Furthermore we use a more
flexible model to quantify the radial dependence of the excess counts.
We also correct the excess source counts for the obscuration by
cluster members \citep{Simet14}. Finally, as described below,
  rather than considering the excess counts, we account for the weight
  provided by the uncertainty in the shape measurement.

H12 used sources as bright as $m_r\sim 20$, for which the level of
contamination is high. This is demonstrated by
Figure~\ref{fig:contam_mag}, which shows the corresponding
  correction factor as a function of distance from the cluster for
  different bins of apparent magnitude. For the Megacam data the
  counts are normalised to the average number density at radii larger
  than $r_{\rm max}=4 h_{70}^{-1}$Mpc, i.e. we assume that the level
  of contamination can be ignored at those large radii. This is
  supported by comparison of the observed counts to those in blank
  fields. In addition, we used predictions based on the halo model
  described in \cite{Cacciato13} to estimate the expected level of
  contamination due to neighbouring structures. In line with our
  comparison of the blank field galaxy counts, we find that the
  contribution from local structures can indeed be ignored. The
  field-of-view of the CFH12k data is smaller and we estimate the
  background level using the number density at radii larger than
  $r_{\rm max}=3 h_{70}^{-1}$Mpc.

\begin{figure}
\centering
\leavevmode \hbox{%
\includegraphics[width=8.5cm]{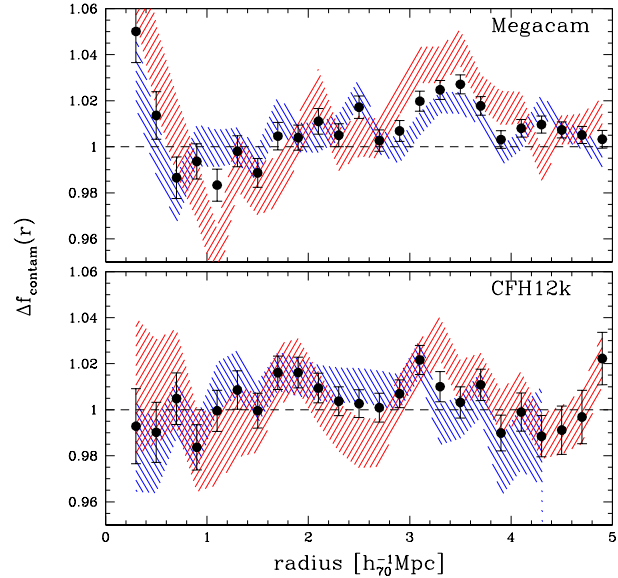}}
\caption{Plot of the ensemble averaged residual contamination as a function
of distance to the cluster center. The
top panel shows the results for sources with $22<m_r<25$ based on
the Megacam data. The hatched regions indicate the 68\% confidence
intervals for the residuals for
clusters with $z<0.25$ (blue) and $z>0.25$ (red). The bottom panel
shows the same, but now for CFH12k data and sources with $22<m_{R_C}<25$.
For both data sets the residual contamination on the scales of
interest ($>0.5 h_{70}^{-1}$Mpc) is at most 2\%.
\label{fig:res_contam}}
\end{figure}

The contamination is much higher for bright galaxies: such galaxies
are rare in the field, but much more common in clusters.  For the
brightest bin $(20<m_r<21)$ the cluster members outnumber source
galaxies 3-to-1 in the inner $\sim 1 h_{70}^{-1}$Mpc. Such a large
level of contamination is difficult to model reliably and for this
reason we decided to increase the bright limit of the source sample to
$m_r=22$ from the typical value of $m_r=20$ used in H12. This leads to
a reduction in excess counts that is comparable to excluding the
galaxies on the red-sequence. Conveniently, the empirical correction
for the multiplicative bias in the shape measurement also performs
better for galaxies with $22<m_r<25$. Furthermore, the lensing signal
is higher for the fainter galaxies. The shapes of brighter
  galaxies are measured better and consequently given more weight in
  the lensing analysis (see Eqn.~\ref{eq:avshear}). Rather than
  correcting for the excess counts, which effectively assumes that the
  weight is uniform, we compute the excess weight as a function of
  radius. This is a minor correction, which increases the masses of 
the parametric NFW fits by $\sim 2-3\%$ (see \S\ref{sec:nfwfit}).

The other important change we make is that we allow the radial profile of the
excess weight to vary from cluster to cluster by introducing a core.
The simple $1/r$ profile used by H12 is not a good description for all
clusters or magnitude bins. Investigation of the ensemble averaged
residuals suggest that it leads to an overestimation of the
contamination in the inner $\sim 500 h_{70}^{-1}$~kpc and an
underestimation by $4-5\%$ at larger radii because the model attempts
to compensate for the poor fit in the cluster cores. To describe the
excess counts we now fit

\begin{equation}
f_{\rm contam}(r)=1+{n_0}\left(\frac{1}{r+r_c} - \frac{1}{r_{\rm max}+r_c}\right),
\end{equation}

\noindent to each cluster, where we take $r_{\rm max}=4
h_{70}^{-1}$Mpc for the Megacam data and $r_{\rm max}=3
h_{70}^{-1}$Mpc for the CFH12k data. The core radius $r_c$ is a free
parameter that we fit for each cluster
separately. Figure~\ref{fig:res_contam} shows the ensemble averaged
residual contamination for sources with $22<m_{r/R_C}<25$ as a
function of radius, suggesting that the systematic uncertainty for the
ensemble of clusters is at most a few percent. Note that the residuals
may be larger for individual clusters, resulting in increased
scatter. However, the results presented in Figure~\ref{fig:res_contam}
suggest that residual contamination will have a minimal impact on the
normalization of scaling relations derived from CCCP measurements and
we adopt a systematic uncertainty in the mass of $2\%$ as a result of
the imperfect correction for contamination by cluster members.

The observed counts are biased low in the inner regions because the
presence of bright cluster members affects our ability to detect and
analyse sources. Although \cite{Simet14} showed that this is an
important source of bias for the measurement of magnification, it may
also lead to a small bias in our estimate of the dillution of the
lensing signal. We simulated the impact of this and as described in
more detail in Appendix~\ref{app:obscure} we find that the impact is
indeed small, boosting the masses from the parametric NFW fits by
$1-2\%$ (see \S\ref{sec:nfwfit}). The aperture masses, which are
discussed in \S\ref{sec:apmass} are not affected, because they rely on
estimates of the lensing signal at large radii where the density of
cluster members is low.

\begin{figure*}
\begin{center}
\leavevmode \hbox{%
\includegraphics[width=8.5cm]{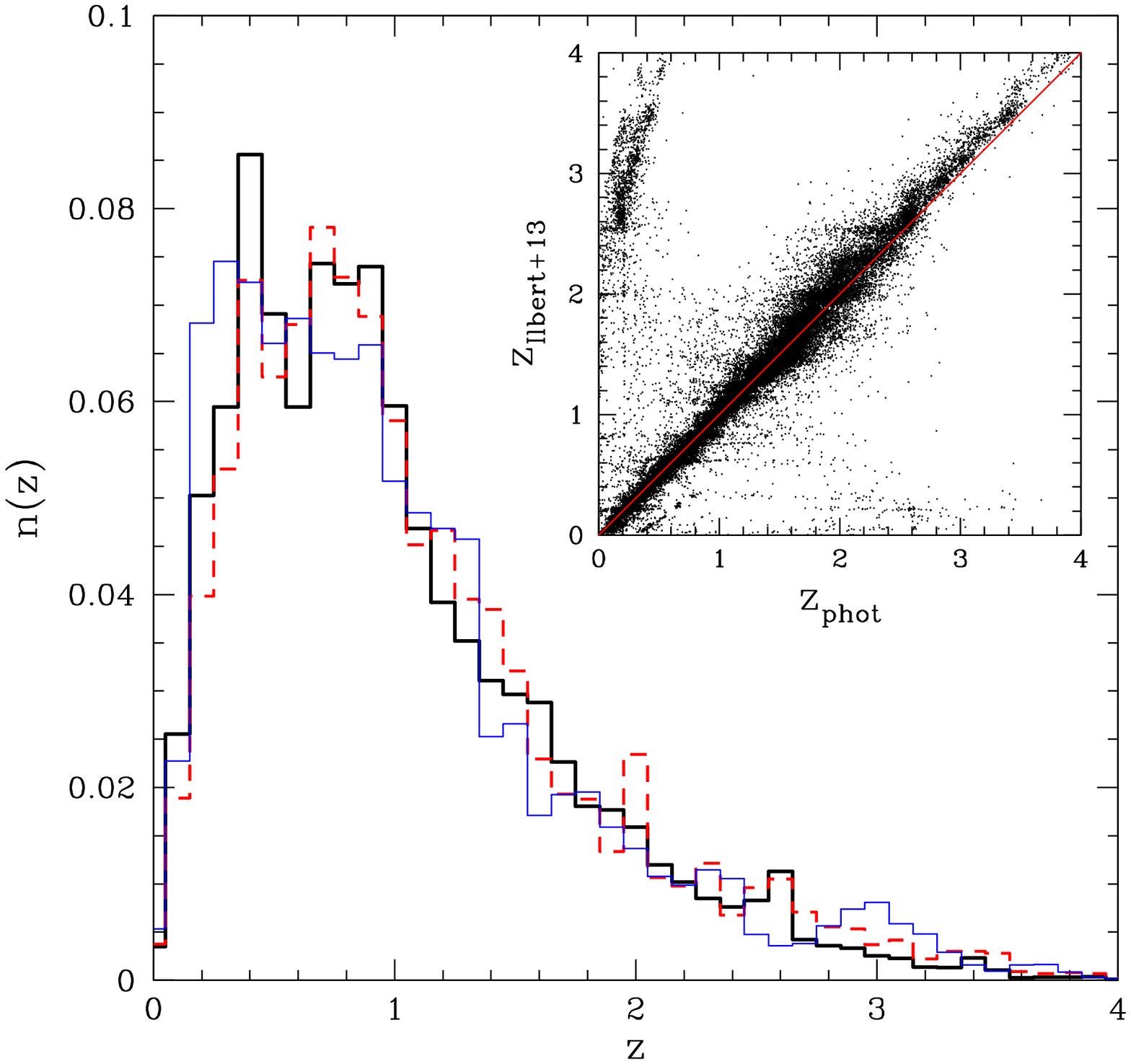}
\includegraphics[width=8.5cm]{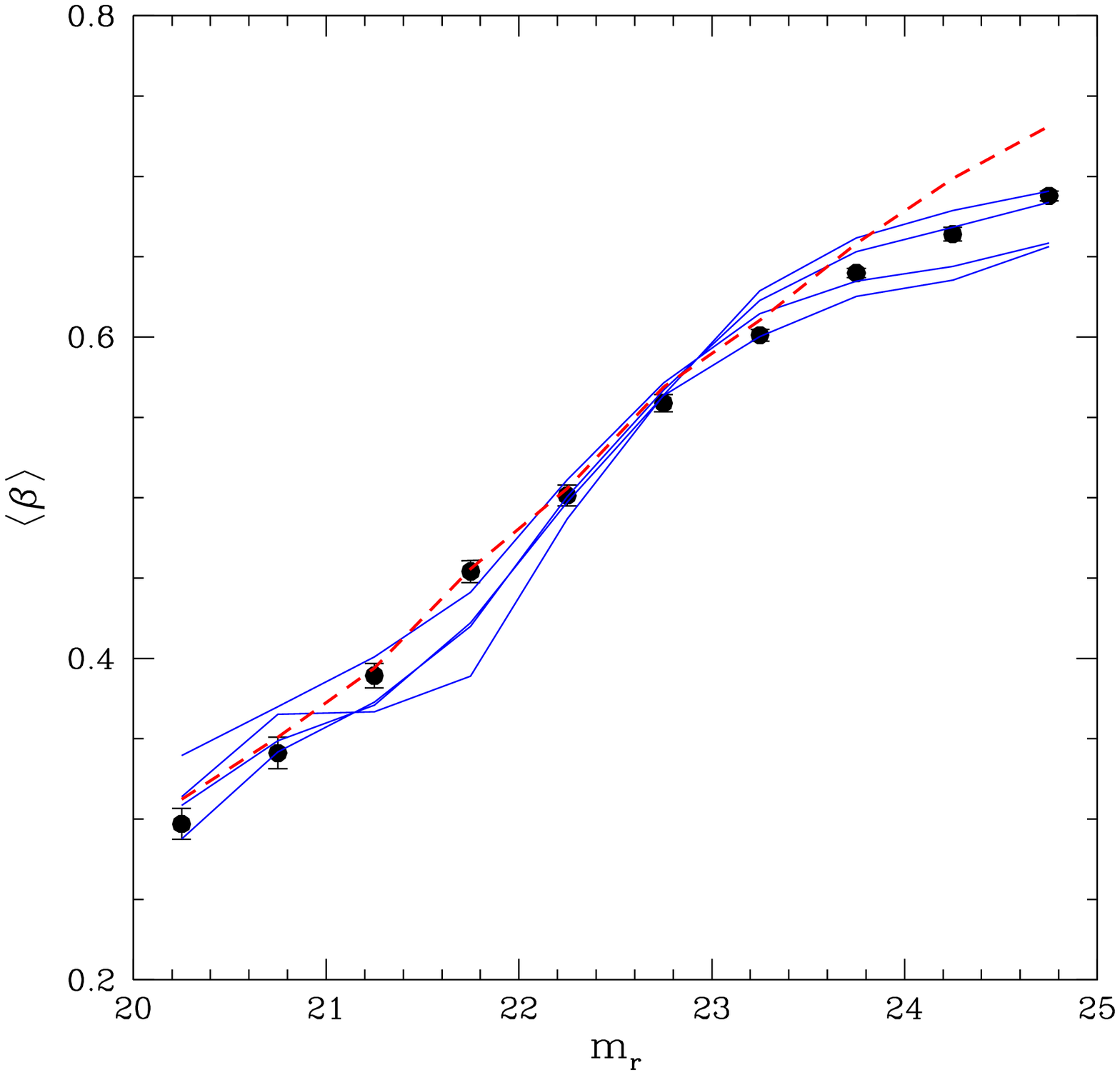}}
\caption{{\it Left panel:} The redshift distribution of galaxies with
  $20<m_r<25$ based on the COSMOS + UltraVISTA photometric redshift
  analysis. The solid black histogram indicates the results from our
  analysis, whereas the red dashed histogram corresponds to the results
  from \citet{Ilbert13}. The inset shows the comparison of the
  photometric redshifts for the galaxies in common. The blue histogram
  shows the redshift distribution used in H12. {\it Right panel:} value of
  $\langle\beta\rangle$ as a function of $m_r$ for a cluster at
  $z=0.2$. The points with error bars are our measurements, which
  agree well with the red dashed line \citep{Ilbert13} for
  $m_r<23$. The four blue lines are the results for the CFHTLS Deep
  fields studied by \citet{Ilbert06} and used in H12.\label{fig:nz_uvista}}
\end{center}
\end{figure*}

\subsection{Photometric redshift catalog}
\label{sec:photoz}

The weak lensing analysis of the initial sample of 20 clusters
observed with the CFH12k camera in \cite{Hoekstra07} used the
photometric redshift distributions derived for the HDF North and South
using the available deep multi-wavelength data \citep{HDF}. However,
the area covered is small, leading to concerns whether the redshift
distributions are representative. For this reason H12 used the
photometric redshift distribution from \cite{Ilbert06}, which is
based on the four CFHT Legacy Survey Deep fields (each field covers
one square degree). However, \cite{Ilbert06} derived photometric
redshifts using observations in 5 optical filters ($ugriz$). Although
these data are very deep, the lack of near-infrared (NIR) data is a
concern for high redshift galaxies. Good quality NIR data are
essential for this purpose because at $z>1.5$ the Balmer and
4000$\AA$~break features in galaxy spectral energy distributions,
which are the strongest features for determining a photometric
redshift, are redshifted into the NIR.

For our analysis we use data from the Cosmic Evolution Survey
\citep[COSMOS;][]{Scoville07}) which observed a single field covering
2 square degree with HST, and for which extensive multi-wavelength and
spectroscopic data are available. \cite{Ilbert09} present photometric
redshifts for these data based on measurements in 30 bands (hereafter
referred to as COSMOS-30). This redshift distribution was used in
\cite{Applegate14} to determine their ``color-cut'' masses. At
the time of the analysis by \cite{Ilbert09}, deep NIR data were not
available. This situation has changed thanks to the UltraVISTA survey,
an ESO public survey performing deep imaging of the COSMOS field in 5
NIR filters \citep[see][for details]{McCracken12}. The UltraVISTA data
are a significant upgrade to the available NIR imaging in the COSMOS
field, and therefore allow for a marked improvement in the quality of
photometric redshifts for galaxies at $z>1.5$. Two public NIR-selected
catalogs have been produced using the UltraVISTA data; one where
galaxies were selected in the $K_{s}$-band \citep{Muzzin13}, and one
where objects were selected using a co-added $\chi^2$ image of the NIR
bands \citep{Ilbert13}. Both of these catalogs provide photometric
redshifts and stellar masses for the galaxies.

\begin{table*}
\begin{center}
\caption{Basic information for the CCCP clusters\label{tabsample}}
\begin{tabular}{lllllllllll}
  \hline
  \hline
  (1) & (2) & (3) & (4) & (5) & (6) & (7) & (8) & (9) & (10) & (11) \\ 
  & name      & $z$ & RA & DEC & mag & $\langle\beta\rangle$ & $\langle\beta\rangle_{\rm I13}$ & $\langle\beta\rangle_{\rm used}$ & $\delta\beta$ & $\langle\beta^2\rangle_{\rm used}$ \\
  &            &     & (J2000.0) & (J2000.0) & [$h_{70}^{-1}$kpc] & & & \\
  \hline
  1  & Abell 68       & 0.255 & $00^{\rm h}37^{\rm m}06.9^{\rm s}$ & $+09^\circ09'24''$ & 22-25   & 0.553 & 0.579 & 0.566 & 0.021 & 0.375 \\
  2  & Abell 209      & 0.206 & $01^{\rm h}31^{\rm m}52.5^{\rm s}$ & $-13^\circ36'40''$ & 22-25   & 0.625 & 0.649 & 0.637 & 0.019 & 0.453 \\
  3  & Abell 267      & 0.230 & $01^{\rm h}52^{\rm m}42.0^{\rm s}$ & $+01^\circ00'26''$ & 22-25   & 0.586 & 0.610 & 0.598 & 0.019 & 0.409 \\
  4  & Abell 370      & 0.375 & $02^{\rm h}39^{\rm m}52.7^{\rm s}$ & $-01^\circ34'18''$ & 22-25   & 0.414 & 0.442 & 0.428 & 0.026 & 0.244 \\
  5  & Abell 383      & 0.187 & $02^{\rm h}48^{\rm m}03.4^{\rm s}$ & $-03^\circ31'44''$ & 22-24.5 & 0.636 & 0.654 & 0.645 & 0.014 & 0.462 \\
  6  & Abell 963      & 0.206 & $10^{\rm h}17^{\rm m}03.8^{\rm s}$ & $+39^\circ02'51''$ & 22-25   & 0.621 & 0.644 & 0.632 & 0.018 & 0.448 \\
  7  & Abell 1689     & 0.183 & $13^{\rm h}11^{\rm m}30.0^{\rm s}$ & $-01^\circ20'30''$ & 22-24.5 & 0.647 & 0.666 & 0.656 & 0.015 & 0.475 \\
  8  & Abell 1763     & 0.223 & $13^{\rm h}35^{\rm m}20.1^{\rm s}$ & $+41^\circ00'04''$ & 22-25   & 0.590 & 0.613 & 0.601 & 0.018 & 0.412 \\
  9  & Abell 2218     & 0.176 & $16^{\rm h}35^{\rm m}48.8^{\rm s}$ & $+66^\circ12'51''$ & 22-24.5 & 0.646 & 0.662 & 0.654 & 0.012 & 0.471 \\
  10 & Abell 2219     & 0.226 & $16^{\rm h}40^{\rm m}19.9^{\rm s}$ & $+46^\circ42'41''$ & 22-25   & 0.596 & 0.621 & 0.609 & 0.020 & 0.421 \\
  11 & Abell 2390     & 0.228 & $21^{\rm h}53^{\rm m}36.8^{\rm s}$ & $+17^\circ41'44''$ & 22-25   & 0.597 & 0.624 & 0.611 & 0.021 & 0.423 \\
  12 & MS 0015.9+1609 & 0.547 & $00^{\rm h}18^{\rm m}33.5^{\rm s}$ & $+16^\circ26'16''$ & 22-25   & 0.277 & 0.304 & 0.291 & 0.025 & 0.138 \\
  13 & MS 0906.5+1110 & 0.170 & $09^{\rm h}09^{\rm m}12.6^{\rm s}$ & $+10^\circ58'28''$ & 22-25   & 0.678 & 0.700 & 0.689 & 0.016 & 0.515 \\
  14 & MS 1224.7+2007 & 0.326 & $12^{\rm h}27^{\rm m}13.5^{\rm s}$ & $+19^\circ50'56''$ & 22-25   & 0.465 & 0.492 & 0.479 & 0.024 & 0.289 \\
  15 & MS 1231.3+1542 & 0.235 & $12^{\rm h}33^{\rm m}55.4^{\rm s}$ & $+15^\circ25'58''$ & 22-25   & 0.587 & 0.614 & 0.600 & 0.021 & 0.412 \\
  16 & MS 1358.4+6245 & 0.329 & $13^{\rm h}59^{\rm m}50.6^{\rm s}$ & $+62^\circ31'05''$ & 22-25   & 0.466 & 0.494 & 0.480 & 0.025 & 0.290 \\
  17 & MS 1455.0+2232 & 0.257 & $14^{\rm h}57^{\rm m}15.1^{\rm s}$ & $+22^\circ20'35''$ & 22-25   & 0.564 & 0.594 & 0.579 & 0.024 & 0.388 \\
  18 & MS 1512.4+3647 & 0.373 & $15^{\rm h}14^{\rm m}22.5^{\rm s}$ & $+36^\circ36'21''$ & 22-25   & 0.427 & 0.458 & 0.442 & 0.027 & 0.256 \\
  19 & MS 1621.5+2640 & 0.428 & $16^{\rm h}23^{\rm m}35.5^{\rm s}$ & $+26^\circ34'14''$ & 22-25   & 0.373 & 0.404 & 0.389 & 0.027 & 0.211 \\
  20 & CL0024.0+1652  & 0.390 & $00^{\rm h}26^{\rm m}35.6^{\rm s}$ & $+17^\circ09'44''$ & 22-25   &  0.393 & 0.420 & 0.407 & 0.025 & 0.226 \\
  \hline
  21 & Abell 115N   & 0.197 & 00$^{\rm h}$55$^{\rm m}$50.6$^{\rm s}$ & $+26^\circ24'38''$ & 22-25 & 0.645 & 0.670 & 0.658 & 0.019 & 0.478 \\
     & Abell 115S   & 0.197 & 00$^{\rm h}$56$^{\rm m}$00.3$^{\rm s}$ & $+26^\circ20'33''$ & 22-25 & 0.645 & 0.670 & 0.658 & 0.019 & 0.478 \\
  22 & Abell 222    & 0.213 & 01$^{\rm h}$37$^{\rm m}$34.0$^{\rm s}$ & $-12^\circ59'29''$ & 22-25 & 0.620 & 0.645 & 0.633 & 0.020 & 0.449 \\
  23 & Abell 223N   & 0.207 & 01$^{\rm h}$38$^{\rm m}$02.3$^{\rm s}$ & $-12^\circ45'20''$ & 22-25 & 0.629 & 0.653 & 0.641 & 0.019 & 0.459 \\
     & Abell 223S   & 0.207 & 01$^{\rm h}$37$^{\rm m}$56.0$^{\rm s}$ & $-12^\circ49'10''$ & 22-25 & 0.629 & 0.653 & 0.641 & 0.019 & 0.459 \\
  24 & Abell 520    & 0.199 & 04$^{\rm h}$54$^{\rm m}$10.1$^{\rm s}$ & $+02^\circ55'18''$ & 22-25 & 0.642 & 0.667 & 0.655 & 0.019 & 0.475 \\
  25 & Abell 521    & 0.253 & 04$^{\rm h}$54$^{\rm m}$06.9$^{\rm s}$ & $-10^\circ13'25''$ & 22-25 & 0.559 & 0.583 & 0.571 & 0.021 & 0.381 \\
  26 & Abell 586    & 0.171 & 07$^{\rm h}$32$^{\rm m}$20.3$^{\rm s}$ & $+31^\circ38'01''$ & 22-25 & 0.668 & 0.687 & 0.678 & 0.014 & 0.501 \\
  27 & Abell 611    & 0.288 & 08$^{\rm h}$00$^{\rm m}$56 8$^{\rm s}$ & $+36^\circ03'24''$ & 22-25 & 0.512 & 0.536 & 0.524 & 0.022 & 0.332 \\
  28 & Abell 697    & 0.282 & 08$^{\rm h}$42$^{\rm m}$57.6$^{\rm s}$ & $+36^\circ21'59''$ & 22-25 & 0.532 & 0.559 & 0.545 & 0.023 & 0.354 \\
  29 & Abell 851    & 0.407 & 09$^{\rm h}$42$^{\rm m}$57.5$^{\rm s}$ & $+46^\circ58'50''$ & 22-25 & 0.391 & 0.418 & 0.405 & 0.026 & 0.224 \\
  30 & Abell 959    & 0.286 & 10$^{\rm h}$17$^{\rm m}$36.0$^{\rm s}$ & $+59^\circ34'02''$ & 22-25 & 0.528 & 0.556 & 0.542 & 0.024 & 0.350 \\
  31 & Abell 1234   & 0.166 & 11$^{\rm h}$22$^{\rm m}$30.0$^{\rm s}$ & $+21^\circ24'22''$ & 22-25 & 0.693 & 0.714 & 0.703 & 0.015 & 0.534 \\
  32 & Abell 1246   & 0.190 & 11$^{\rm h}$23$^{\rm m}$58.8$^{\rm s}$ & $+21^\circ28'50''$ & 22-25 & 0.651 & 0.673 & 0.662 & 0.017 & 0.483 \\
  33 & Abell 1758  & 0.279 &  13$^{\rm h}$32$^{\rm m}$43.5$^{\rm s}$ & $+50^\circ32'38''$ & 22-25 & 0.539 & 0.567 & 0.553 & 0.024 & 0.361 \\
  34 & Abell 1835   & 0.253 & 14$^{\rm h}$01$^{\rm m}$02.1$^{\rm s}$ & $+02^\circ52'43''$ & 22-25 & 0.562 & 0.588 & 0.575 & 0.021 & 0.385 \\
  35 & Abell 1914   & 0.171 & 14$^{\rm h}$26$^{\rm m}$02.8$^{\rm s}$ & $+37^\circ49'28''$ & 22-25 & 0.685 & 0.708 & 0.697 & 0.017 & 0.525 \\
  36 & Abell 1942   & 0.224 & 14$^{\rm h}$38$^{\rm m}$21.9$^{\rm s}$ & $+03^\circ40'13''$ & 22-25 & 0.607 & 0.633 & 0.620 & 0.021 & 0.434 \\
  37 & Abell 2104   & 0.153 & 15$^{\rm h}$40$^{\rm m}$07.9$^{\rm s}$ & $-03^\circ18'16''$ & 22-25 & 0.707 & 0.727 & 0.717 & 0.014 & 0.552 \\
  38 & Abell 2111   & 0.229 & 15$^{\rm h}$39$^{\rm m}$40.5$^{\rm s}$ & $+34^\circ25'27''$ & 22-25 & 0.599 & 0.625 & 0.612 & 0.021 & 0.425 \\
  39 & Abell 2163   & 0.203 & 16$^{\rm h}$15$^{\rm m}$49.0$^{\rm s}$ & $-06^\circ08'41''$ & 22-25 & 0.619 & 0.639 & 0.629 & 0.016 & 0.445 \\
  40 & Abell 2204   & 0.152 & 16$^{\rm h}$32$^{\rm m}$47.0$^{\rm s}$ & $+05^\circ34'33''$ & 22-25 & 0.708 & 0.728 & 0.718 & 0.014 & 0.554 \\
  41 & Abell 2259   & 0.164 & 17$^{\rm h}$20$^{\rm m}$09.7$^{\rm s}$ & $+27^\circ40'08''$ & 22-25 & 0.690 & 0.711 & 0.700 & 0.015 & 0.530 \\
  42 & Abell 2261   & 0.224 & 17$^{\rm h}$22$^{\rm m}$27.2$^{\rm s}$ & $+32^\circ07'58''$ & 22-25 & 0.606 & 0.632 & 0.619 & 0.020 & 0.433 \\
  43 & Abell 2537   & 0.295 & 23$^{\rm h}$08$^{\rm m}$22.2$^{\rm s}$ & $-02^\circ11'32''$ & 22-25 & 0.511 & 0.537 & 0.524 & 0.023 & 0.332 \\
  44 & MS0440.5+0204     & 0.190 & 04$^{\rm h}$43$^{\rm m}$09.9$^{\rm s}$ & $+02^\circ10'19''$ & 22-25 & 0.646 & 0.667 & 0.656 & 0.017 & 0.476 \\
  45 & MS0451.6-0305     & 0.550 & 04$^{\rm h}$54$^{\rm m}$10.8$^{\rm s}$ & $-03^\circ00'51''$ & 22-25 & 0.283 & 0.307 & 0.295 & 0.024 & 0.140 \\
  46 & MS1008.1-1224     & 0.301 & 10$^{\rm h}$10$^{\rm m}$32.3$^{\rm s}$ & $-12^\circ39'53''$ & 22-25 & 0.504 & 0.531 & 0.517 & 0.024 & 0.326 \\
  47 & RXJ1347.5-1145    & 0.451 & 13$^{\rm h}$47$^{\rm m}$30.1$^{\rm s}$ & $-11^\circ45'09''$ & 22-25 & 0.346 & 0.371 & 0.358 & 0.024 & 0.187 \\
  48 & RXJ1524.6+0957    & 0.516 & 15$^{\rm h}$24$^{\rm m}$38.3$^{\rm s}$ & $+09^\circ57'43''$ & 22-25 & 0.297 & 0.321 & 0.309 & 0.024 & 0.150 \\
  49 & MACS J0717.5+3745 & 0.548 & 07$^{\rm h}$17$^{\rm m}$30.4$^{\rm s}$ & $+37^\circ45'38''$ & 22-25 & 0.269 & 0.291 & 0.280 & 0.022 & 0.131 \\
  50 & MACS J0913.7+4056 & 0.442 & 09$^{\rm h}$13$^{\rm m}$45.5$^{\rm s}$ & $+40^\circ56'29''$ & 22-25 & 0.366 & 0.393 & 0.380 & 0.026 & 0.203 \\
  51 & CIZA J1938+54     & 0.260 & 19$^{\rm h}$38$^{\rm m}$18.1$^{\rm s}$ & $+54^\circ09'40''$ & 22-25 & 0.550 & 0.574 & 0.562 & 0.021 & 0.371 \\
  52 & 3C295             & 0.460 & 14$^{\rm h}$11$^{\rm m}$20.6$^{\rm s}$ & $+52^\circ12'10''$ & 22-25 & 0.343 & 0.368 & 0.356 & 0.025 & 0.185 \\
  \hline
  \hline
\end{tabular}
\bigskip
\begin{minipage}{\linewidth}
{\footnotesize Column 2: cluster name; Column 3: cluster redshift;
  Column 4,5: right ascension and declination (J2000.0) of the adopted
  cluster center. In all but four cases (Abell~520, Abell~851, Abell 1758 and
  Abell 1914) we take this to be the position of the brightest cluster
  galaxy (BCG). Column 6: magnitude range used for the source galaxies. 
  For clusters $1-20$ this is the $R_C$ filter and $r'$ for the remaining clusters; 
  Column 7: the average value of $\beta=D_{ls}/D_{s}$ based on the photo-z analysis presented
  here; Column 8: the values for $\beta$ measured as 
  described in \cite{Ilbert13}; Column~9: the value for $\beta$ we use
  to estimate masses, which is the average of the two measurements;
  Column~10: estimate for the systematic uncertainty in $\beta$ as described in the text;
  Column~11: average value for $\langle\beta^2\rangle$.}
\end{minipage}
\end{center}
\end{table*}

Galaxies have a wide range of optical-NIR colors, and therefore
NIR-selected samples of galaxies are typically quite different from
$r$-selected samples, particularly for the high-redshift end of the
distribution. Consequently we cannot simply use the available
photometric redshift catalogs, because our source galaxies are
selected from the deep CCCP $r$-band imaging. Furthermore, many of the
sources that are of interest for the lensing analysis may be missing
because the UltraVISTA data ($K_{s} = 23.9$ AB) are 1.5 magnitudes
shallower than the CCCP optical imaging.

In order to construct a representative photometric redshift
distribution for the sources, we created a new $r^{+}$-selected
catalog of the COSMOS/UltraVISTA field using the Subaru imaging of the
field from \cite{Capak07}. The Subaru $r^{+}$ imaging has good image
quality (FWHM $\sim$ 0$\farcs$8) and reaches a 5$\sigma$ depth of
$\sim$ 26.5 AB.  This is approximately a full magnitude deeper than
the CCCP imaging, and therefore provides a complete sample of galaxies
to $m_r$ $\sim$ 25.5 with well-measured spectral energy distributions, a
prerequisite for calculating photometric redshifts. The
$r^{+}$-selected catalog was constructed in the identical manner as
the $K_{s}$-selected catalog described in \cite{Muzzin13} and we refer
the reader to that paper for complete details of the catalog
construction.  In brief, the catalog consists of photometry in 29
photometric bands ranging from $0.15\micron - 24\micron$ and
incorporates the available GALEX \citep{Martin05}, Subaru
\citep{Capak07}, UltraVISTA \citep{McCracken12}, and Spitzer data
\citep{Sanders07}.  Images were PSF-matched and photometry was
performed in fixed 2$\farcs$1 diameter apertures.

Photometric redshifts for all galaxies were calculated using the EAZY
photometric redshift code \citep{Brammer08}, which determines
photometric redshifts using linear combinations of multiple templates
as well as a template error function to account for data/template
mismatch.  EAZY is well-tested and performs well amongst the best of
publicly-available photometric redshift codes
\citep[e.g.][]{Hildebrandt10}.  The photometric redshifts were further
refined by determining small offsets to the photometric zeropoints
using the $\sim$ 5000 spectroscopic redshifts available in the COSMOS
field from the zCOSMOS-10k sample \citep{Lilly09}.  The process is
iterative, and the final photometric catalog contains photometric
redshifts accurate to $\Delta z/(1 + z) = 0.01$, with a catastrophic
outlier fraction of $\sim 1\%$.  This estimate of the accuracy is
based on the zCOSMOS spectroscopic redshifts, which are primarily
bright galaxies at $z < 1.5$.  At $z > 1.5$ it is more difficult to
assess how accurate the photometric redshifts are, due to the lack of
a spectroscopic calibration sample.  Small numbers of spectroscopic
redshifts are available from various NIR spectroscopic surveys, and
those suggest that the accuracy at $z > 1.5$ for bright galaxies is
only slightly worse, of order a few percent. We have made the full
$r^{+}$-selected catalog publicly available on the $K_{s}$-selected
catalog
website\footnote{http://www.strw.leidenuniv.nl/galaxyevolution/ULTRAVISTA/}.
Also included is a simplified version of that catalog that can be used
for a quick calculation of photometric redshift distributions for
future lensing analyses.

The left panel in Figure~\ref{fig:nz_uvista} shows the resulting
redshift distribution for galaxies with $20<m_r<25$ using our $r$-band
selected photometric redshift catalog. The red dashed histogram shows
the redshift distribution of the galaxies matched to the NIR-selected
catalog from \cite{Ilbert13}, which makes use of the $\sim$ 3000
unpublished zCOSMOS-deep spectroscopic redshifts. For reference, the
blue histogram shows the corresponding redshift distribution for the
CFHTLS Deep fields from \cite{Ilbert06}, which was used by H12. The
inset panel shows a direct comparison of the photometric redshifts
derived here and those from \cite{Ilbert13} who used a different
algorithm and measured the photometry independently. The overall
agreement is remarkably good for the galaxies in common; only for
$m_r>24$ are some of the galaxies assigned a high redshift in the
catalog from \cite{Ilbert13} and a low redshift by EAZY. These
represent $2\%$ of the total sample of sources and about $20\%$ of the
galaxies for which \cite{Ilbert13} find $z>2$. The excellent agreement
for most of the galaxies is a demonstration of the quality of the
data, both in terms of depth and wavelength coverage.

The impact of the differences in source redshift distributions on the
cluster mass estimates is quantified in the right panel of
Figure~\ref{fig:nz_uvista}, which shows the lensing efficiency
$\langle\beta\rangle$ as a function of apparent magnitude. Our results
are indicated by the black points, whereas the red dashed curve
corresponds to the redshift distribution from \cite{Ilbert13}. For
$m_r<23$ the agreement is very good, whereas the higher number of
$z>2.5$ galaxies in the \cite{Ilbert13} catalog results in a higher
value for $\langle\beta\rangle$, and thus a lower mass.  We list
$\langle\beta\rangle$ for the sources in Table~\ref{tabsample};
Column~7 lists the values for our analysis of the COSMOS and
UltraVISTA data, and Column~8 lists the results using the results from
\cite{Ilbert13}. To determine cluster masses we use
$\langle\beta\rangle_{\rm used}$ provided in Column~9, which is the
average of the values obtained for the two redshift distributions. The
value for $\langle\beta^2\rangle_{\rm used}$, which is a measure of
the width of the distribution \citep[see][for
details]{Hoekstra00,Hoekstra07} is also an average of the two
estimates\footnote{The values listed here are corrected for an error
  in the calculation of $\langle \beta^2\rangle$ that reduced the
  senstitivity to the convergence in our previous work.}. These
  numbers include a size cut similar to the one applied to our lensing
  data. The red line in Figure~\ref{fig:re_bgsize} shows how
  $\langle\beta\rangle$ depends on the observed half-light radius; we
  find that the dependence on size is very small, and we therefore
  conclude that the size cuts do not introduce a significant bias.

The significant differences in redshift distribution demonstrate that
the lack of reliable photometric redshift estimates remains a key
source of error. The unique range in wavelength and quality of the
COSMOS+UltraVISTA data are a major step forward, but without complete
spectroscopic coverage, the uncertainty at the highest redshifts
remains. Furthermore, cosmic variance can still be important for a
single field \citep[see e.g.][for estimates]{Hoekstra11}. Assigning a
systematic uncertainty remains difficult, but we use the difference
between our photometric redshifts and those from \cite{Ilbert13}
(indicated as I13 in Table~\ref{tabsample}) as an estimate for the
systematic uncertainty.

The blue lines in the right panel of Figure~\ref{fig:nz_uvista} show
$\langle\beta\rangle$ for the four CFHTLS Deep fields
\citep{Ilbert06}.  The main difference occurs at $m_r\sim 21.5$,
although the average is also lower than the new estimates for faint
galaxies. We use the variation between the fields to estimate the
contribution of cosmic variance for our redshift distribution. This is
conservative, because the COSMOS field itself is larger than each of
the CFHTLS Deep fields. The systematic uncertainty $\delta\beta$
listed in Table~\ref{tabsample} is the sum of the dispersion measured
from the four fields and $|\beta_{\rm I13}-\beta|/2$. These amount to
$\sim 3\%$ for clusters at $z=0.2$, increasing to $\sim 8\%$ for
$z=0.55$.

Variation in the actual source redshift distribution behind a cluster
leads to an increase in the statistical uncertainty. The impact
of this was studied by \cite{Hoekstra11} using simulations. Their
findings suggest that the lack of redshift information for the
individual sources increases the statistical uncertainty in the mass
by $\sim 3\%$ for a cluster at $z=0.2$ and by $\sim 10\%$ at
$z=0.6$. 

\section{Updated Mass Estimates}
\label{sec:update}

We use our new insights in the shape measurement bias and the source
redshift distribution to update the mass estimates provided by
H12. The only change we made to the original shape measurements is a
corrected estimate for $\sigma_e$, the uncertainty in the
polarization\footnote{The older version of the code included an
  incorrect treatment of the Poisson noise.} because we use this
quantity in our estimate of the multiplicative bias. This correction
also affects the weight defined by Eqn.~\ref{eq:avshear}, although the
impact is very minor. The star-galaxy separation is somewhat more
conservative: previously we included faint galaxies with sizes
comparable to the PSF in the lensing analysis, although they were
severely downweighted in practice. As the correction scheme requires
sizes larger than the PSF, and the fact that these objects do not
contribute much to the average signal, we now only select objects
larger than the PSF in the object catalog. The star selection itself
was not changed, and consequently we use the same PSF model
parameters. 

\begin{figure}
\centering
\leavevmode \hbox{%
\includegraphics[width=8.5cm]{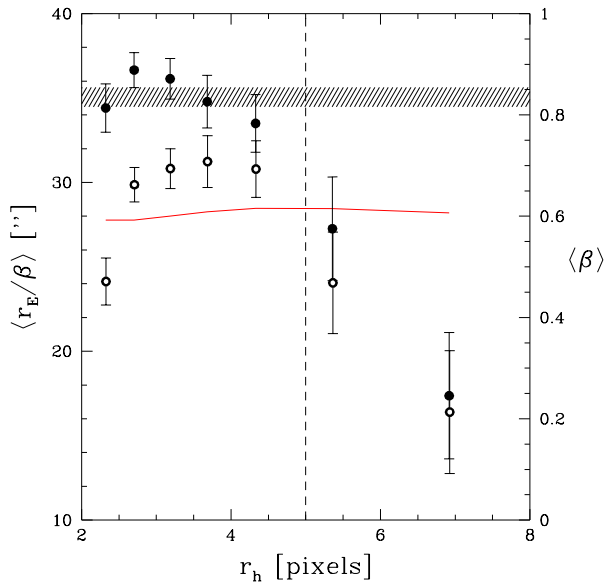}}
\caption{The ensemble averaged lensing signal $\langle
    r_E/\beta\rangle$ as a function of observed source half-light
    radius $r_h$. The value for the Einstein radius $r_E$ was obtained
    by fitting a SIS model to the lensing signal at radii
    $0.75-4h_{70}^{-1}$~Mpc.  The solid points, which should not
    depend on source size, indicate the results when we apply our
    correction for the multiplicative bias, whereas the open points
    are the uncorrected values. The lensing signal is biased low for
    galaxies with $r_h>5$, similar to what we observed in our
    simulated data. The red line indicates the estimate of
    $\langle\beta\rangle$ as a function of source size (the values are
    indicated on the right-most vertical axis), which does not vary
    significantly with object size, suggesting that the adopted
    redshift distribution is not very sensitive to the size cuts
    applied.
  \label{fig:re_bgsize}}
\end{figure}

We apply our empirical correction given by Eqn.~\ref{eq:correct} to
the object catalogs using the observed values of $r_h$ and $\nu$ and
recompute the tangential shear profile as a function of radius, taking
the cluster centers used in H12 (Table~1 of that paper). Based on our
image simulations we also apply a size cut $r_h<5$ pixels. The solid
points in Figure~\ref{fig:re_bgsize} show the lensing signal, averaged
over the full ensemble of clusters, as a function of the observed
half-light radius of the sources. The signal is quantified by the
Einstein radius $r_E$ obtained from a fit to the measurements between
$0.5-4h_{70}^{-1}$~Mpc. We correct each bin for the mean $\beta$,
which does not vary significantly with source size (indicated by the
red line), suggesting that our results are not sensitive to the size
cuts. The resulting value of $r_E/\beta$ should not depend on the
source size.

The observations support our finding from the image simulations that
the signal is biased low for galaxies with $r_h>5$ pixels. For
reference we also show the results if we do not apply our empirical
correction for multiplicative bias (open points). The corrected
results are consistent with a constant signal, but the uncorrected
measurements are bias low for the smallest sources.

We update the correction for Galactic extinction using the
\cite{Schlafly11} recalibration of the \cite{Schlegel98}
infrared-based dust map. This reduces the correction for clusters in
highly extinguished regions.  To ensure a more robust correction for
the contamination by cluster members we do not apply a color cut, but
instead limit the source sample to galaxies with $22<m_{r/R_C}<25$.
In the case of the CFH12k data, H12 used galaxies fainter than
$R_C=25$, for which the redshift distribution is not well
constrained. We now limit the analysis to galaxies brighter than
$R_C=25$.

\begin{table*}
\begin{center}
\caption{Weak lensing mass estimates for the CCCP sample\label{tab:masses}}
\begin{tabular}{lccccccccccc}
\hline
(1) & (2) & (3) & (4) & (5) & (6) & (7) & (8) & (9) & (10) & (11) \\ 
name   &         $\sigma$ & $M^{\rm proj}_{0.5}$ & $M^{\rm proj}_{1.0}$ & $r_{2500}$ & $M^{\rm ap}_{2500}$ & $r_{500}$ & $M^{\rm ap}_{500}$ & $M_{\rm vir}^{\rm NFW}$ & $M_{2500}^{\rm NFW}$ & $M_{500}^{\rm NFW}$ \\   
\hline
Abell~68 & $1117^{+67}_{-71}$ & $4.5\pm0.5$ & $8.3\pm1.3$ & $552$ & $3.1^{+0.4}_{-0.4}$ & $1380$ & $9.7^{+1.9}_{-2.0}$ & $12.9^{+2.7}_{-2.7}$ & $3.0^{+0.6}_{-0.6}$ & $7.5^{+1.6}_{-1.6}$ \\
Abell~209 & $970^{+78}_{-84}$ & $3.2\pm0.5$ & $7.3\pm1.2$ & $490$ & $2.1^{+0.4}_{-0.4}$ & $1230$ & $6.5^{+1.4}_{-1.3}$ & $8.4^{+2.2}_{-2.4}$ & $2.0^{+0.5}_{-0.6}$ & $4.9^{+1.3}_{-1.4}$ \\
Abell~267 & $1006^{+85}_{-92}$ & $3.9\pm0.5$ & $7.2\pm1.4$ & $525$ & $2.6^{+0.4}_{-0.4}$ &$1203$ & $6.3^{+1.8}_{-1.6}$ &$7.9^{+2.4}_{-2.3}$ &$1.9^{+0.6}_{-0.6}$ & $4.7^{+1.4}_{-1.3}$ \\
Abell~370 & $1489^{+75}_{-79}$ & $6.9\pm0.7$ & $15.3\pm1.6$ & $638$ & $5.5^{+0.8}_{-0.8}$ &$1637$ & $18.5^{+2.3}_{-2.3}$ &$30.4^{+5.5}_{-5.3}$ &$6.5^{+1.2}_{-1.1}$ & $17.5^{+3.2}_{-3.1}$  \\
Abell~383 & $821^{+117}_{-135}$ & $2.5\pm0.6$ & $6.1\pm1.3$ & $430$ & $1.4^{+0.6}_{-0.5}$ &$1217$ & $6.2^{+2.2}_{-2.2}$ &$5.8^{+2.6}_{-2.4}$ &$1.4^{+0.6}_{-0.6}$ & $3.4^{+1.5}_{-1.4}$  \\
Abell~963 & $1106^{+76}_{-82}$ & $3.4\pm0.5$ & $6.6\pm1.4$ & $506$ & $2.3^{+0.5}_{-0.4}$ &$1185$ & $5.8^{+1.5}_{-1.5}$ &$12.3^{+3.0}_{-3.0}$ &$2.9^{+0.7}_{-0.7}$ & $7.1^{+1.7}_{-1.7}$ \\
Abell~1689 & $1429^{+59}_{-62}$ & $7.0\pm0.5$ & $13.2\pm1.4$ & $702$ & $5.9^{+0.7}_{-0.7}$ &$1571$ & $13.3^{+2.4}_{-2.2}$ &$30.9^{+5.0}_{-4.8}$ &$6.6^{+1.1}_{-1.0}$ & $17.3^{+2.8}_{-2.7}$ \\
Abell~1763 & $1229^{+70}_{-75}$ & $4.9\pm0.6$ & $10.0\pm1.4$ & $604$ & $3.9^{+0.6}_{-0.6}$ &$1511$ & $12.3^{+3.2}_{-2.9}$ &$16.9^{+3.5}_{-3.5}$ &$3.8^{+0.8}_{-0.8}$ & $9.7^{+2.0}_{-2.0}$ \\
Abell~2218 & $1181^{+77}_{-82}$ & $5.0\pm0.6$ & $8.7\pm1.4$ & $630$ & $4.2^{+0.7}_{-0.7}$ &$1379$ & $8.9^{+2.2}_{-2.1}$ &$16.4^{+3.8}_{-3.6}$ &$3.7^{+0.9}_{-0.8}$ & $9.4^{+2.2}_{-2.1}$ \\
Abell~2219 & $1041^{+75}_{-80}$ & $4.1\pm0.5$ & $9.5\pm1.4$ & $552$ & $3.0^{+0.5}_{-0.5}$ &$1408$ & $10.0^{+2.0}_{-1.8}$ &$11.3^{+2.5}_{-2.5}$ &$2.7^{+0.6}_{-0.6}$ & $6.6^{+1.5}_{-1.5}$ \\
Abell~2390 & $1331^{+61}_{-64}$ & $4.9\pm0.5$ & $9.9\pm1.3$ & $602$ & $3.9^{+0.5}_{-0.5}$ &$1351$ & $8.8^{+1.5}_{-1.5}$ &$23.1^{+3.8}_{-3.6}$ &$5.1^{+0.8}_{-0.8}$ & $13.2^{+2.1}_{-2.1}$ \\
MS~0015.9+1609 & $1456^{+117}_{-127}$ & $6.8\pm0.6$ & $17.9\pm1.9$ & $601$ & $5.6^{+0.7}_{-0.8}$ &$1617$ & $21.8^{+3.2}_{-3.2}$ &$28.9^{+7.0}_{-6.8}$ &$6.2^{+1.5}_{-1.5}$ & $16.9^{+4.1}_{-4.0}$ \\
MS~0906.5+1110 & $1077^{+70}_{-74}$ & $3.8\pm0.5$ & $8.7\pm1.2$ & $549$ & $2.8^{+0.5}_{-0.5}$ &$1423$ & $9.7^{+1.4}_{-1.6}$ &$12.6^{+2.9}_{-2.8}$ &$2.9^{+0.7}_{-0.6}$ & $7.3^{+1.7}_{-1.6}$ \\
MS~1224.7+2007 & $860^{+118}_{-136}$ & $1.8\pm0.6$ & $2.4\pm1.8$ & $345$ & $0.8^{+0.4}_{-0.7}$ &$782$ & $1.9^{+1.0}_{-0.9}$ &$4.9^{+2.3}_{-2.1}$ &$1.2^{+0.6}_{-0.5}$ & $3.0^{+1.4}_{-1.3}$ \\
MS~1231.3+1542 & $590^{+115}_{-141}$ & $1.0\pm0.5$ & $0.4\pm1.3$ & $344$ & $0.7^{+0.2}_{-0.2}$ &$565$ & $0.6^{+0.5}_{-0.4}$ &$1.9^{+1.3}_{-1.1}$ &$0.5^{+0.3}_{-0.3}$ & $1.1^{+0.8}_{-0.7}$ \\
MS~1358.4+6245 & $1167^{+74}_{-79}$ & $4.3\pm0.5$ & $8.7\pm1.5$ & $529$ & $3.0^{+0.5}_{-0.5}$ &$1291$ & $8.6^{+2.0}_{-2.0}$ &$13.4^{+3.2}_{-3.1}$ &$3.1^{+0.7}_{-0.7}$ & $7.8^{+1.9}_{-1.8}$ \\
MS~1455.0+2232 & $1131^{+63}_{-66}$ & $3.7\pm0.4$ & $7.2\pm1.2$ & $510$ & $2.5^{+0.4}_{-0.4}$ &$1158$ & $5.7^{+1.2}_{-1.1}$ &$13.2^{+2.4}_{-2.6}$ &$3.1^{+0.6}_{-0.6}$ & $7.7^{+1.4}_{-1.5}$ \\
MS~1512.4+3647 & $733^{+111}_{-130}$ & $1.5\pm0.6$ & $4.5\pm1.4$ & $282$ & $0.5^{+0.3}_{-0.3}$ &$853$ & $2.6^{+1.5}_{-1.6}$ &$3.9^{+1.7}_{-1.7}$ &$1.0^{+0.4}_{-0.4}$ & $2.4^{+1.1}_{-1.1}$ \\
MS~1621.5+2640 & $1300^{+83}_{-89}$ & $4.9\pm0.6$ & $11.4\pm1.7$ & $543$ & $3.6^{+0.8}_{-0.8}$ &$1286$ & $9.5^{+2.0}_{-1.9}$ &$19.1^{+4.2}_{-4.1}$ &$4.3^{+0.9}_{-0.9}$ & $11.2^{+2.5}_{-2.4}$ \\
CL~0024.0+1652 & $1311^{+94}_{-101}$ & $5.6\pm0.6$ & $11.4\pm1.7$ & $571$ & $4.0^{+0.6}_{-0.6}$ &$1333$ & $10.2^{+2.4}_{-2.2}$ &$24.4^{+5.7}_{-5.4}$ &$5.3^{+1.3}_{-1.2}$ & $14.1^{+3.3}_{-3.1}$ \\
\hline
Abell~115N & $833^{+89}_{-99}$ & $1.4\pm0.4$ & $5.3\pm1.1$ & $283$ & $0.4^{+0.3}_{-0.4}$ &$1098$ & $4.6^{+1.0}_{-1.1}$ &$5.9^{+2.0}_{-2.0}$ &$1.5^{+0.5}_{-0.5}$ & $3.5^{+1.2}_{-1.2}$\\
Abell~115S & $859^{+82}_{-91}$ & $2.6\pm0.4$ & $5.9\pm1.2$ & $416$ & $1.2^{+0.4}_{-0.5}$ &$1127$ & $5.0^{+1.3}_{-1.2}$ &$7.0^{+2.0}_{-2.0}$ &$1.7^{+0.5}_{-0.5}$ & $4.1^{+1.2}_{-1.2}$\\
Abell~222 & $916^{+85}_{-93}$ & $2.9\pm0.5$ & $6.9\pm1.4$ & $450$ & $1.6^{+0.6}_{-0.8}$ &$1174$ & $5.7^{+1.5}_{-1.3}$ &$6.4^{+2.1}_{-2.1}$ &$1.6^{+0.5}_{-0.5}$ & $3.8^{+1.3}_{-1.3}$\\
Abell~223N & $989^{+80}_{-86}$ & $3.0\pm0.5$ & $7.5\pm1.3$ & $463$ & $1.7^{+0.5}_{-0.6}$ &$1236$ & $6.6^{+1.3}_{-1.3}$ &$8.9^{+2.5}_{-2.5}$ &$2.1^{+0.6}_{-0.6}$ & $5.2^{+1.5}_{-1.5}$\\
Abell~223S & $923^{+90}_{-99}$ & $3.0\pm0.5$ & $8.3\pm1.1$ & $466$ & $1.8^{+0.5}_{-0.5}$ &$1370$ & $9.0^{+1.5}_{-1.5}$ &$7.8^{+2.6}_{-2.4}$ &$1.9^{+0.6}_{-0.6}$ & $4.6^{+1.5}_{-1.4}$\\
Abell~520 & $1144^{+64}_{-67}$ & $3.6\pm0.4$ & $7.3\pm1.1$ & $526$ & $2.5^{+0.5}_{-0.4}$ &$1208$ & $6.1^{+1.2}_{-1.1}$ &$15.3^{+3.0}_{-3.0}$ &$3.5^{+0.7}_{-0.7}$ & $8.8^{+1.7}_{-1.7}$\\
Abell~521 & $944^{+94}_{-103}$ & $3.1\pm0.5$ & $9.0\pm1.4$ & $448$ & $1.7^{+0.7}_{-0.8}$ &$1335$ & $8.8^{+2.0}_{-1.9}$ &$10.7^{+3.0}_{-3.0}$ &$2.5^{+0.7}_{-0.7}$ & $6.3^{+1.7}_{-1.7}$\\
Abell~586 & $804^{+107}_{-122}$ & $2.5\pm0.6$ & $6.1\pm1.6$ & $441$ & $1.4^{+0.5}_{-0.4}$ &$1221$ & $6.1^{+2.6}_{-2.6}$ &$4.6^{+2.1}_{-2.0}$ &$1.2^{+0.5}_{-0.5}$ & $2.8^{+1.3}_{-1.2}$\\
Abell~611 & $995^{+94}_{-103}$ & $3.7\pm0.5$ & $9.0\pm1.4$ & $502$ & $2.4^{+0.5}_{-0.5}$ &$1236$ & $7.2^{+1.5}_{-1.4}$ &$9.4^{+2.8}_{-2.8}$ &$2.2^{+0.7}_{-0.7}$ & $5.5^{+1.6}_{-1.6}$\\
Abell~697 & $1146^{+74}_{-79}$ & $4.6\pm0.5$ & $10.5\pm1.4$ & $565$ & $3.4^{+0.6}_{-0.6}$ &$1431$ & $11.2^{+1.5}_{-1.7}$ &$14.1^{+3.1}_{-3.1}$ &$3.3^{+0.7}_{-0.7}$ & $8.2^{+1.8}_{-1.8}$\\
Abell~851 & $1328^{+91}_{-98}$ & $5.4\pm0.6$ & $12.2\pm1.6$ & $553$ & $3.7^{+0.5}_{-0.5}$ &$1362$ & $11.1^{+2.2}_{-2.1}$ &$21.4^{+5.3}_{-4.9}$ &$4.7^{+1.2}_{-1.1}$ & $12.5^{+3.1}_{-2.8}$\\
Abell~959 & $1257^{+70}_{-74}$ & $5.0\pm0.5$ & $10.8\pm1.4$ & $596$ & $4.0^{+0.6}_{-0.6}$ &$1343$ & $9.2^{+1.6}_{-1.6}$ &$19.7^{+3.8}_{-3.6}$ &$4.4^{+0.8}_{-0.8}$ & $11.4^{+2.2}_{-2.1}$\\
Abell~1234 & $969^{+77}_{-84}$ & $2.5\pm0.5$ & $4.4\pm1.3$ & $447$ & $1.5^{+0.3}_{-0.3}$ &$989$ & $3.2^{+1.2}_{-1.0}$ &$7.6^{+2.3}_{-2.1}$ &$1.9^{+0.6}_{-0.5}$ & $4.5^{+1.3}_{-1.2}$\\
Abell~1246 & $921^{+78}_{-85}$ & $2.7\pm0.4$ & $5.6\pm1.1$ & $440$ & $1.5^{+0.3}_{-0.4}$ &$1089$ & $4.4^{+1.0}_{-1.0}$ &$8.7^{+2.4}_{-2.2}$ &$2.1^{+0.6}_{-0.5}$ & $5.1^{+1.4}_{-1.3}$\\
Abell~1758 & $1278^{+60}_{-62}$ & $5.5\pm0.5$ & $12.1\pm1.4$ & $651$ & $5.2^{+0.7}_{-0.7}$ &$1507$ & $12.9^{+1.9}_{-1.9}$ &$19.4^{+3.2}_{-3.2}$ &$4.3^{+0.7}_{-0.7}$ & $11.2^{+1.9}_{-1.9}$\\
Abell~1835 & $1295^{+65}_{-68}$ & $5.3\pm0.5$ & $10.7\pm1.3$ & $618$ & $4.3^{+0.5}_{-0.5}$ &$1398$ & $10.0^{+1.6}_{-1.6}$ &$19.9^{+3.7}_{-3.5}$ &$4.4^{+0.8}_{-0.8}$ & $11.4^{+2.1}_{-2.0}$\\
Abell~1914 & $1098^{+57}_{-60}$ & $3.7\pm0.5$ & $7.8\pm1.2$ & $531$ & $2.5^{+0.4}_{-0.4}$ &$1293$ & $7.3^{+1.3}_{-1.3}$ &$13.5^{+2.4}_{-2.4}$ &$3.1^{+0.6}_{-0.6}$ & $7.8^{+1.4}_{-1.4}$\\
Abell~1942 & $1080^{+70}_{-74}$ & $3.8\pm0.6$ & $7.5\pm1.3$ & $531$ & $2.7^{+0.6}_{-0.5}$ &$1212$ & $6.4^{+1.4}_{-1.3}$ &$13.5^{+2.8}_{-2.6}$ &$3.1^{+0.6}_{-0.6}$ & $7.8^{+1.6}_{-1.5}$\\
Abell~2104 & $1135^{+71}_{-76}$ & $4.2\pm0.5$ & $10.2\pm1.2$ & $596$ & $3.5^{+0.6}_{-0.6}$ &$1426$ & $9.6^{+1.7}_{-1.4}$ &$15.7^{+3.5}_{-3.3}$ &$3.6^{+0.8}_{-0.8}$ & $9.0^{+2.0}_{-1.9}$\\
Abell~2111 & $996^{+77}_{-83}$ & $3.9\pm0.5$ & $6.6\pm1.5$ & $528$ & $2.6^{+0.5}_{-0.5}$ &$1170$ & $5.7^{+1.9}_{-1.8}$ &$9.4^{+2.4}_{-2.2}$ &$2.3^{+0.6}_{-0.5}$ & $5.5^{+1.4}_{-1.3}$\\
Abell~2163 & $1188^{+74}_{-79}$ & $4.4\pm0.4$ & $9.4\pm1.2$ & $574$ & $3.3^{+0.4}_{-0.4}$ &$1466$ & $11.0^{+2.0}_{-2.0}$ &$17.4^{+3.8}_{-3.6}$ &$3.9^{+0.9}_{-0.8}$ & $10.0^{+2.2}_{-2.1}$\\
Abell~2204 & $1229^{+56}_{-58}$ & $4.8\pm0.5$ & $10.8\pm1.2$ & $631$ & $4.2^{+0.5}_{-0.6}$ &$1491$ & $11.0^{+1.6}_{-1.5}$ &$19.9^{+3.2}_{-3.1}$ &$4.4^{+0.7}_{-0.7}$ & $11.3^{+1.8}_{-1.7}$\\
Abell~2259 & $932^{+89}_{-98}$ & $2.4\pm0.6$ & $5.6\pm1.2$ & $427$ & $1.3^{+0.5}_{-0.4}$ &$1113$ & $4.6^{+1.2}_{-1.1}$ &$7.9^{+2.6}_{-2.4}$ &$1.9^{+0.6}_{-0.6}$ & $4.6^{+1.5}_{-1.4}$\\
Abell~2261 & $1307^{+65}_{-68}$ & $6.0\pm0.5$ & $14.1\pm1.4$ & $682$ & $5.7^{+0.6}_{-0.6}$ &$1663$ & $16.4^{+1.9}_{-1.9}$ &$24.4^{+4.1}_{-3.9}$ &$5.3^{+0.9}_{-0.9}$ & $13.9^{+2.3}_{-2.2}$\\
Abell~2537 & $1285^{+71}_{-75}$ & $5.4\pm0.6$ & $10.0\pm1.4$ & $599$ & $4.1^{+0.6}_{-0.6}$ &$1312$ & $8.7^{+1.6}_{-1.5}$ &$20.9^{+4.0}_{-3.8}$ &$4.6^{+0.9}_{-0.9}$ & $12.1^{+2.3}_{-2.2}$\\
MS~0440.5+0204 & $780^{+112}_{-130}$ & $2.9\pm0.6$ & $2.8\pm1.3$ & $468$ & $1.8^{+0.6}_{-0.5}$ &$896$ & $2.5^{+0.7}_{-0.7}$ &$3.5^{+1.8}_{-1.8}$ &$0.9^{+0.5}_{-0.5}$ & $2.1^{+1.1}_{-1.1}$\\
MS~0451.6-0305 & $1302^{+129}_{-142}$ & $4.4\pm0.6$ & $8.4\pm2.1$ & $466$ & $2.6^{+0.6}_{-0.5}$ &$1082$ & $6.5^{+1.7}_{-1.9}$ &$17.4^{+5.7}_{-5.3}$ &$3.9^{+1.3}_{-1.2}$ & $10.3^{+3.4}_{-3.1}$\\
MS~1008.1-1224 & $1218^{+74}_{-79}$ & $4.1\pm0.4$ & $8.2\pm1.4$ & $520$ & $2.7^{+0.4}_{-0.4}$ &$1176$ & $6.3^{+1.2}_{-1.2}$ &$16.3^{+3.5}_{-3.3}$ &$3.7^{+0.8}_{-0.8}$ & $9.5^{+2.0}_{-1.9}$\\
RX~J1347.5-1145 & $1358^{+95}_{-102}$ & $5.2\pm0.7$ & $10.1\pm2.2$ & $547$ & $3.8^{+0.9}_{-0.9}$ &$1323$ & $10.7^{+3.5}_{-3.7}$ &$19.9^{+5.2}_{-5.0}$ &$4.4^{+1.2}_{-1.1}$ & $11.7^{+3.0}_{-3.0}$\\
RX~J1524.6+0957 & $839^{+199}_{-258}$ & $2.1\pm0.7$ & $6.5\pm2.1$ & $249$ & $0.4^{+0.3}_{-12.2}$ &$977$ & $4.6^{+2.2}_{-2.0}$ &$6.2^{+4.2}_{-3.7}$ &$1.5^{+1.0}_{-0.9}$ & $3.8^{+2.5}_{-2.2}$\\
MACS~J0717.5+3745 & $1617^{+119}_{-128}$ & $6.4\pm0.8$ & $19.3\pm2.3$ & $612$ & $5.9^{+1.2}_{-1.4}$ &$1489$ & $17.1^{+3.2}_{-3.1}$ &$38.4^{+8.9}_{-8.6}$ &$8.0^{+1.8}_{-1.8}$ & $22.3^{+5.2}_{-5.0}$\\
MACS~J0913.7+4056 & $919^{+143}_{-168}$ & $3.1\pm0.8$ & $5.3\pm1.8$ & $397$ & $1.4^{+0.9}_{-0.7}$ &$945$ & $3.9^{+1.3}_{-1.2}$ &$6.5^{+3.4}_{-2.9}$ &$1.6^{+0.8}_{-0.7}$ & $3.9^{+2.0}_{-1.8}$\\
CIZA~J1938+54 & $1186^{+84}_{-90}$ & $5.3\pm0.6$ & $11.3\pm1.5$ & $601$ & $4.0^{+0.6}_{-0.6}$ &$1573$ & $14.4^{+2.4}_{-2.4}$ &$17.2^{+4.2}_{-4.0}$ &$3.9^{+0.9}_{-0.9}$ & $9.9^{+2.4}_{-2.3}$\\
3C295 & $1076^{+113}_{-125}$ & $4.6\pm0.7$ & $8.1\pm1.9$ & $501$ & $2.9^{+0.7}_{-0.6}$ &$1101$ & $6.2^{+1.6}_{-1.7}$ &$12.0^{+3.9}_{-3.8}$ &$2.8^{+0.9}_{-0.9}$ & $7.1^{+2.3}_{-2.2}$\\
\hline
\hline
\end{tabular}
\bigskip
\begin{minipage}{\linewidth} {\footnotesize Column 1: cluster name;
    Column 2: line-of-sight velocity dispersion (in units of km/s) of
    the best fit SIS model; Columns~3 \& 4: projected mass within an
    aperture of $0.5h_{70}^{-1}$ and $1h_{70}^{-1}$Mpc, resp.;
    Columns~5 \& 7: $r_{\Delta}$ (in units of $h_{70}^{-1}$kpc)
    determined using aperture masses; Columns~6 \& 8: deprojected
    aperture masses within $r_\Delta$; Columns~9-11: masses from best
    fit NFW model. All masses are listed in units of
    $[10^{14}h_{70}^{-1}$\msun]}
\end{minipage}
\end{center}
\end{table*}

\subsection{Parametric mass models}
\label{sec:nfwfit}

One approach to infer masses is to fit parametrized models to the
lensing signal. The most commonly used profile is the
Navarro-Frenk-White (NFW) fitting function proposed by \cite{NFW},
which is a good description of the average density profiles of halos
in numerical simulations of structure formation in cold dark matter
dominated universes. It also describes the stacked lensing signal for
ensembles of clusters well \citep[e.g.][]{Okabe13,Umetsu14}. The NFW
profile is characterized by two parameters. We use the mass of the
halo and the concentration $c$, although we do not fit for these
parameters simultaneously: we use the fact that simulations show that
the mass and concentraton are correlated.  However, as the
concentration depends on the formation redshift of the halo, this
relation depends on cosmology.  H12 used the results from
\cite{Duffy08}, which are based on the cosmological parameters from
the five-year {\it Wilkinson Microwave Anisotropy Probe} observations
\citep[WMAP5;][]{Komatsu09}. These have since been superseded by the
measurements of the {\it Planck} satellite
\citep{PlanckXVI}. \cite{Dutton14} present fitting functions for the
mass-concentration relation for this cosmology, which we use when we
estimate cluster masses (cf. Table~\ref{tab:masses}).

\cite{Becker11} have shown that fitting an NFW model to the observed
lensing signal can lead to mass estimates that are biased low when
measurements at large radii are included. For this reason we restrict
the fit to $0.5-2 h_{70}^{-1}$ Mpc from the cluster, where biases
should be negligible. The resulting masses\footnote{$M_\Delta$ is the
  mass enclosed within a radius where the mean density of the halo is
  $\Delta$ times the critical density at the redshift of the cluster;
  the virial mass is defined relative to the background density. See
  \cite{Hoekstra07} for more details about our choice of definition.},
for different overdensities $\Delta$ are presented in columns~$9-11$
in Table~\ref{tab:masses}. The statistical uncertainties on the
measurements are estimated as described in \cite{Hoekstra07} and
H12. The uncertainties in the mass estimates include the contribution
from distant large-scale structure
\citep{Hoekstra01,Hoekstra03,Hoekstra11}. For reference with other
studies we also present the velocity dispersion corresponding to the
best-fit singular isothermal sphere (SIS).

For reference we note that if we had used the mass-concentration
relation from \cite{Duffy08}, which yields concentrations that are
$\sim 20\%$ lower compared to the values used here, our masses would
change as follows: $M_{2500}$ decreases on average by $7\%$, while
$M_{500}$ and $M_{200}$ increase by $5\%$ and $9\%$, respectively. The
relative change in mass does not depend significantly on the cluster
redshift. However, for a direct comparison with the existing
literature, we present mass measurements using the \cite{Duffy08}
mass-concentration relation for the WMAP5 cosmology in
Appendix~\ref{app:duffy}.

\subsubsection{Comparison to other weak lensing studies}
\label{sec:comparison}

Several studies have determined weak lensing masses for large samples
of clusters using observations with the Subaru telescope. The most
relevant for the comparison with CCCP is the Weighing the Giants (WtG)
project, described in \cite{vonderLinden14}, which targeted 51 massive
clusters. For a subset of the clusters WtG determined photometric
redshifts for the sources \citep[see][for details]{Kelly14}. However,
these are predominantly the high redshift systems where the overlap
with CCCP is limited. For this reason we compare with the
``color-cut'' masses, which are presented in \cite{Applegate14}.

A closer inspection of the sample studied by \cite{vonderLinden14} and
\cite{Applegate14} shows that they associated MS0906.5+1110 with the
cluster Abell~750 which is located only 3' away in
projection. However, as discussed by \cite{Rines13} the latter is a
different cluster, which is clearly separated in redshift.  The
location of A750 provided in \cite{vonderLinden14} is in fact that of
MS0906.5+1110, and we therefore include this cluster in the
comparison. Abell~1758 is a merging cluster and therefore H12
considered the Eastern and Western component separately \citep[also
see][]{Ragozzine12}. However, other studies consider this a single
cluster and we therefore decided to provide results for the location
listed by \cite{vonderLinden14}, who refer to this cluster as
Abell~1758N.  As a result we have 18 clusters in common with WtG.

\begin{figure}
\centering
\leavevmode \hbox{%
\includegraphics[width=8.5cm]{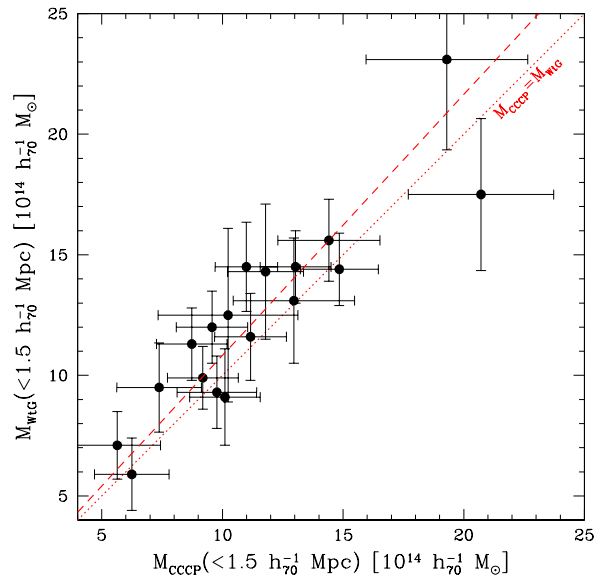}}
\caption{Comparison with the mass estimates from
  \citet{Applegate14}. The CCCP masses are computed from the best fit
  NFW model to the lensing measurements within $0.75-3
  h_{70}^{-1}$Mpc, adopting a concentration $c_{200}=4$. This matches
  the procedure described in \citet{Applegate14}, although our
    source redshift distribution is somewhat different, as explained
    in the text. The dotted line indicates the line of equality,
  whereas the dashed line is the best fit, which has a slope 1.08.
\label{fig:comp_wtg}}
\end{figure}

To compare the results for these clusters, we follow
\cite{Applegate14} and fit an NFW model with a concentration
$c_{200}=4$ to the tangential distortion within $0.75-3
h_{70}^{-1}$Mpc and compute the mass within a sphere of radius
$1.5h_{70}^{-1}$Mpc. The results are presented in
Figure~\ref{fig:comp_wtg}. We find that the WtG masses are somewhat
larger: the dashed line indicates the best fit linear relation $M_{\rm
  WtG}=(1.082\pm 0.038)M_{\rm CCCP}$.  Repeating the comparison using
the results from H12 yields $M_{\rm WtG}=(1.263\pm 0.048)M_{\rm
  H12}$. Hence the analysis presented here reduces the
  discrepancy considerably. We note that differences in the fitting
  procedure can lead to additional uncertaintly, and it is therefore
  not clear whether the difference is significant. Furthermore, the
  ``color-cut'' masses from \cite{Applegate14} are derived using the
  photometric redshift catalog from \cite{Ilbert09}, which are based
  on the original COSMOS-30 data. Using this redshift distribution we
  find $M_{\rm WtG}=(1.063\pm 0.038)M_{\rm CCCP}$. Interestingly, when
  we compute deprojected aperture masses within a radius of $1
  h_{70}^{-1}$Mpc (in this case adopting $c_{200}=4$; see
  \S\ref{sec:apmass} for details), the agreement with the
  corresponding masses from \cite{Applegate14} is excellent: $M_{\rm
    WtG}=(1.018\pm 0.036)M^{\rm ap}_{\rm CCCP}$.

The NFW model is fit to relatively small radii, where the
contamination by cluster members is large (although the inner
$750h_{70}^{-1}$~kpc are not used): if we omit the correction for the
contamination of cluster members our masses decrease, as expected, and
$M_{\rm WtG} \sim 1.28 \times M_{\rm CCCP}$. Although the correction
is substantial, Figure~\ref{fig:res_contam} suggests that the bias
after correction should be $<2\%$. 

We investigated this further by restricting the fit to small ($r_{\rm
  in}=0.75-1.5 h_{70}^{-1}$ Mpc) and large ($r_{\rm out}=1.5-3
h_{70}^{-1}$ Mpc) radii. If the contamination correction is adequate,
the resulting average masses should agree, whereas a ratio $M_{\rm
  out}/M_{\rm in}>1$ would imply residual contamination by cluster
members. For the 18 clusters in common with WtG we find $M_{\rm
  out}/M_{\rm in}=1.05\pm0.05$, suggesting that the correction has
worked well (the ratio is $1.16\pm0.05$ if we do not correct for
contamination). For the full CCCP sample we find $M_{\rm out}/M_{\rm
  in}=1.00\pm0.03$.

\begin{figure*}
\centering
\leavevmode \hbox{%
\includegraphics[width=8.5cm]{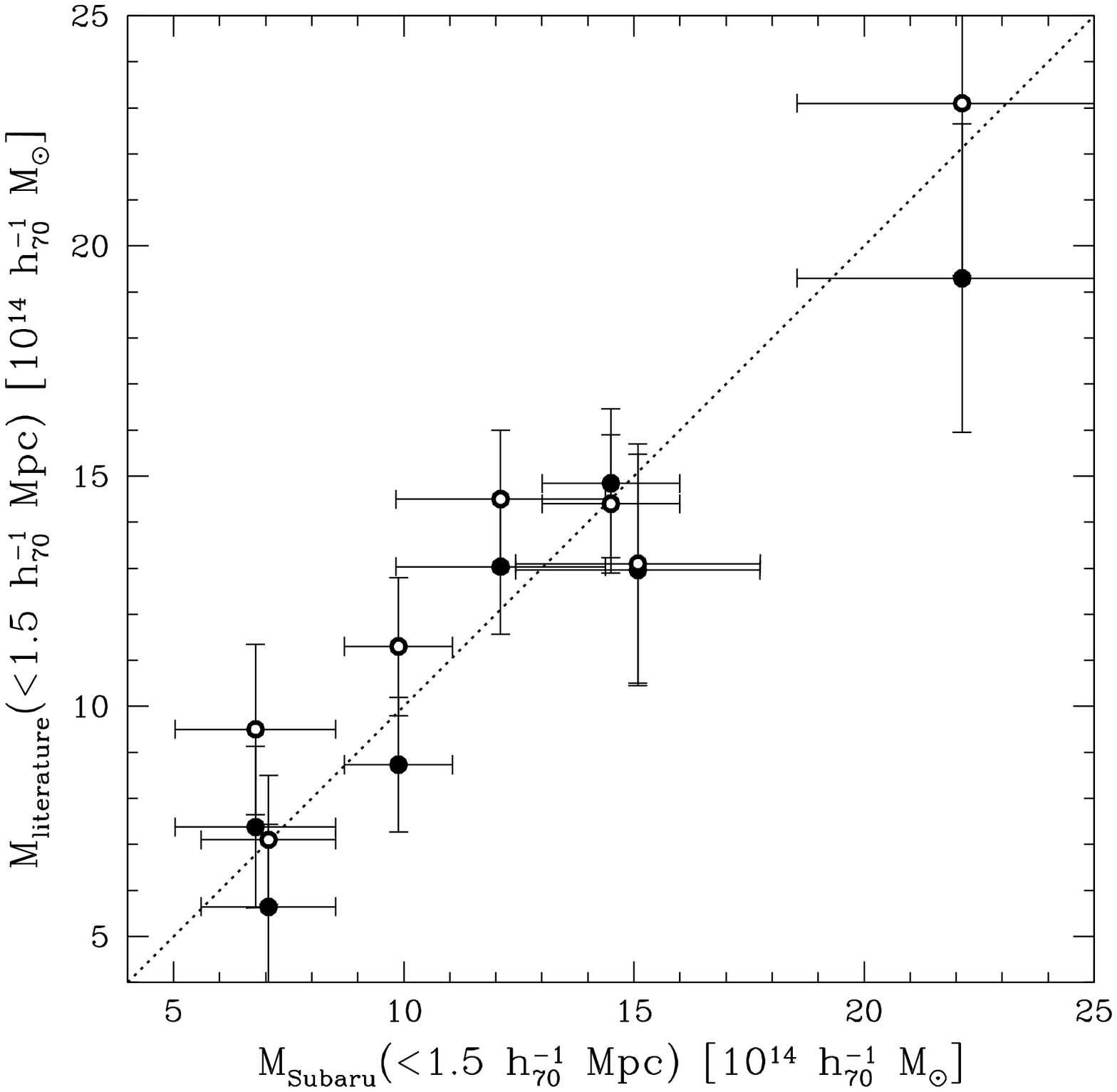}
\includegraphics[width=8.5cm]{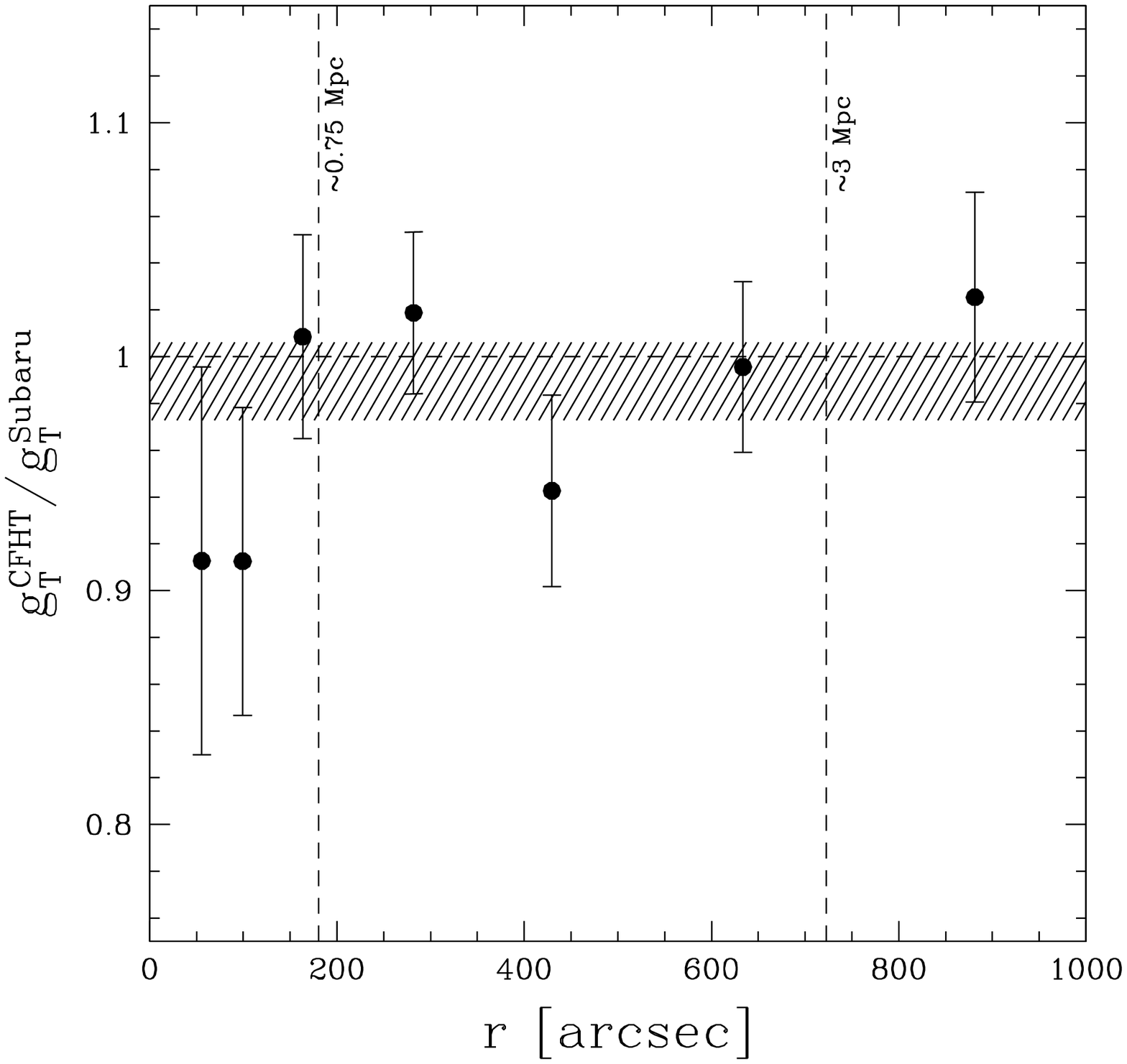}}
\caption{{\it Left panel:} Black points indicate the mass based on our
  analysis of CFHT data as a function of the mass obtained from the
  Subaru data.  As opposed to the results presented in the right
  panel, the object catalogs were not matched. We find that the
  Subaru-based masses are $4\pm 6\%$ higher than the CFHT-based
  results we report in this paper. The open points indicate the
  comparison of our Subaru-based masses with those from
  \citet{Applegate14}, which are $7\pm 6\%$ higher. The dotted line
  indicates the line of equality. {\it Right panel:} Ratio of the
  ensemble averaged tangential distortion as a function of radius
  measured from CFHT and Subaru data.  The ratio is computed by
  combining the measurements for the sample of 7 clusters, where the
  catalogs are matched, such that they contain essentially the same
  objects. The hatched region indicates the 68\% confidence region
for the average ratio $\langle
  g_T^{\rm CFHT}/g_T^{\rm Subaru}\rangle=0.99\pm0.02$.  For reference we
  also indicate the fitting range used in the comparison with
  \citet{Applegate14}.
  \label{fig:comp_subaru}}
\end{figure*}

\cite{Umetsu14} present results for 20 clusters studied as part of the
Cluster Lensing and Supernova survey with Hubble (CLASH). Of these 17
clusters overlap with WtG, but only 6 overlap with CCCP. For the
  six clusters we have in common, the CLASH masses are $12\pm5\%$
  higher than the CCCP results. However, we note that \cite{Umetsu14}
use a different fit range, while leaving the concentration a free
parameter. Although they find a best fit concentration of $\sim 4$
when they stack the clusters in their sample, cluster-to-cluster
variation complicates a more direct comparison. As discussed below, we
analysed the CLASH data using our pipeline and find better agreement
with our masses derived from CFHT observations.

Another large study that does overlap considerably with CCCP is the
Local Cluster Substructure Survey (LoCuSS). Results for 30 clusters,
of which 13 overlap with CCCP are presented in
\cite{Okabe10}. \cite{Okabe13} suggest that a revised analysis leads
to higher masses, but only present results for an ensemble stacked
lensing signal and do not provide updated masses for the individual
clusters.

\subsubsection{Direct comparison with Subaru data}
\label{sec:subaru}

The processed Subaru imaging data used by CLASH have been publicly
released\footnote{http://archive.stsci.edu/prepds/clash/} for nine of
the twenty clusters. We retrieved the data for the four clusters that
overlap with CCCP. To extend the comparison sample, Keiichi Umetsu
kindly provided us with the data for Abell~209 and
Abell~611. Observation of Abell~1758 were provided by James Jee. We
analysed these data using our CCCP weak lensing pipeline. We made no
modifications and thus assume that our empirical correction for noise
bias also applies to the typically deeper Subaru data (note that we
also apply the size cut of $r_h<5$ pixels in this case). The results
presented in Appendix~\ref{sec:rcs2} suggest that our approach, which
is a function of signal-to-noise ratio and galaxy size, is
sufficiently flexible.

The data that were provided are stacks of dithered exposures. As a
consequence multiple chips can contribute to a given location, which
can lead to a more complex PSF. Our observing strategy with Megacam
allowed us to avoid this, but we note that the same problem occurs for
our CFH12k data. However, we did not measure noticeable differences in
the scaling relations based on CFH12k or Megacam data. As we did for
the CFH12k data, we use the weight images to split the data into
regions that more or less correspond to the chips of the camera, and
analyse the resulting images using the pipeline described earlier.
This analysis is done completely independently from the analysis based
on the CFHT data. Hence we redo the object detection and masking,
identify the stars which are used to model the PSF, etc.

The results are presented in Figure~\ref{fig:comp_subaru}. The left
panel shows a comparison of the weak lensing masses when the source
catalog is determined independently from the CFHT analysis. Following
\cite{Applegate14}, these masses are based on the best fit NFW model
to the lensing measurements within $0.75-3h_{70}^{-1}$Mpc, adopting a
concentration $c_{200}=4$. For most clusters the Subaru data are
deeper, resulting in a different effective source redshift
distribution, which we account for. Consequently the correction for
contamination by cluster members also differs somewhat. It is
important to stress this correction works well on average (as shown in
Figure~\ref{fig:res_contam}), but the statistical uncertainty is
larger when comparing a small sample of clusters.

The filled points in the left panel of Figure~\ref{fig:comp_subaru}
compare the masses based on the CFHT data ($M_{\rm literature}$) to
those determined from our analysis of the Subaru data ($M_{\rm
  Subaru}$). We find excellent agreement with a best fit $M_{\rm
  CFHT}=(0.97\pm0.06)M_{\rm Subaru}$. We can use this result as a
measure of the systematic uncertainty when cluster masses are
determined independently using different instruments, albeit with the
same shape measurement pipeline. This comparison is made by fitting a
parametric model to the lensing signal at relatively small radii,
where our correction for contamination by cluster members is
largest. As discussed in more detail in \S\ref{sec:errorbudget}, we
expect our aperture masses to be more reliable. We find that the
masses from \cite{Applegate14}, indicated by the open points, are
somewhat higher. We obtain $M_{\rm WtG}=(1.07\pm0.06)M_{\rm
  Subaru}$. A similar result is obtained if we compare to the six
clusters in common with \cite{Umetsu14}, where we note that our Subaru
based masses are in excellent agreement (they are $\sim 2.4\%$ lower
on average).

To examine the performance of the shape measurement algorithm further,
we created a source catalog where the objects were matched by position
(note that we do apply the size cut before matching). Although
blending in the inner regions may cause some misidentification, such a
comparison should eliminate differences in the source redshift
distribution and the contamination by cluster members. The shape
measurements for individual galaxies are too noisy, and we therefore
compare the tangential shear profiles from the two telescopes. To
improve the signal-to-noise ratio even further we combine the signals
from the seven clusters. This allowed us to measure the ratio of the
lensing signal as a function of distance from the cluster center. The
results are presented in the right panel of
Figure~\ref{fig:comp_subaru}. We find an average ratio $\langle
g_T^{\rm CFHT}/g_T^{\rm Subaru}\rangle=0.99\pm0.02$, indicated by the
hatched region. This result suggests that our pipeline is able to
recover the shapes to within $1\pm2\%$ for different data sets.

In addition we have matched our CFHT-based measurements to catalogs
provided by the WtG team (Von der Linden, private communication). This
direct comparison for the overlapping sample of clusters showed a
remarkable agreement with $\langle g_T^{\rm CCCP}/g_T^{\rm
  WtG}\rangle=0.991\pm0.018$. These direct comparisons of shear
catalogs obtained from observations using different telescopes and
teams suggest that the pipelines are robust. The differences we
observe are consistent with the statistical uncertainties associated
with comparing such a small sample of clusters.

\subsubsection{Comparison to Hectospec Cluster Survey}

\begin{figure}
\centering
\leavevmode \hbox{%
\includegraphics[width=8.5cm]{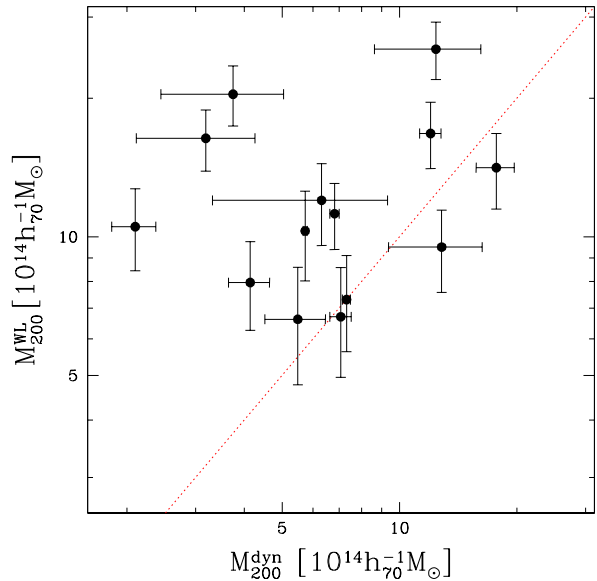}}
\caption{Comparison of the dynamical estimates for $M_{200}$ from
the Hectospec Cluster Survey from \citet{Rines13} with the best 
fit NFW model to the lensing data. The dotted line indicates the
line of equality.
\label{fig:comp_rines}}
\end{figure}

The infall regions of galaxy clusters provide an interesting
alternative way to estimate cluster masses at relatively large radii.
The Hectospec Cluster Survey targeted a sample of 58 clusters, 14 of
which overlap with our study. This survey is described in
\cite{Rines13}, who measured cluster mass profiles using the caustic
technique \citep[e.g.][]{Diaferio99}. We compare the estimates for
$M_{200}$ from the best fit NFW models to those obtained by
\cite{Rines13}. The results are presented in Figure~\ref{fig:comp_rines}.
We note that a comparison to the velocity dispersions yields similar
results.

The lensing masses are higher than the dynamical masses; the average
ratio is $1.22\pm0.07$. If we adopt the mass-concentration relation
from \cite{Duffy08} the agreement is worse with an average ratio of
$1.33\pm0.08$. Although the comparison sample is small, we observe
substantial scatter. The most significant outliers in
Figure~\ref{fig:comp_rines} are \hbox{MS0906.5+1110}, Abell~1758 and
Abell~2261.  In \S3.6 of their paper, \cite{Rines13} comment on
indivual clusters, including these outliers. The first is part of a
pair of clusters, but the infall patterns can be
separated. \cite{Rines13} find the higher mass for the other
component, Abell~750, but comment that \hbox{MS0906.5+1110} has the
higher X-ray luminosity.  \cite{Okabe10} also find a higher lensing
mass for this cluster, suggesting that the dynamical mass is too
low. \cite{Coe12} present a detailed study of Abell~2261, suggesting
that the dynamical mass estimate for Abell~2261 is low compared to the
lensing and X-ray estimates. We note the large range in hydrostatic
mass estimates for this cluster, including the measurement from
\cite{Mahdavi13}; although lower than the lensing mass, the X-ray mass
is still larger than the dynamical mass from \cite{Rines13}.  Finally
Abell~1758 is a merging system, at the high redshift end of the sample
studied by \cite{Rines13}. As a result the dynamical mass is not that
well constrained.

\subsection{Aperture masses}
\label{sec:apmass}

Although the NFW profile is a good description for an ensemble of
clusters, as is suggested by stacked weak lensing studies
\citep{Okabe13,Umetsu14}, it may not be a good model to fit to
individual systems. In particular the presence of substructure may
lead to incorrect masses \citep[see e.g.][for a clear
example]{Hoekstra00}. The CCCP sample contains several complex, or
merging clusters, of which Abell~520 has been studied in particular
detail \citep{Mahdavi07,Jee12,Jee14}. In these cases a more direct
estimate of the projected mass, without having to rely on a particular
profile, would be preferable.

It is possible to estimate the projected mass within an aperture with
minimal assumptions about the actual mass distribution. However,
comparison to other proxies typically still relies on deprojected
masses, which do depend on the assumed density profile. Various
estimators are available, but we prefer to use the one proposed by
\cite{Clowe98}:

\begin{equation}
\zeta_c(r_1)=2\int_{r_1}^{r_2}d\ln r\langle\gamma_t\rangle+
\frac{2r_{\rm max}^2}{r_{\rm max}^2-r_2^2} \int_{r_2}^{r_{\rm max}}
d\ln r \langle\gamma_t\rangle,
\end{equation}

\noindent which can be expressed in terms of the mean dimensionless
surface density interior to $r_1$ relative to the mean
surface density in an annulus from $r_2$ to $r_{\rm max}$

\begin{equation}
\zeta_c(r_1)=\bar\kappa(r'<r_1)-\bar\kappa(r_2<r'<r_{\rm max}).
\end{equation}

\noindent Hence we can determine the mass up to constant, which is
determined by the mean convergence in the annulus $r_2<r'<r_{\rm
  max}$. Assumptions about the mass distribution enter in two
ways. First of all, we do not measure the tangential shear $\gamma_T$,
but the reduced shear $g_T=\gamma_T/(1-\kappa)$. For this conversion
we use the best-fit NFW model. However, the estimate of $\zeta_c(r)$
depends on the lensing measurements at large radii and consequently
this correction is small.

The more important dependence on the density profile is through the
need to estimate the average convergence in the annulus. Although the
contribution is relatively small if we consider large radii, it cannot
be ignored for our analysis. H12 used an outer radius $r_{\rm
  max}=1000''$ for the CFH12k data. However, for these data the
azimuthal coverage is incomplete. We therefore keep the inner radius
of 600'', but reduce the outer radius to 800''. The annuli are
unchanged for the Megacam data, i.e. we use $r_2=900''$ and $r_{\rm
  max}=1500''$.

As was done by H12, we use the best-fit NFW model to estimate the mean
convergence in the annulus. We quantify the sensitivity to the profile
by varying the normalization of the $c(M)$ relation by $20\%$. This
should capture the variation in results for different cosmologies
\citep[e.g.][]{Bhattacharya13}. The resulting change in projected mass within
an aperture of $1 h_{70}^{-1}$Mpc depends on redshift, with a
reduction (increase) of $<1\%$ for clusters with $z>0.3$ if the
normalization is increased (decreased). The changes are somewhat
larger at lower redshifts but at most only $\sim 2\%$. Hence the
systematic uncertainty in the estimate of the projected mass at this
radius is remarkably small.

We do make one small change with respect to H12: we include a
contribution from neighbouring halos. \cite{Oguri11} show that such a
two-halo term dominates over the NFW profile on large scales. A
convenient way to describe such contributions is provided by the
so-called {\it halo model} \cite[see e.g.,][for a review]{Cooray02}.
The implementation we use here is described in \cite{Cacciato13}. It
was used in \cite{Cacciato14} to model the lensing signal around
galaxies. In this context, the cluster lensing signal is simply the
lensing signal around the brightest cluster galaxy of a very massive
halo. The contribution from the clustering of halos, the two-halo
term, is in its most basic implementation proportional to the linear
matter power spectrum \citep[but see e.g.][for a more sophisticated
implementation]{vandenBosch13}. The constant of proportionality is
determined by the product of the bias of the halo of interest and an
average of the halo bias over all halo masses in the range of interest
weighted by the halo mass function \citep[see e.g.][]{Cacciato09}.

To compute the contribution to the average convergence in the annulus
we use the functions for the halo mass and bias provided by
\cite{Tinker10}.  We find that the correction is small, less than
$1\%$ for the radii we are interested in. This also depends somewhat
on the assumed density profile, but the main source of theoretical
uncertainty is the halo bias function, especially at the high mass end
considered here. We gauge the systematic uncertainty in the correction
by considering different fitting functions from the literature, namely
those from \cite{Sheth99}, \cite{Sheth01} and
\cite{Seljak04}. Although these may differ by up to a factor of a few
at the highest masses of interest, the corresponding cluster bias
varies by at most $\sim 20\%$ because of the exponential drop-off of
the halo mass function. We find that the resulting systematic
uncertainty in the correction itself is $\sim 10\%$, which can be safely
ignored.

\begin{figure}
\centering
\leavevmode \hbox{%
\includegraphics[width=8.5cm]{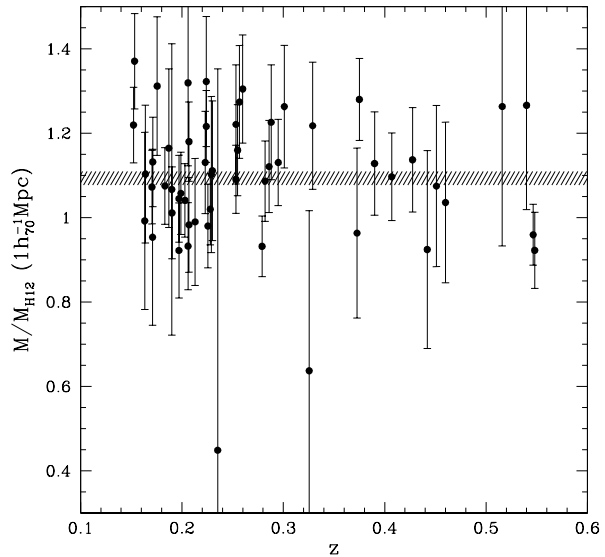}}
\caption{Ratio of the projected aperture mass within an aperture of
  radius $1h_{70}^{-1}$Mpc and the mass obtained by H12. The hatched region indicates the weighted average
  ratio $\langle M/M_{\rm H12}\rangle=1.093\pm0.016$. Note that the
  error bars indicate only the uncertainty in the updated mass
  estimates. \label{fig:comp_old}}
\end{figure}

The resulting projected masses within fixed apertures of $0.5$ and $1
h_{70}^{-1}$Mpc are listed in Table~\ref{tab:masses}. The
uncertainties include the contributions from cosmic noise and shape
noise. Although the variation in the source redshift distribution is
in principle an additional source of noise \citep{Hoekstra11}, the
impact is smaller for the larger angular scales used to compute the
aperture masses. Furthermore \cite{Hoekstra11} find that the
combination of cosmic noise and the variation in the redshift
distribution leads to a slight reduction in the uncertainty, compared
to the situation where only cosmic noise is considered.

In Figure~\ref{fig:comp_old} we compare the projected mass to the
results from H12 as a function of cluster redshift. We find no
significant trend with redshift and obtain a weighted average ratio
$\langle M/M_{\rm H12}\rangle=1.093\pm0.016$. The increase in the
amplitude of the lensing signal because of the correction for the
effects of noise in the images is partly offset by the increase in the
mean source redshift and the change in the mass-concentration relation.

\begin{figure}
\centering
\leavevmode \hbox{%
\includegraphics[width=8.5cm]{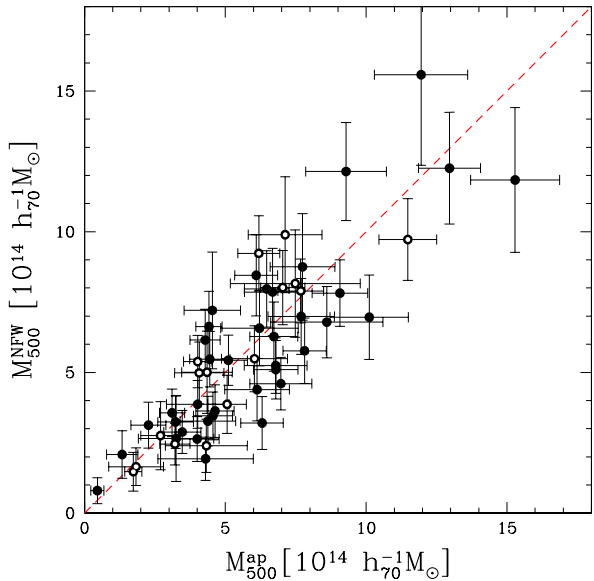}}
\caption{Comparison of $M^{\rm NFW}_{500}$ from the best-fit NFW model
  and the value for $M^{\rm ap}_{500}$ using the deprojected aperture
  mass. The former is based on measurements of the lensing signal at
  radii $0.5-2 h_{70}^{-1}$~Mpc, whereas the latter uses data from
  radii larger than $r_{500}$ which is typically larger than $1
  h_{70}^{-1}$~Mpc.  As a consequence the measurements are almost
  independent. The red dashed line indicates the line of equality. The
  open points indicate clusters for which \citet{Mahdavi13} measured a
  central entropy $K_0<70~\rm{keV cm}^2$. \label{fig:comp_map_nfw}}
\end{figure}

\subsubsection{Deprojected masses}

Although the projected masses can be determined robustly within an
aperture of fixed radius, comparison with other observations requires
the deprojection of the mass estimates. To do so we follow the
procedure described in \cite{Hoekstra07}, which was also used in H12:
at each radius we find the NFW model that yields the same projected
mass. We take the mass of this model, which depends on the adopted
mass-concentration relation, to be the estimate for the deprojected
mass. The results for $M^{\rm ap}_{2500}$ and $M^{\rm ap}_{500}$, using the relation
between concentration and mass from \cite{Dutton14}, are listed in
columns~6~\& 8 of Table~\ref{tab:masses} (for reference we also list
the corresponding radius $r_{\Delta}$). We present results for the
$c(M)$ relation from \cite{Duffy08} in Appendix~\ref{app:duffy}.

In figure~\ref{fig:comp_map_nfw} we compare $M^{\rm ap}_{500}$ to the estimate
obtained from the best-fit NFW model to the lensing signal at radii
$0.5-2 h_{70}^{-1}$~Mpc. The two measurements are nearly independent,
because the aperture mass is based on the lensing signal at radii
larger than $r_{500}$ which is larger than $1 h_{70}^{-1}$~Mpc for
most of the clusters. From a linear fit to the masses we find that
$M_{500}^{\rm NFW}/M_{500}^{\rm ap}=0.97\pm0.03 $, i.e. the estimates
from the NFW model are in good agreement on average.

It is interesting to compare the results for clusters with a low and
high central entropy $K_0$, the value of the deprojected entropy
profile at a radius of $20 h_{70}^{-1}$~kpc, as measured by
\cite{Mahdavi13}. The open points in Figure~\ref{fig:comp_map_nfw}
indicate the clusters with $K_0<70~\rm{keV cm}^2$, for which we find a
ratio $1.04\pm0.06$; for the remaining clusters $M_{500}^{\rm NFW}$ is
on average $0.95\pm0.04$ times smaller than the aperture mass
estimate. In neither case do we observe a signifant difference between
the two mass estimates.

As discussed in H12, the deprojection depends on the
mass-concentration relation. The results listed in
Table~\ref{tab:masses} are based on the mass-concentration relation
from \cite{Dutton14}. If we instead consider the concentrations from
\cite{Duffy08}, which are $\sim 20\%$ lower, we find that the
resulting $M^{\rm ap}_{2500}$ is 8\% lower, and $M^{\rm ap}_{500}$ is
3\% lower. Hence, the value for $M^{\rm ap}_{500}$ is fairly robust against
changes in the concentration.

If we compare to the deprojected masses from H12 we find that $M^{\rm
  ap}_{500}$ has increased by 19\% on average, whereas the
increase in $M_{2500}$ is 28\%. The larger change in $M_{2500}$ is due
to the higher concentrations from \cite{Dutton14}, which affect the
deprojection. As discussed in H12 (see Fig.~4 in that paper), lowering
the concentration increases the ratio $M_{500}^{\rm NFW}/M_{500}^{\rm
  ap}$.

\subsection{Error budget}
\label{sec:errorbudget}

\begin{figure}
\centering
\leavevmode \hbox{%
\includegraphics[width=8.5cm]{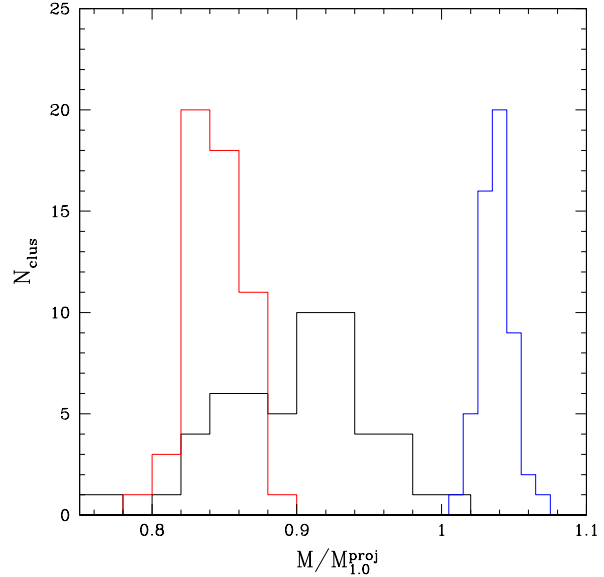}}
\caption{Ratio of the projected aperture mass within an aperture of
  radius $1h_{70}^{-1}$Mpc when one of the corrections is not included,
and the final mass estimate. The red histogram shows the distribution
if we ignore the correction for the multiplicative bias in the shape
measurement. The black histogram shows the decrease in mass if we
ignore the contamination by cluster members. The blue histogram
shows the change in mass if we use the source redshift distribution
used by H12. \label{fig:comp_cor}}
\end{figure}

Figure~\ref{fig:comp_cor} shows the distribution of the relative
change in the projected mass within an aperture of radius
$1h_{70}^{-1}$Mpc when one of the corrections is not included. The red
histogram shows that the average mass would be on average reduced by a
factor $0.84$ if we do not include the correction for the
multiplicative bias in the shape measurement. Ignoring the correction
for contamination by cluster members reduces the mass by a factor
$0.90$. As indicated by the black histogram, this correction varies
more from cluster to cluster. This is expected, as the correction
depends on the cluster redshift, the spatial distribution of cluster
galaxies and the richness of the cluster. If we use the source
redshift distribution used by H12 the masses would increase by 4\%, as
indicated by the blue histogram.

Although each of these corrections is substantial and cannot be
ignored, they are well-determined. The associated remaining systematic
errors are much smaller than the statistical uncertainties for
individual clusters. However, the precision that is afforded by the
full CCCP sample is much higher, and we therefore summarize the
systematic error budget in this section. These are relevant for the
discussion in \S\ref{sec:SZ} where we examine the scaling relation
between the lensing mass and the SZ measurements from {\it Planck}.

We assume a 2\% systematic uncertainty in the calibration of the
multiplicative bias. Although Figure~\ref{fig:bias_cor_e} suggests
that the empirical correction results in biases $<1\%$ if we adopt
$\epsilon_0=0.25$, the variation in seeing and the dependence of the
bias on morphology are expected to lead to addtional error. The
latter will need to be investigated in more detail, but the observed
variation in bias as a function of Sersic index (see
Fig.~\ref{fig:bias_profile}) suggests that the contribution from
morphology is $\sim 1\%$.

As shown by the black points in Figure~\ref{fig:res_contam} the mean
residual contamination by cluster members is at most $\sim 2\%$,
except for the innermost regions. Although the contamination exceeds
this value at some radii for clusters with $z>0.25$ (red hatched
region), the contamination is still smaller than $2\%$ when averaged
over the radii of interest. For the projected masses, the uncertainty
in the extrapolated density profile changes the masses by less than
$1\%$.

The largest contribution comes from the uncertainty in the source
redshift distribution, despite the wavelength coverage and depth of
the COSMOS data: for clusters at $z=0.2$ the systematic uncertainty in
$\langle\beta\rangle$ is $\sim 3\%$, increasing to $\sim 8\%$ for the
highest redshift clusters in our sample ($z=0.55$). These estimates
include the difference in $\langle\beta\rangle$ between our
photometric redshift distribution and that of \cite{Ilbert13}, and the
field-to-field variation in the four CFHTLS Deep fields. These
systematic errors are independent from one-another and thus can be added
in quadrature, resulting in a total systematic uncertainty of $4.2\%$
at $z=0.2$ and $8.5\%$ at $z=0.55$ for the projected masses in a fixed
aperture.

The deprojected masses are more sensitive to the assumed
mass-concentration relation. For instance, $M^{\rm ap}_{500}$ decreases by
$\sim 3\%$ if we lower the concentration by $20\%$ (i.e., by switching
to the \cite{Duffy08} values). This shift in concentration is caused
by the change in the cosmological parameters determined by WMAP5, used
by \cite{Duffy08}, and the more precise {\it Planck} values used by
\cite{Dutton14}.  However, these estimates are based on simulations
that only include dark matter and thus ignore the additional effects
of baryon physics. The impact of this has been studied by
\cite{Duffy10} using hydrodynamic simulations. \cite{Duffy10} found
that the change in concentration is $<10\%$ for cluster-mass halos. We
therefore adopt a similar uncertainty in the concentrations, which
implies a systematic contribution of $\sim 2\%$ for $M^{\rm ap}_{500}$.

Importantly the systematic errors for the deprojected masses are
increased because a change in the lensing signal also affects the
radius corresponding to a particular overdensity. This increases the
uncertainty compared to a fixed aperture because the enclosed mass
increases with radius. As a result we estimate a total systematic
uncertainty in $M^{\rm ap}_{500}$ of $6\%$ for clusters at $z=0.2$,
which increases to $12\%$ at $z=0.55$. Given the observational cost of
calibrating scaling relations, it makes sense to avoid introducing
such unnecessary sources of uncertainty: numerical simulations should
instead be used to make predictions for the observed lensing
measurements.

In \S\ref{sec:subaru} we determined masses for Abell~1758 and the six
clusters in common with \cite{Umetsu14} using available Subaru
observations. We found that these masses were $4\pm 6\%$ higher than
our estimates based on the CFHT data. Given the differences between
the data from the two telescopes (e.g., depth, masked areas), the
results agree very well. Moreover, when we match the object catalogs
we find that we recover the average tangential distortion within
$1\pm2\%$, suggesting our shape measurement pipeline is robust. 

\begin{figure*}
\centering
\leavevmode \hbox{%
\includegraphics[width=8.5cm]{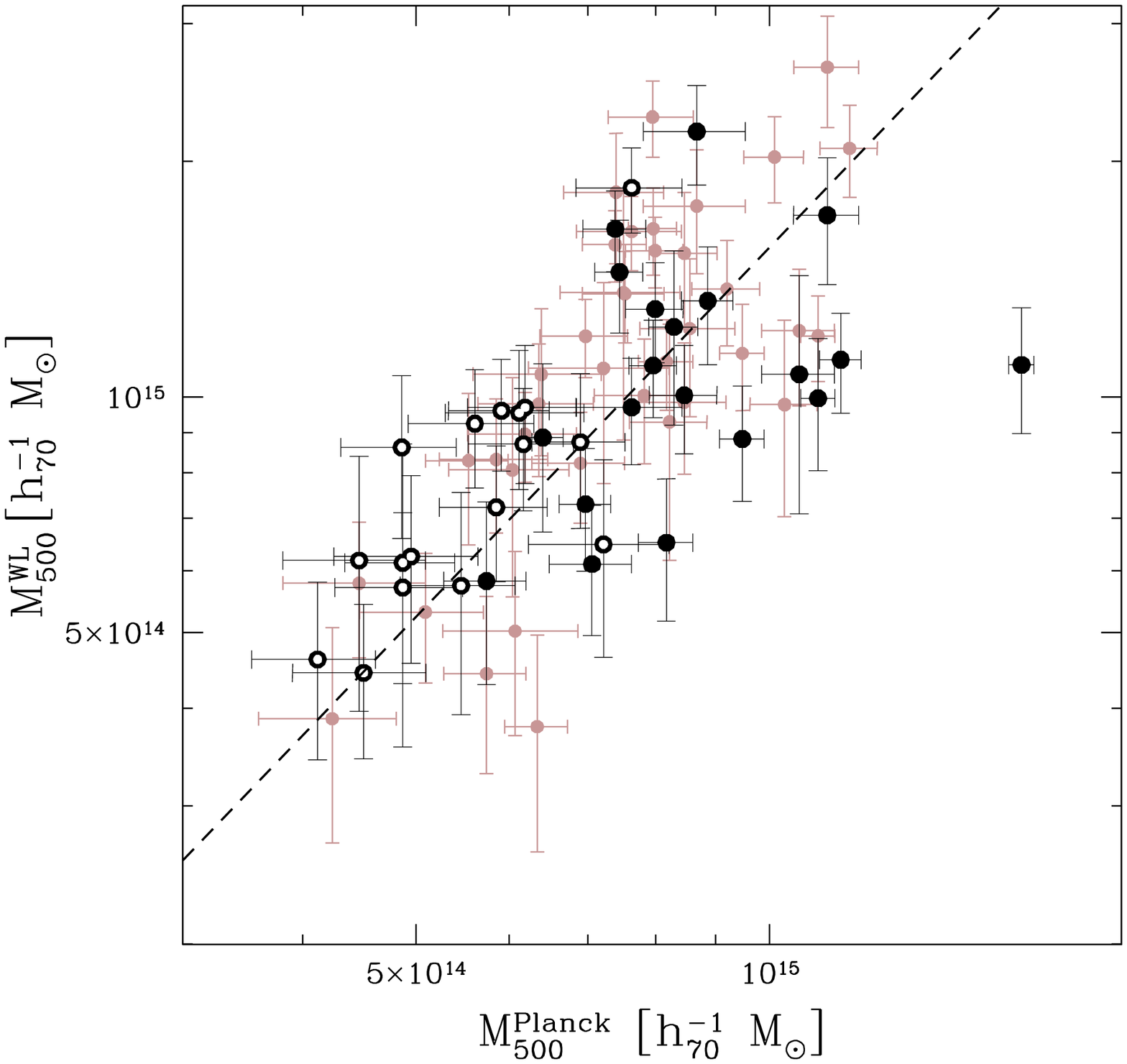}
\includegraphics[width=8.5cm]{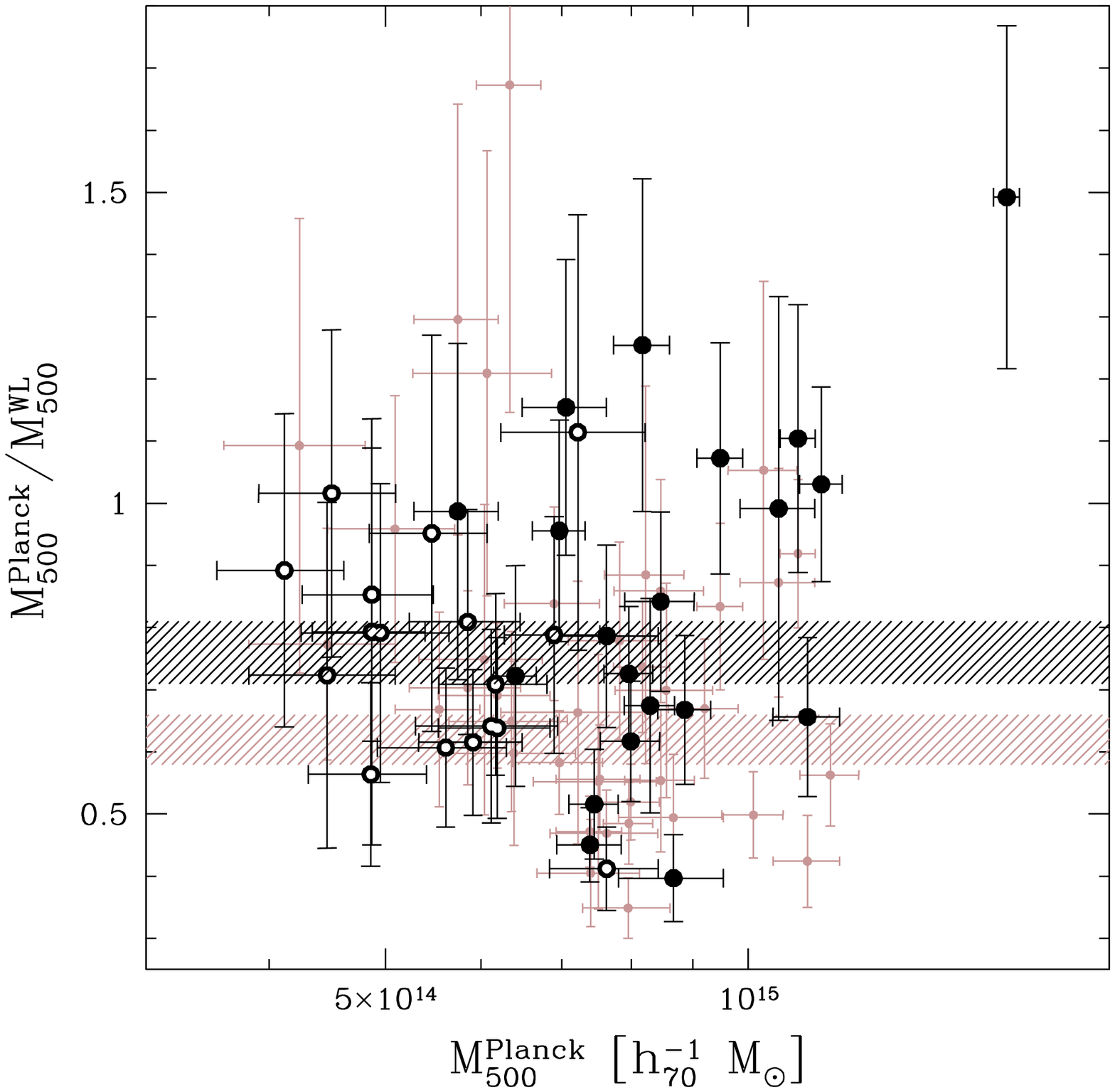}}
\caption{{\it Left panel}: the deprojected aperture mass $M_{500}$ from 
  weak lensing as a function of the hydrostatic mass from \citet{PlanckXXIX}.  
  Note that $M^{\rm Planck}_{500}$ is measured using $r_{500}$ from the
  estimate of $Y_X$, and $M^{\rm WL}_{500}$ is determined using the
  lensing derived value for $r_{500}$. The black points show our CCCP
  measurements, with the filled symbols indicating the clusters detected
  by {\it Planck} with a signal-to-noise ratio $SNR>7$ and the open points
  the remainder of the sample. The dashed line shows the best-fit
  power law model. The WtG results are shown as rosy brown colored
  points. {\it Right panel}: ratio of the hydrostatic and the weak
  lensing mass as a function of mass. The dark hatched area indicates
  the average value of $0.76\pm0.05$ for the CCCP sample, whereas the
  rosy brown colored hatched region is the average for the published WtG
  measurements, for which we find $0.62\pm0.04$.}
 \label{fig:planck}
\end{figure*}

\section{Planck SZE scaling relation}\label{sec:SZ}

\cite{PlanckXX} present constraints on cosmological parameters using
the cluster number counts as a function of redshift for a sample of
189 clusters of galaxies detected through the SZE by {\it Planck}. 
The estimates for the masses from {\it Planck} are based on the
scaling relation between the X-ray hydrostatic mass and $Y_X$, the
product of the X-ray temperature and the gas mass. This relation is
calibrated using measurements from \cite{Arnaud10}, who studied a
sample of 20 nearby relaxed clusters. Consequently a measurement of
$Y_X$ can be converted into a hydrostatic mass estimate
$M_{500}^{Y_X}$. These results are then used to establish the relation
between the SZ signal $Y_{500}$ and $M_{500}^{Y_X}$ \citep[see
Appendix A.2.2 of][for details]{PlanckXX}. We denote this hydrodynamic
mass estimate as $M_{500}^{Planck}$.

Numerical simulations \citep[e.g.][]{Rasia06, Nagai07,Lau09} suggest
that such mass estimates are biased low. Similarly, \cite{Mahdavi13}
studied the scaling relations between X-ray observations and the weak
lensing masses from H12 and found that hydrostatic masses
underestimate the weak lensing masses by $10-15\%$ at $r_{500}$. Our
updated masses do not change this conclusion, and in fact strengthen
it. In their analysis \cite{PlanckXX} assume that the hydrostatic
masses are biased low by a factor $(1-b)=0.8$ based on a comparison
with numerical simulations. They find that their best fit parameters
for $\sigma_8$ and $\Omega_m$ are in tension with the measurements
obtained from the analysis of the primary CMB by \cite{PlanckXVI}. The
results can be reconciled by considering a low value of $(1-b)\sim
0.6$.

Recently, \cite{vonderLinden14b} estimated the bias using the lensing
masses for the 38 clusters in common between {\it Planck} and WtG.
They compared their estimates for $M_{500}$ based on the NFW fits with
$c_{200}=4$ from \cite{Applegate14} to the hydrostatic mass estimates
from \cite{PlanckXXIX}. They obtained an average ratio
$(1-b)=0.69\pm0.07$, which alleviates the tension. As our
comparison in \S\ref{sec:comparison} and Fig.~\ref{fig:comp_wtg}
shows, the WtG masses are slightly higher than our estimates when we
follow the same approach, but when we compare the masses from WtG
to our deprojected aperture masses, which are more robust and
therefore used here, we find that the agreement is excellent.

There are 38 clusters in common between CCCP and the catalog provided
by \cite{PlanckXXIX}, although we omit Abell~115 from the comparison
as we determine masses for the two separate components of this merging
cluster. The left panel in Figure~\ref{fig:planck} shows the
deprojected aperture mass $M^{\rm WL}_{500}$ as a function of the
hydrostatic mass $M^{\rm Planck}_{500}$ from \cite{PlanckXXIX}. Note
that the observed value for $Y_X$ was used to estimate the radius
$r_{500}$ used to determine $M^{Planck}_{500}$, whereas $M^{\rm WL}_{500}$
is based on the value for $r_{500}$ listed in
Table~\ref{tab:masses}. For the cosmological analysis, \cite{PlanckXX}
restricted the sample to clusters above a SNR threshold of 7 in
unmasked areas. In our case, the mask only impacts the merging cluster
Abell~2163, which corresponds to the right-most point in
Figure~\ref{fig:planck}. There are 20 SNR$>7$ clusters in common with
CCCP and these are indicated as filled points in
Figure~\ref{fig:planck}, whereas the remaining clusters are indicated
by the open points. We find that the SNR threshold is essentially a
selection by mass. For reference, the measurements from
\cite{vonderLinden14b} are indicated by the rosy brown colored points.

The right panel shows the ratio of the hydrostatic masses from {\it
  Planck} and our weak lensing estimates for all 37 clusters in
common. The hatched region indicates our estimate for
$(1-b)=0.76\pm0.05{\rm~(stat)}\pm0.06{\rm~(syst)}$, which was obtained
from a linear fit to $M^{\it Planck}_{500}$ as a function of $M^{\rm
  WL}_{500}$ that accounts for intrinsic scatter \citep{Hogg10}.  The
systematic error is based on the estimates presented in
\S\ref{sec:errorbudget}. We measure an intrinsic scatter of
$(28\pm6)\%$, most of which can be attributed to the triaxial nature
of dark matter halos \citep[e.g.][]{Corless07,Meneghetti10}. If we
restrict the comparison to the clusters with SNR$>7$ (black points) we
obtain $(1-b)=0.78\pm0.07$, whereas $(1-b)=0.69\pm0.05$ for the
remaining clusters.  For reference, the rosy brown colored points and
hatched region indicate the results for WtG, used in
\cite{vonderLinden14b}. We refit these measurements, which
  yields $(1-b)=0.62\pm0.04$ and an intrinsic scatter of
$(26\pm5)\%$.  Our measurement of the bias is in agreement with the
nominal value adopted by \cite{PlanckXX} and we conclude that a large
bias in the hydrostatic mass estimate is unlikely to be the
explanation of the tension of the cluster counts and the primary CMB.

\cite{vonderLinden14b} find modest evidence for a mass dependence of
the bias, with $M_{\it Planck}\propto M_{\rm WtG}^{0.68}$. It is
therefore interesting to repeat this for our measurements. If we
restrict the fit to the clusters with a SNR$>7$, the range is too
small to obtain a useful constraint on the slope. We therefore fit a
power law to the CCCP measurements of the 37 clusters that overlap
with \cite{PlanckXXIX}, which yields
$$\frac{M_{\it Planck}}{10^{15} h_{70}^{-1}{\rm M}_\odot}=(0.76\pm0.04)\times 
\left(\frac{M_{\rm CCCP}}{10^{15} h_{70}^{-1}{\rm M}_\odot}\right)^{0.64\pm0.17},$$
\noindent and an intrinsic scatter of $(21\pm 4)\%$. The slope is
similar to that found by \cite{vonderLinden14b} and our results
therefore support their conclusion that the bias in the hydrostatic
masses used by {\it Planck} depends on the cluster mass, but our
normalization is $9\%$ higher.

As noted above, \cite{PlanckXX} use X-ray data to relate the observed
SZ-signal to cluster mass. It is, however, more convenient to directly
constrain the scaling relation between the lensing mass and the
observed SZ signal. H12 presented results for the early data release
from \cite{Planck11}, but here we expand the sample to the 37 clusters
in common with CCCP and use the measurements for $Y_{500}$ provided by
\cite{PlanckXXIX}.

\begin{figure}
\centering
\leavevmode \hbox{%
\includegraphics[width=8.5cm]{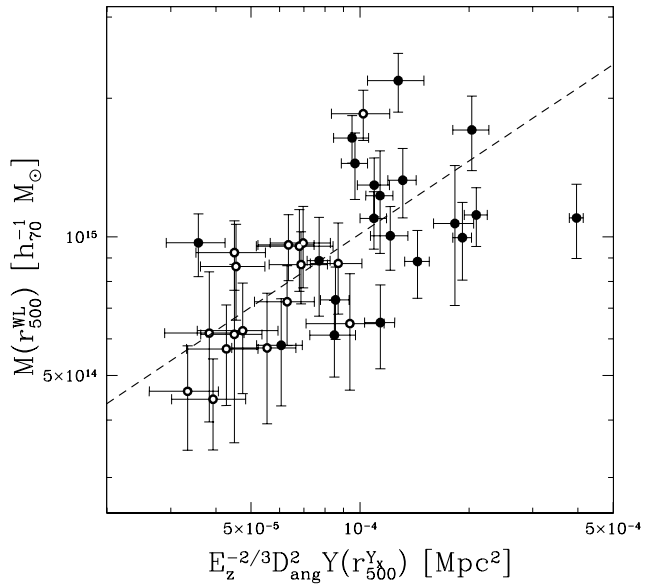}}
\caption{Plot of $M_{500}$, the aperture mass estimate within an
aperture $r_{500}^{\rm WL}$, as a function of the SZ signal measured within an 
aperture $r_{500}^{Y_X}$ from \citet{PlanckXXIX}. The dashed line indicates the 
best fit power law, which has a slope of $0.53\pm0.13$. We measure an intrinsic
scatter of $(27\pm 6)\%$.  The open circles indicate clusters which
{\it Planck} detected with a SNR$<7$.\label{fig:m_y}}
\end{figure}

Assuming a constant gas fraction and self-similarity, the SZ signal
$Y$ scales with mass as $M_{500}^{5/3}\propto D_{\rm
  ang}^2E(z)^{-2/3}Y_{500}$ \citep[e.g.][]{McCarthy03}, where $D_{\rm
  ang}$ is the angular diameter distance to the cluster and
$E(z)=H(z)/H_0$ is the normalized Hubble parameter. The results are
presented in Figure~\ref{fig:m_y}, where the open points indicate the
clusters that {\it Planck} detected with a SNR$<7$. The dashed line
indicates the best-fit power law relation to all clusters in common,
for which we find
$$\frac{M_{500}^{\rm WL}}{10^{15} h_{70}^{-1}{\rm M}_\odot}=(1.01\pm 0.06)\times 
\left(\frac{10^4 D_{\rm ang}^2 Y_{500}}{E(z)^{2/3}{\rm
      Mpc}^2}\right)^{0.53\pm0.13},$$
\noindent and an intrinsic scatter of $(27\pm 6)\%$. For this
comparison we used measurements of the SZ signal and the lensing mass
in apertures that were determined independently. Although this is what
one needs for the cosmological interpretation of the {\it Planck}
cluster sample, the use of different apertures introduces additional
noise as well as an offset because the lensing aperture is larger on
average.

In the case of hydrodynamic simulations of clusters the comparison can
be made at a common radius, as $r_{500}$ is known. It is therefore
useful to consider also the scaling relation for the SZ signal within
$r_{500}^{\rm WL}$. The SZ measurements within $r_{500}^{\rm WL}$ were
estimated (Arnaud \& Pratt, private communication) using the
two-dimensional marginal probability distribution between the SZ
signal and size available from the 2013 Planck SZ catalog
\citep{PlanckXXIX}. They correlate very well with $Y_{500}$.  The best
fit power-law scaling relation is given by
$$\frac{M_{500}^{\rm WL}}{10^{15} h_{70}^{-1}{\rm M}_\odot}=(0.98\pm 0.05)\times 
\left(\frac{10^4 D_{\rm ang}^2 Y(r_{500}^{\rm WL})}{E(z)^{2/3}{\rm
      Mpc}^2}\right)^{0.64\pm0.12},$$
\noindent and the intrinsic scatter is reduced to $(22\pm 7)\%$. Note
that for both scaling relations the slopes are consistent with the
value of 0.6 expected for a self-similar model.

As a caveat, we note that the combination of relatively low
significance of the SZ detections and large intrinsic scatter leads to
Malmquist bias. As a result the average SZ signal of the observed
sample is biased high compared to the average of the parent population
the clusters are drawn from. Consequently, the best fit parameters
for the scaling relations are expected to be slightly biased. If the
CCCP sample were a well-defined sample one could account for this,
which was done by \cite{PlanckXX}. Although \cite{Mahdavi13} show that
the CCCP sample is representative of more carefully selected samples
of X-ray luminous clusters, the selection may impact the scaling
relation. We therefore do not attempt to correct our results, but note
that, based on the findings of \cite{PlanckXX} and \cite{Mahdavi13},
we expect the bias to be small.

\section{Conclusions}

Accurate cluster masses are necessary to interpret the cluster counts
from wide-area surveys. In particular the scaling relations and their
scatter need to be determined observationally. Weak lensing masses are
particularly well-suited as they provide a direct estimate of the
projected mass and do not depend on the dynamical state of the cluster.
In this paper we revisited the analysis of a sample of 50 massive
clusters by \cite{Hoekstra12}, with a particular focus on improving
the corrections for various sources of systematic error in the cluster
masses.

We use extensive image simulations to quantify the bias in our shape
measurement algorithm. The bias is a strong function of
signal-to-noise ratio and size, but depends relatively weakly on
surface brightness profile and ellipticity distribution.  We
demonstrate that the inferred bias depends on the input parameters
used to create the simulated data. For instance we find that the bias
is underestimated if faint galaxies are lacking from the simulations:
the bias converges if the simulation includes galaxies that are at
least 1.5 magnitude fainter than the limiting magnitude of the sample
of sources used for the lensing analysis. The large number of
simulated galaxies enables us to determine an empirical correction,
which is found to be accurate and robust to the main uncertainties. We
estimate that the systematic uncertainty due to the shape measurements
is at most $2\%$.

The dominant source of systematic error is the source redshift
distribution, which is needed to convert the lensing signal into an estimate of
the mass. We use the latest state-of-the-art photometric redshift
catalogs that are based on measurements in 29 bands in the COSMOS
field \citep{Scoville07,Capak07}, including new deep near-infrared
observations from UltraVISTA \citep{McCracken12}. Compared to our
previous analysis, this leads to a small increase in mean source
redshift, or a modest ($\sim 4\%$) reduction in the cluster mass.
Despite the unprecedented quality of the data, different analyses show
variations in the source redshift distributions that result in
systematic uncertainties that are substantial compared to the
statistical uncertainties for the full sample of clusters.

We find that the projected aperture masses within apertures of fixed
radius provide the most robust measurements. The dependence on the
assumed density profile is minimal and the systematic error is
dominated by the uncertainty in the source redshift distribution. We
estimate a total systematic error of $4.2\%$ for clusters at $z=0.2$,
which increases to $8.5\%$ at $z=0.55$. To compare to measurements at
other wavelenghts, we deproject the masses. This leads to an increased
sensitivity to the assumed density profile, although this is modest in
the case of $M_{500}$: we estimate a systematic uncertainty of
$2\%$. Compared to the masses within a fixed aperture, the additional
uncertainty in determining $r_{500}$ increases both the statistical
and systematic errors. Comparison of the deprojected aperture masses
within $r_{500}$ and the corresponding mass for the best fit NFW
profile shows good agreement, even though the two estimates are nearly
independent from each other.

We compare masses for the 18 clusters in common with
\cite{Applegate14}. To do so we fit an NFW model with $c=4$ to our
lensing signal within $0.75-3 h_{70}^{-1}$Mpc. We find that the
resulting CCCP masses are on average $8\pm 4$ per cent lower, which
reduces to $6\%$ if we use the same source redshift distribution as
\cite{Applegate14}. Given the limitations of the comparison, it is not
clear whether this difference is a sign of residual systematics or
merely statistical in nature.

We did examine the robustness of our analysis. To this end we
determined masses for Abell~1758 and the six clusters in common with
\cite{Umetsu14} using available Subaru observations. We found good
agreement with our mass estimates based on the CFHT data, suggesting
our pipeline yields robust results when data from different telescopes
are analysed.  Interestingly, a direct comparison of our CFHT
measurements to catalogs provided by the WtG team (Von der Linden,
private communication) also showed remarkable agreement with $\langle
g_T^{\rm CCCP}/g_T^{\rm WtG}\rangle=0.991\pm0.018$.

The overlap of 37 clusters with the sample of clusters for which {\it
  Planck} detected the SZ signal \citep{PlanckXXIX} enables us to
calibrate the bias in the hydrostatic masses used by \cite{PlanckXX}
to infer cosmological parameters. The resulting estimates for
$\sigma_8$ and $\Omega_m$ are in tension with the measurements from
the primary CMB \cite{PlanckXVI}. Our measurements for the overlapping
clusters yield $1-b=0.76\pm0.05{\rm (stat)}\pm0.06{\rm (syst)}$, in
good agreement with the nominal value used by \cite{PlanckXX}. Our
results do not support a large bias in the hydrostatic masses, which
could alleviate the tension. We also directly calibrate the scaling
relation between the SZ signal $Y_{500}$ and the weak lensing mass.
When we compare the lensing mass to the SZ signal measured in the same
aperture, we find a best fit slope of $0.64\pm0.12$ which is in good
agreement with the expectation of a self-similar model
\citep[e.g.][]{McCarthy03}.

The constraints from the current CCCP sample are already limited by
systematic uncertainties. The most dominant of these is our limited
knowledge of the source redshift distribution. Although this can be
alleviated by measuring photometric redshifts for the sources in the
cluster fields \citep[e.g.][]{Applegate14,Umetsu14}, biases may remain
due to limited wavelength coverage. Improving this situation is
critical to calibrate cluster scaling relations to the level of
accuracy afforded by the next generation of cluster surveys.

\vspace{0.5cm} 

We thank Monique Arnaud, Gabriel Pratt, Keiichi Umetsu, James Jee, Edo
van Uitert, Anja von der Linden, Doug Applegate, Ken Rines and Douglas
Scott for useful discussions. We are also grateful to Keiichi Umetsu
and James Jee for providing some of their Subaru images, and to 
Anja von der Linden for sharing the WtG lensing catalogs. HH and MC
were supported by the Netherlands organisation for Scientific Research
(NWO) Vidi grant 639.042.814; HH, RH and MV acknowledge support from
the European Research Council FP7 grant number 279396. AB would like
to acknowledge research funding from NSERC Canada through its
Discovery Grant program. AM acknowledges support from NASA ADAP grant
NNX12AE45G. This research used the facilities of the Canadian
Astronomy Data Centre operated by the National Research Council of
Canada with the support of the Canadian Space Agency.

\bibliographystyle{mn2e}
\bibliography{cal}

\begin{thebibliography}{111}
\expandafter\ifx\csname natexlab\endcsname\relax\def\natexlab#1{#1}\fi

\bibitem[{{Allen} {et~al}\mbox{.}(2011){Allen}, {Evrard}, \& {Mantz}}]{Allen11}
{Allen} S.~W., {Evrard} A.~E., {Mantz} A.~B., 2011, \araa, 49, 409

\bibitem[{{Applegate} {et~al}\mbox{.}(2014){Applegate}, {von der Linden},
  {Kelly}, {Allen}, {Allen}, {Burchat}, {Burke}, {Ebeling}, {Mantz}, \&
  {Morris}}]{Applegate14}
{Applegate} D.~E. {et~al.}, 2014, \mnras, 439, 48

\bibitem[{{Arnaud} {et~al}\mbox{.}(2010){Arnaud}, {Pratt}, {Piffaretti},
  {B{\"o}hringer}, {Croston}, \& {Pointecouteau}}]{Arnaud10}
{Arnaud} M., {Pratt} G.~W., {Piffaretti} R., {B{\"o}hringer} H., {Croston}
  J.~H., {Pointecouteau} E., 2010, \aap, 517, A92

\bibitem[{{Bahcall} \& {Fan}(1998)}]{Bahcall98}
{Bahcall} N.~A., {Fan} X., 1998, \apj, 504, 1

\bibitem[{{Bartelmann} \& {Schneider}(2001)}]{Bartelmann01}
{Bartelmann} M., {Schneider} P., 2001, \physrep, 340, 291

\bibitem[{{Becker} \& {Kravtsov}(2011)}]{Becker11}
{Becker} M.~R., {Kravtsov} A.~V., 2011, \apj, 740, 25

\bibitem[{{Bernstein}(2010)}]{Bernstein10}
{Bernstein} G.~M., 2010, \mnras, 406, 2793

\bibitem[{{Bernstein} \& {Armstrong}(2014)}]{Bernstein14}
{Bernstein} G.~M., {Armstrong} R., 2014, \mnras, 438, 1880

\bibitem[{{Bertin} \& {Arnouts}(1996)}]{Bertin96}
{Bertin} E., {Arnouts} S., 1996, \aaps, 117, 393

\bibitem[{{Bhattacharya} {et~al}\mbox{.}(2013){Bhattacharya}, {Habib},
  {Heitmann}, \& {Vikhlinin}}]{Bhattacharya13}
{Bhattacharya} S., {Habib} S., {Heitmann} K., {Vikhlinin} A., 2013, \apj, 766,
  32

\bibitem[{{Brammer} {et~al}\mbox{.}(2008){Brammer}, {van Dokkum}, \&
  {Coppi}}]{Brammer08}
{Brammer} G.~B., {van Dokkum} P.~G., {Coppi} P., 2008, \apj, 686, 1503

\bibitem[{{Bridle} {et~al}\mbox{.}(2010){Bridle}, {Balan}, {Bethge}, {Gentile},
  {Harmeling}, {Heymans}, {Hirsch}, {Hosseini}, {Jarvis}, {Kirk}, {Kitching},
  {Kuijken}, {Lewis}, {Paulin-Henriksson}, {Sch{\"o}lkopf}, {Velander},
  {Voigt}, {Witherick}, {Amara}, {Bernstein}, {Courbin}, {Gill}, {Heavens},
  {Mandelbaum}, {Massey}, {Moghaddam}, {Rassat}, {R{\'e}fr{\'e}gier}, {Rhodes},
  {Schrabback}, {Shawe-Taylor}, {Shmakova}, {van Waerbeke}, \&
  {Wittman}}]{GREAT08}
{Bridle} S. {et~al.}, 2010, \mnras, 405, 2044

\bibitem[{{Cacciato} {et~al}\mbox{.}(2009){Cacciato}, {van den Bosch}, {More},
  {Li}, {Mo}, \& {Yang}}]{Cacciato09}
{Cacciato} M., {van den Bosch} F.~C., {More} S., {Li} R., {Mo} H.~J., {Yang}
  X., 2009, \mnras, 394, 929

\bibitem[{{Cacciato} {et~al}\mbox{.}(2013){Cacciato}, {van den Bosch}, {More},
  {Mo}, \& {Yang}}]{Cacciato13}
{Cacciato} M., {van den Bosch} F.~C., {More} S., {Mo} H., {Yang} X., 2013,
  \mnras, 430, 767

\bibitem[{{Cacciato} {et~al}\mbox{.}(2014){Cacciato}, {van Uitert}, \&
  {Hoekstra}}]{Cacciato14}
{Cacciato} M., {van Uitert} E., {Hoekstra} H., 2014, \mnras, 437, 377

\bibitem[{{Capak} {et~al}\mbox{.}(2007){Capak}, {Aussel}, {Ajiki}, {McCracken},
  {Mobasher}, {Scoville}, {Shopbell}, {Taniguchi}, {Thompson}, {Tribiano},
  {Sasaki}, {Blain}, {Brusa}, {Carilli}, {Comastri}, {Carollo}, {Cassata},
  {Colbert}, {Ellis}, {Elvis}, {Giavalisco}, {Green}, {Guzzo}, {Hasinger},
  {Ilbert}, {Impey}, {Jahnke}, {Kartaltepe}, {Kneib}, {Koda}, {Koekemoer},
  {Komiyama}, {Leauthaud}, {Le Fevre}, {Lilly}, {Liu}, {Massey}, {Miyazaki},
  {Murayama}, {Nagao}, {Peacock}, {Pickles}, {Porciani}, {Renzini}, {Rhodes},
  {Rich}, {Salvato}, {Sanders}, {Scarlata}, {Schiminovich}, {Schinnerer},
  {Scodeggio}, {Sheth}, {Shioya}, {Tasca}, {Taylor}, {Yan}, \&
  {Zamorani}}]{Capak07}
{Capak} P. {et~al.}, 2007, \apjs, 172, 99

\bibitem[{{Clowe} {et~al}\mbox{.}(1998){Clowe}, {Luppino}, {Kaiser}, {Henry},
  \& {Gioia}}]{Clowe98}
{Clowe} D., {Luppino} G.~A., {Kaiser} N., {Henry} J.~P., {Gioia} I.~M., 1998,
  \apjl, 497, L61

\bibitem[{{Coe} {et~al}\mbox{.}(2012){Coe}, {Umetsu}, {Zitrin}, {Donahue},
  {Medezinski}, {Postman}, {Carrasco}, {Anguita}, {Geller}, {Rines},
  {Diaferio}, {Kurtz}, {Bradley}, {Koekemoer}, {Zheng}, {Nonino}, {Molino},
  {Mahdavi}, {Lemze}, {Infante}, {Ogaz}, {Melchior}, {Host}, {Ford}, {Grillo},
  {Rosati}, {Jim{\'e}nez-Teja}, {Moustakas}, {Broadhurst}, {Ascaso}, {Lahav},
  {Bartelmann}, {Ben{\'{\i}}tez}, {Bouwens}, {Graur}, {Graves}, {Jha},
  {Jouvel}, {Kelson}, {Moustakas}, {Maoz}, {Meneghetti}, {Merten}, {Riess},
  {Rodney}, \& {Seitz}}]{Coe12}
{Coe} D. {et~al.}, 2012, \apj, 757, 22

\bibitem[{{Cooray} \& {Sheth}(2002)}]{Cooray02}
{Cooray} A., {Sheth} R., 2002, \physrep, 372, 1

\bibitem[{{Corless} \& {King}(2007)}]{Corless07}
{Corless} V.~L., {King} L.~J., 2007, \mnras, 380, 149

\bibitem[{{Diaferio}(1999)}]{Diaferio99}
{Diaferio} A., 1999, \mnras, 309, 610

\bibitem[{{Duffy} {et~al}\mbox{.}(2008){Duffy}, {Schaye}, {Kay}, \& {Dalla
  Vecchia}}]{Duffy08}
{Duffy} A.~R., {Schaye} J., {Kay} S.~T., {Dalla Vecchia} C., 2008, \mnras, 390,
  L64

\bibitem[{{Duffy} {et~al}\mbox{.}(2010){Duffy}, {Schaye}, {Kay}, {Dalla
  Vecchia}, {Battye}, \& {Booth}}]{Duffy10}
{Duffy} A.~R., {Schaye} J., {Kay} S.~T., {Dalla Vecchia} C., {Battye} R.~A.,
  {Booth} C.~M., 2010, \mnras, 405, 2161

\bibitem[{{Dutton} \& {Macci{\`o}}(2014)}]{Dutton14}
{Dutton} A.~A., {Macci{\`o}} A.~V., 2014, \mnras, 441, 3359

\bibitem[{{Fern{\'a}ndez-Soto} {et~al}\mbox{.}(1999){Fern{\'a}ndez-Soto},
  {Lanzetta}, \& {Yahil}}]{HDF}
{Fern{\'a}ndez-Soto} A., {Lanzetta} K.~M., {Yahil} A., 1999, \apj, 513, 34

\bibitem[{{Gruen} {et~al}\mbox{.}(2014){Gruen}, {Seitz}, {Brimioulle},
  {Kosyra}, {Koppenhoefer}, {Lee}, {Bender}, {Riffeser}, {Eichner},
  {Weidinger}, \& {Bierschenk}}]{Gruen14}
{Gruen} D. {et~al.}, 2014, \mnras, 442, 1507

\bibitem[{{Hasselfield} {et~al}\mbox{.}(2013){Hasselfield}, {Hilton},
  {Marriage}, {Addison}, {Barrientos}, {Battaglia}, {Battistelli}, {Bond},
  {Crichton}, {Das}, {Devlin}, {Dicker}, {Dunkley}, {D{\"u}nner}, {Fowler},
  {Gralla}, {Hajian}, {Halpern}, {Hincks}, {Hlozek}, {Hughes}, {Infante},
  {Irwin}, {Kosowsky}, {Marsden}, {Menanteau}, {Moodley}, {Niemack}, {Nolta},
  {Page}, {Partridge}, {Reese}, {Schmitt}, {Sehgal}, {Sherwin}, {Sievers},
  {Sif{\'o}n}, {Spergel}, {Staggs}, {Swetz}, {Switzer}, {Thornton}, {Trac}, \&
  {Wollack}}]{Hasselfield13}
{Hasselfield} M. {et~al.}, 2013, \jcap, 7, 8

\bibitem[{{Henry}(2000)}]{Henry00}
{Henry} J.~P., 2000, \apj, 534, 565

\bibitem[{{Henry} {et~al}\mbox{.}(2009){Henry}, {Evrard}, {Hoekstra}, {Babul},
  \& {Mahdavi}}]{Henry09}
{Henry} J.~P., {Evrard} A.~E., {Hoekstra} H., {Babul} A., {Mahdavi} A., 2009,
  \apj, 691, 1307

\bibitem[{{Heymans} {et~al}\mbox{.}(2006){Heymans}, {Van Waerbeke}, {Bacon},
  {Berge}, {Bernstein}, {Bertin}, {Bridle}, {Brown}, {Clowe}, {Dahle}, {Erben},
  {Gray}, {Hetterscheidt}, {Hoekstra}, {Hudelot}, {Jarvis}, {Kuijken},
  {Margoniner}, {Massey}, {Mellier}, {Nakajima}, {Refregier}, {Rhodes},
  {Schrabback}, \& {Wittman}}]{STEP1}
{Heymans} C. {et~al.}, 2006, \mnras, 368, 1323

\bibitem[{{Hildebrandt} {et~al}\mbox{.}(2010){Hildebrandt}, {Arnouts}, {Capak},
  {Moustakas}, {Wolf}, {Abdalla}, {Assef}, {Banerji}, {Ben{\'{\i}}tez},
  {Brammer}, {Budav{\'a}ri}, {Carliles}, {Coe}, {Dahlen}, {Feldmann}, {Gerdes},
  {Gillis}, {Ilbert}, {Kotulla}, {Lahav}, {Li}, {Miralles}, {Purger},
  {Schmidt}, \& {Singal}}]{Hildebrandt10}
{Hildebrandt} H. {et~al.}, 2010, \aap, 523, A31

\bibitem[{{Hoekstra}(2001)}]{Hoekstra01}
{Hoekstra} H., 2001, \aap, 370, 743

\bibitem[{{Hoekstra}(2003)}]{Hoekstra03}
{Hoekstra} H., 2003, \mnras, 339, 1155

\bibitem[{{Hoekstra}(2004)}]{Hoekstra04}
{Hoekstra} H., 2004, \mnras, 347, 1337

\bibitem[{{Hoekstra}(2007)}]{Hoekstra07}
{Hoekstra} H., 2007, \mnras, 379, 317

\bibitem[{{Hoekstra} {et~al}\mbox{.}(2013){Hoekstra}, {Bartelmann}, {Dahle},
  {Israel}, {Limousin}, \& {Meneghetti}}]{Hoekstra13}
{Hoekstra} H., {Bartelmann} M., {Dahle} H., {Israel} H., {Limousin} M.,
  {Meneghetti} M., 2013, \ssr, 177, 75

\bibitem[{{Hoekstra} {et~al}\mbox{.}(2000){Hoekstra}, {Franx}, \&
  {Kuijken}}]{Hoekstra00}
{Hoekstra} H., {Franx} M., {Kuijken} K., 2000, \apj, 532, 88

\bibitem[{{Hoekstra} {et~al}\mbox{.}(1998){Hoekstra}, {Franx}, {Kuijken}, \&
  {Squires}}]{Hoekstra98}
{Hoekstra} H., {Franx} M., {Kuijken} K., {Squires} G., 1998, \apj, 504, 636

\bibitem[{{Hoekstra} {et~al}\mbox{.}(2011){Hoekstra}, {Hartlap}, {Hilbert}, \&
  {van Uitert}}]{Hoekstra11}
{Hoekstra} H., {Hartlap} J., {Hilbert} S., {van Uitert} E., 2011, \mnras, 412,
  2095

\bibitem[{{Hoekstra} {et~al}\mbox{.}(2012){Hoekstra}, {Mahdavi}, {Babul}, \&
  {Bildfell}}]{Hoekstra12}
{Hoekstra} H., {Mahdavi} A., {Babul} A., {Bildfell} C., 2012, \mnras, 427, 1298

\bibitem[{{Hogg} {et~al}\mbox{.}(2010){Hogg}, {Bovy}, \& {Lang}}]{Hogg10}
{Hogg} D.~W., {Bovy} J., {Lang} D., 2010, ArXiv:1008.4686

\bibitem[{{Hogg} {et~al}\mbox{.}(1997){Hogg}, {Pahre}, {McCarthy}, {Cohen},
  {Blandford}, {Smail}, \& {Soifer}}]{Hogg97}
{Hogg} D.~W., {Pahre} M.~A., {McCarthy} J.~K., {Cohen} J.~G., {Blandford} R.,
  {Smail} I., {Soifer} B.~T., 1997, \mnras, 288, 404

\bibitem[{{Ilbert} {et~al}\mbox{.}(2006){Ilbert}, {Arnouts}, {McCracken},
  {Bolzonella}, {Bertin}, {Le F{\`e}vre}, {Mellier}, {Zamorani}, {Pell{\`o}},
  {Iovino}, {Tresse}, {Le Brun}, {Bottini}, {Garilli}, {Maccagni}, {Picat},
  {Scaramella}, {Scodeggio}, {Vettolani}, {Zanichelli}, {Adami}, {Bardelli},
  {Cappi}, {Charlot}, {Ciliegi}, {Contini}, {Cucciati}, {Foucaud}, {Franzetti},
  {Gavignaud}, {Guzzo}, {Marano}, {Marinoni}, {Mazure}, {Meneux}, {Merighi},
  {Paltani}, {Pollo}, {Pozzetti}, {Radovich}, {Zucca}, {Bondi}, {Bongiorno},
  {Busarello}, {de La Torre}, {Gregorini}, {Lamareille}, {Mathez}, {Merluzzi},
  {Ripepi}, {Rizzo}, \& {Vergani}}]{Ilbert06}
{Ilbert} O. {et~al.}, 2006, \aap, 457, 841

\bibitem[{{Ilbert} {et~al}\mbox{.}(2009){Ilbert}, {Capak}, {Salvato}, {Aussel},
  {McCracken}, {Sanders}, {Scoville}, {Kartaltepe}, {Arnouts}, {Le Floc'h},
  {Mobasher}, {Taniguchi}, {Lamareille}, {Leauthaud}, {Sasaki}, {Thompson},
  {Zamojski}, {Zamorani}, {Bardelli}, {Bolzonella}, {Bongiorno}, {Brusa},
  {Caputi}, {Carollo}, {Contini}, {Cook}, {Coppa}, {Cucciati}, {de la Torre},
  {de Ravel}, {Franzetti}, {Garilli}, {Hasinger}, {Iovino}, {Kampczyk},
  {Kneib}, {Knobel}, {Kovac}, {Le Borgne}, {Le Brun}, {F{\`e}vre}, {Lilly},
  {Looper}, {Maier}, {Mainieri}, {Mellier}, {Mignoli}, {Murayama}, {Pell{\`o}},
  {Peng}, {P{\'e}rez-Montero}, {Renzini}, {Ricciardelli}, {Schiminovich},
  {Scodeggio}, {Shioya}, {Silverman}, {Surace}, {Tanaka}, {Tasca}, {Tresse},
  {Vergani}, \& {Zucca}}]{Ilbert09}
{Ilbert} O. {et~al.}, 2009, \apj, 690, 1236

\bibitem[{{Ilbert} {et~al}\mbox{.}(2013){Ilbert}, {McCracken}, {Le F{\`e}vre},
  {Capak}, {Dunlop}, {Karim}, {Renzini}, {Caputi}, {Boissier}, {Arnouts},
  {Aussel}, {Comparat}, {Guo}, {Hudelot}, {Kartaltepe}, {Kneib}, {Krogager},
  {Le Floc'h}, {Lilly}, {Mellier}, {Milvang-Jensen}, {Moutard}, {Onodera},
  {Richard}, {Salvato}, {Sanders}, {Scoville}, {Silverman}, {Taniguchi},
  {Tasca}, {Thomas}, {Toft}, {Tresse}, {Vergani}, {Wolk}, \& {Zirm}}]{Ilbert13}
{Ilbert} O. {et~al.}, 2013, \aap, 556, A55

\bibitem[{{Jee} {et~al}\mbox{.}(2014){Jee}, {Hoekstra}, {Mahdavi}, \&
  {Babul}}]{Jee14}
{Jee} M.~J., {Hoekstra} H., {Mahdavi} A., {Babul} A., 2014, \apj, 783, 78

\bibitem[{{Jee} {et~al}\mbox{.}(2012){Jee}, {Mahdavi}, {Hoekstra}, {Babul},
  {Dalcanton}, {Carroll}, \& {Capak}}]{Jee12}
{Jee} M.~J., {Mahdavi} A., {Hoekstra} H., {Babul} A., {Dalcanton} J.~J.,
  {Carroll} P., {Capak} P., 2012, \apj, 747, 96

\bibitem[{{Kacprzak} {et~al}\mbox{.}(2014){Kacprzak}, {Bridle}, {Rowe},
  {Voigt}, {Zuntz}, {Hirsch}, \& {MacCrann}}]{Kacprzak14}
{Kacprzak} T., {Bridle} S., {Rowe} B., {Voigt} L., {Zuntz} J., {Hirsch} M.,
  {MacCrann} N., 2014, \mnras, 441, 2528

\bibitem[{{Kacprzak} {et~al}\mbox{.}(2012){Kacprzak}, {Zuntz}, {Rowe},
  {Bridle}, {Refregier}, {Amara}, {Voigt}, \& {Hirsch}}]{Kacprzak12}
{Kacprzak} T., {Zuntz} J., {Rowe} B., {Bridle} S., {Refregier} A., {Amara} A.,
  {Voigt} L., {Hirsch} M., 2012, \mnras, 427, 2711

\bibitem[{{Kaiser} {et~al}\mbox{.}(1995){Kaiser}, {Squires}, \&
  {Broadhurst}}]{KSB95}
{Kaiser} N., {Squires} G., {Broadhurst} T., 1995, \apj, 449, 460

\bibitem[{{Kelly} {et~al}\mbox{.}(2014){Kelly}, {von der Linden}, {Applegate},
  {Allen}, {Allen}, {Burchat}, {Burke}, {Ebeling}, {Capak}, {Czoske},
  {Donovan}, {Mantz}, \& {Morris}}]{Kelly14}
{Kelly} P.~L. {et~al.}, 2014, \mnras, 439, 28

\bibitem[{{Kitching} {et~al}\mbox{.}(2012){Kitching}, {Balan}, {Bridle},
  {Cantale}, {Courbin}, {Eifler}, {Gentile}, {Gill}, {Harmeling}, {Heymans},
  {Hirsch}, {Honscheid}, {Kacprzak}, {Kirkby}, {Margala}, {Massey}, {Melchior},
  {Nurbaeva}, {Patton}, {Rhodes}, {Rowe}, {Taylor}, {Tewes}, {Viola},
  {Witherick}, {Voigt}, {Young}, \& {Zuntz}}]{GREAT10}
{Kitching} T.~D. {et~al.}, 2012, \mnras, 423, 3163

\bibitem[{{Komatsu} {et~al}\mbox{.}(2009){Komatsu}, {Dunkley}, {Nolta},
  {Bennett}, {Gold}, {Hinshaw}, {Jarosik}, {Larson}, {Limon}, {Page},
  {Spergel}, {Halpern}, {Hill}, {Kogut}, {Meyer}, {Tucker}, {Weiland},
  {Wollack}, \& {Wright}}]{Komatsu09}
{Komatsu} E. {et~al.}, 2009, \apjs, 180, 330

\bibitem[{{Lau} {et~al}\mbox{.}(2009){Lau}, {Kravtsov}, \& {Nagai}}]{Lau09}
{Lau} E.~T., {Kravtsov} A.~V., {Nagai} D., 2009, \apj, 705, 1129

\bibitem[{{Lilly} {et~al}\mbox{.}(2009){Lilly}, {Le Brun}, {Maier}, {Mainieri},
  {Mignoli}, {Scodeggio}, {Zamorani}, {Carollo}, {Contini}, {Kneib}, {Le
  F{\`e}vre}, {Renzini}, {Bardelli}, {Bolzonella}, {Bongiorno}, {Caputi},
  {Coppa}, {Cucciati}, {de la Torre}, {de Ravel}, {Franzetti}, {Garilli},
  {Iovino}, {Kampczyk}, {Kovac}, {Knobel}, {Lamareille}, {Le Borgne}, {Pello},
  {Peng}, {P{\'e}rez-Montero}, {Ricciardelli}, {Silverman}, {Tanaka}, {Tasca},
  {Tresse}, {Vergani}, {Zucca}, {Ilbert}, {Salvato}, {Oesch}, {Abbas},
  {Bottini}, {Capak}, {Cappi}, {Cassata}, {Cimatti}, {Elvis}, {Fumana},
  {Guzzo}, {Hasinger}, {Koekemoer}, {Leauthaud}, {Maccagni}, {Marinoni},
  {McCracken}, {Memeo}, {Meneux}, {Porciani}, {Pozzetti}, {Sanders},
  {Scaramella}, {Scarlata}, {Scoville}, {Shopbell}, \& {Taniguchi}}]{Lilly09}
{Lilly} S.~J. {et~al.}, 2009, \apjs, 184, 218

\bibitem[{{Luppino} \& {Kaiser}(1997)}]{LK97}
{Luppino} G.~A., {Kaiser} N., 1997, \apj, 475, 20

\bibitem[{{Mahdavi} {et~al}\mbox{.}(2007){Mahdavi}, {Hoekstra}, {Babul},
  {Balam}, \& {Capak}}]{Mahdavi07}
{Mahdavi} A., {Hoekstra} H., {Babul} A., {Balam} D.~D., {Capak} P.~L., 2007,
  \apj, 668, 806

\bibitem[{{Mahdavi} {et~al}\mbox{.}(2013){Mahdavi}, {Hoekstra}, {Babul},
  {Bildfell}, {Jeltema}, \& {Henry}}]{Mahdavi13}
{Mahdavi} A., {Hoekstra} H., {Babul} A., {Bildfell} C., {Jeltema} T., {Henry}
  J.~P., 2013, \apj, 767, 116

\bibitem[{{Mahdavi} {et~al}\mbox{.}(2008){Mahdavi}, {Hoekstra}, {Babul}, \&
  {Henry}}]{Mahdavi08}
{Mahdavi} A., {Hoekstra} H., {Babul} A., {Henry} J.~P., 2008, \mnras, 384, 1567

\bibitem[{{Mandelbaum} {et~al}\mbox{.}(2014{\natexlab{a}}){Mandelbaum}, {Rowe},
  {Armstrong}, {Bard}, {Bertin}, {Bosch}, {Boutigny}, {Courbin}, {Dawson},
  {Donnarumma}, {Fenech Conti}, {Gavazzi}, {Gentile}, {Gill}, {Hogg}, {Huff},
  {Jee}, {Kacprzak}, {Kilbinger}, {Kuntzer}, {Lang}, {Luo}, {March},
  {Marshall}, {Meyers}, {Miller}, {Miyatake}, {Nakajima}, {Ngole Mboula},
  {Nurbaeva}, {Okura}, {Paulin-Henriksson}, {Rhodes}, {Schneider}, {Shan},
  {Sheldon}, {Simet}, {Starck}, {Sureau}, {Tewes}, {Zarb Adami}, {Zhang}, \&
  {Zuntz}}]{GREAT3}
{Mandelbaum} R. {et~al.}, 2014{\natexlab{a}}, ArXiv:1412.1825

\bibitem[{{Mandelbaum} {et~al}\mbox{.}(2014{\natexlab{b}}){Mandelbaum}, {Rowe},
  {Bosch}, {Chang}, {Courbin}, {Gill}, {Jarvis}, {Kannawadi}, {Kacprzak},
  {Lackner}, {Leauthaud}, {Miyatake}, {Nakajima}, {Rhodes}, {Simet}, {Zuntz},
  {Armstrong}, {Bridle}, {Coupon}, {Dietrich}, {Gentile}, {Heymans}, {Jurling},
  {Kent}, {Kirkby}, {Margala}, {Massey}, {Melchior}, {Peterson}, {Roodman}, \&
  {Schrabback}}]{Mandelbaum14}
{Mandelbaum} R. {et~al.}, 2014{\natexlab{b}}, \apjs, 212, 5

\bibitem[{{Mantz} {et~al}\mbox{.}(2010){Mantz}, {Allen}, {Rapetti}, \&
  {Ebeling}}]{Mantz10}
{Mantz} A., {Allen} S.~W., {Rapetti} D., {Ebeling} H., 2010, \mnras, 406, 1759

\bibitem[{{Mantz} {et~al}\mbox{.}(2015){Mantz}, {von der Linden}, {Allen},
  {Applegate}, {Kelly}, {Morris}, {Rapetti}, {Schmidt}, {Adhikari}, {Allen},
  {Burchat}, {Burke}, {Cataneo}, {Donovan}, {Ebeling}, {Shandera}, \&
  {Wright}}]{Mantz15}
{Mantz} A.~B. {et~al.}, 2015, \mnras, 446, 2205

\bibitem[{{Martin} {et~al}\mbox{.}(2005){Martin}, {Fanson}, {Schiminovich},
  {Morrissey}, {Friedman}, {Barlow}, {Conrow}, {Grange}, {Jelinsky},
  {Milliard}, {Siegmund}, {Bianchi}, {Byun}, {Donas}, {Forster}, {Heckman},
  {Lee}, {Madore}, {Malina}, {Neff}, {Rich}, {Small}, {Surber}, {Szalay},
  {Welsh}, \& {Wyder}}]{Martin05}
{Martin} D.~C. {et~al.}, 2005, \apjl, 619, L1

\bibitem[{{Massey} {et~al}\mbox{.}(2007){Massey}, {Heymans}, {Berg{\'e}},
  {Bernstein}, {Bridle}, {Clowe}, {Dahle}, {Ellis}, {Erben}, {Hetterscheidt},
  {High}, {Hirata}, {Hoekstra}, {Hudelot}, {Jarvis}, {Johnston}, {Kuijken},
  {Margoniner}, {Mandelbaum}, {Mellier}, {Nakajima}, {Paulin-Henriksson},
  {Peeples}, {Roat}, {Refregier}, {Rhodes}, {Schrabback}, {Schirmer}, {Seljak},
  {Semboloni}, \& {van Waerbeke}}]{STEP2}
{Massey} R. {et~al.}, 2007, \mnras, 376, 13

\bibitem[{{Massey} {et~al}\mbox{.}(2013){Massey}, {Hoekstra}, {Kitching},
  {Rhodes}, {Cropper}, {Amiaux}, {Harvey}, {Mellier}, {Meneghetti}, {Miller},
  {Paulin-Henriksson}, {Pires}, {Scaramella}, \& {Schrabback}}]{Massey13}
{Massey} R. {et~al.}, 2013, \mnras, 429, 661

\bibitem[{{Massey} {et~al}\mbox{.}(2004){Massey}, {Refregier}, {Conselice},
  {David}, \& {Bacon}}]{Massey04}
{Massey} R., {Refregier} A., {Conselice} C.~J., {David} J., {Bacon} J., 2004,
  \mnras, 348, 214

\bibitem[{{McCarthy} {et~al}\mbox{.}(2003){McCarthy}, {Babul}, {Holder}, \&
  {Balogh}}]{McCarthy03}
{McCarthy} I.~G., {Babul} A., {Holder} G.~P., {Balogh} M.~L., 2003, \apj, 591,
  515

\bibitem[{{McCracken} {et~al}\mbox{.}(2012){McCracken}, {Milvang-Jensen},
  {Dunlop}, {Franx}, {Fynbo}, {Le F{\`e}vre}, {Holt}, {Caputi}, {Goranova},
  {Buitrago}, {Emerson}, {Freudling}, {Hudelot}, {L{\'o}pez-Sanjuan},
  {Magnard}, {Mellier}, {M{\o}ller}, {Nilsson}, {Sutherland}, {Tasca}, \&
  {Zabl}}]{McCracken12}
{McCracken} H.~J. {et~al.}, 2012, \aap, 544, A156

\bibitem[{{Melchior} \& {Viola}(2012)}]{Melchior12}
{Melchior} P., {Viola} M., 2012, \mnras, 424, 2757

\bibitem[{{Mellier}(1999)}]{Mellier99}
{Mellier} Y., 1999, \araa, 37, 127

\bibitem[{{Meneghetti} {et~al}\mbox{.}(2010){Meneghetti}, {Rasia}, {Merten},
  {Bellagamba}, {Ettori}, {Mazzotta}, {Dolag}, \& {Marri}}]{Meneghetti10}
{Meneghetti} M., {Rasia} E., {Merten} J., {Bellagamba} F., {Ettori} S.,
  {Mazzotta} P., {Dolag} K., {Marri} S., 2010, \aap, 514, A93

\bibitem[{{Miller} {et~al}\mbox{.}(2013){Miller}, {Heymans}, {Kitching}, {van
  Waerbeke}, {Erben}, {Hildebrandt}, {Hoekstra}, {Mellier}, {Rowe}, {Coupon},
  {Dietrich}, {Fu}, {Harnois-D{\'e}raps}, {Hudson}, {Kilbinger}, {Kuijken},
  {Schrabback}, {Semboloni}, {Vafaei}, \& {Velander}}]{Miller13}
{Miller} L. {et~al.}, 2013, \mnras, 429, 2858

\bibitem[{{Muzzin} {et~al}\mbox{.}(2013){Muzzin}, {Marchesini}, {Stefanon},
  {Franx}, {Milvang-Jensen}, {Dunlop}, {Fynbo}, {Brammer}, {Labb{\'e}}, \& {van
  Dokkum}}]{Muzzin13}
{Muzzin} A. {et~al.}, 2013, \apjs, 206, 8

\bibitem[{{Nagai} {et~al}\mbox{.}(2007){Nagai}, {Vikhlinin}, \&
  {Kravtsov}}]{Nagai07}
{Nagai} D., {Vikhlinin} A., {Kravtsov} A.~V., 2007, \apj, 655, 98

\bibitem[{{Navarro} {et~al}\mbox{.}(1997){Navarro}, {Frenk}, \& {White}}]{NFW}
{Navarro} J.~F., {Frenk} C.~S., {White} S.~D.~M., 1997, \apj, 490, 493

\bibitem[{{Oguri} \& {Hamana}(2011)}]{Oguri11}
{Oguri} M., {Hamana} T., 2011, \mnras, 414, 1851

\bibitem[{{Okabe} {et~al}\mbox{.}(2013){Okabe}, {Smith}, {Umetsu}, {Takada}, \&
  {Futamase}}]{Okabe13}
{Okabe} N., {Smith} G.~P., {Umetsu} K., {Takada} M., {Futamase} T., 2013,
  \apjl, 769, L35

\bibitem[{{Okabe} {et~al}\mbox{.}(2010){Okabe}, {Takada}, {Umetsu}, {Futamase},
  \& {Smith}}]{Okabe10}
{Okabe} N., {Takada} M., {Umetsu} K., {Futamase} T., {Smith} G.~P., 2010,
  \pasj, 62, 811

\bibitem[{{Peng} {et~al}\mbox{.}(2002){Peng}, {Ho}, {Impey}, \& {Rix}}]{Peng02}
{Peng} C.~Y., {Ho} L.~C., {Impey} C.~D., {Rix} H.-W., 2002, \aj, 124, 266

\bibitem[{{Planck Collaboration} {et~al}\mbox{.}(2014{\natexlab{a}}){Planck
  Collaboration}, {Ade}, {Aghanim}, {Armitage-Caplan}, {Arnaud}, {Ashdown},
  {Atrio-Barandela}, {Aumont}, {Aussel}, {Baccigalupi}, \& et~al.}]{PlanckXXIX}
{Planck Collaboration} {et~al.}, 2014{\natexlab{a}}, \aap, 571, A29

\bibitem[{{Planck Collaboration} {et~al}\mbox{.}(2014{\natexlab{b}}){Planck
  Collaboration}, {Ade}, {Aghanim}, {Armitage-Caplan}, {Arnaud}, {Ashdown},
  {Atrio-Barandela}, {Aumont}, {Baccigalupi}, {Banday}, \& et~al.}]{PlanckXVI}
{Planck Collaboration} {et~al.}, 2014{\natexlab{b}}, \aap, 571, A16

\bibitem[{{Planck Collaboration} {et~al}\mbox{.}(2014{\natexlab{c}}){Planck
  Collaboration}, {Ade}, {Aghanim}, {Armitage-Caplan}, {Arnaud}, {Ashdown},
  {Atrio-Barandela}, {Aumont}, {Baccigalupi}, {Banday}, \& et~al.}]{PlanckXX}
{Planck Collaboration} {et~al.}, 2014{\natexlab{c}}, \aap, 571, A20

\bibitem[{{Planck Collaboration} {et~al}\mbox{.}(2011){Planck Collaboration},
  {Ade}, {Aghanim}, {Arnaud}, {Ashdown}, {Aumont}, {Baccigalupi}, {Balbi},
  {Banday}, {Barreiro}, \& et~al.}]{Planck11}
{Planck Collaboration} {et~al.}, 2011, \aap, 536, A8

\bibitem[{{Ragozzine} {et~al}\mbox{.}(2012){Ragozzine}, {Clowe}, {Markevitch},
  {Gonzalez}, \& {Brada{\v c}}}]{Ragozzine12}
{Ragozzine} B., {Clowe} D., {Markevitch} M., {Gonzalez} A.~H., {Brada{\v c}}
  M., 2012, \apj, 744, 94

\bibitem[{{Rasia} {et~al}\mbox{.}(2006){Rasia}, {Ettori}, {Moscardini},
  {Mazzotta}, {Borgani}, {Dolag}, {Tormen}, {Cheng}, \& {Diaferio}}]{Rasia06}
{Rasia} E. {et~al.}, 2006, \mnras, 369, 2013

\bibitem[{{Reichardt} {et~al}\mbox{.}(2013){Reichardt}, {Stalder}, {Bleem},
  {Montroy}, {Aird}, {Andersson}, {Armstrong}, {Ashby}, {Bautz}, {Bayliss},
  {Bazin}, {Benson}, {Brodwin}, {Carlstrom}, {Chang}, {Cho}, {Clocchiatti},
  {Crawford}, {Crites}, {de Haan}, {Desai}, {Dobbs}, {Dudley}, {Foley},
  {Forman}, {George}, {Gladders}, {Gonzalez}, {Halverson}, {Harrington},
  {High}, {Holder}, {Holzapfel}, {Hoover}, {Hrubes}, {Jones}, {Joy}, {Keisler},
  {Knox}, {Lee}, {Leitch}, {Liu}, {Lueker}, {Luong-Van}, {Mantz}, {Marrone},
  {McDonald}, {McMahon}, {Mehl}, {Meyer}, {Mocanu}, {Mohr}, {Murray}, {Natoli},
  {Padin}, {Plagge}, {Pryke}, {Rest}, {Ruel}, {Ruhl}, {Saliwanchik}, {Saro},
  {Sayre}, {Schaffer}, {Shaw}, {Shirokoff}, {Song}, {Spieler}, {Staniszewski},
  {Stark}, {Story}, {Stubbs}, {{\v S}uhada}, {van Engelen}, {Vanderlinde},
  {Vieira}, {Vikhlinin}, {Williamson}, {Zahn}, \& {Zenteno}}]{Reichardt13}
{Reichardt} C.~L. {et~al.}, 2013, \apj, 763, 127

\bibitem[{{Reiprich} \& {B{\"o}hringer}(2002)}]{Reiprich02}
{Reiprich} T.~H., {B{\"o}hringer} H., 2002, \apj, 567, 716

\bibitem[{{Rines} {et~al}\mbox{.}(2013){Rines}, {Geller}, {Diaferio}, \&
  {Kurtz}}]{Rines13}
{Rines} K., {Geller} M.~J., {Diaferio} A., {Kurtz} M.~J., 2013, \apj, 767, 15

\bibitem[{{Rix} {et~al}\mbox{.}(2004){Rix}, {Barden}, {Beckwith}, {Bell},
  {Borch}, {Caldwell}, {H{\"a}ussler}, {Jahnke}, {Jogee}, {McIntosh},
  {Meisenheimer}, {Peng}, {Sanchez}, {Somerville}, {Wisotzki}, \&
  {Wolf}}]{Rix04}
{Rix} H.-W. {et~al.}, 2004, \apjs, 152, 163

\bibitem[{{Robin} {et~al}\mbox{.}(2003){Robin}, {Reyl{\'e}}, {Derri{\`e}re}, \&
  {Picaud}}]{Robin03}
{Robin} A.~C., {Reyl{\'e}} C., {Derri{\`e}re} S., {Picaud} S., 2003, \aap, 409,
  523

\bibitem[{{Rowe} {et~al}\mbox{.}(2014){Rowe}, {Jarvis}, {Mandelbaum},
  {Bernstein}, {Bosch}, {Simet}, {Meyers}, {Kacprzak}, {Nakajima}, {Zuntz},
  {Miyatake}, {Dietrich}, {Armstrong}, {Melchior}, \& {Gill}}]{Rowe14}
{Rowe} B. {et~al.}, 2014, ArXiv:1407.7676

\bibitem[{{Sanders} {et~al}\mbox{.}(2007){Sanders}, {Salvato}, {Aussel},
  {Ilbert}, {Scoville}, {Surace}, {Frayer}, {Sheth}, {Helou}, {Brooke},
  {Bhattacharya}, {Yan}, {Kartaltepe}, {Barnes}, {Blain}, {Calzetti}, {Capak},
  {Carilli}, {Carollo}, {Comastri}, {Daddi}, {Ellis}, {Elvis}, {Fall},
  {Franceschini}, {Giavalisco}, {Hasinger}, {Impey}, {Koekemoer}, {Le
  F{\`e}vre}, {Lilly}, {Liu}, {McCracken}, {Mobasher}, {Renzini}, {Rich},
  {Schinnerer}, {Shopbell}, {Taniguchi}, {Thompson}, {Urry}, \&
  {Williams}}]{Sanders07}
{Sanders} D.~B. {et~al.}, 2007, \apjs, 172, 86

\bibitem[{{Schlafly} \& {Finkbeiner}(2011)}]{Schlafly11}
{Schlafly} E.~F., {Finkbeiner} D.~P., 2011, \apj, 737, 103

\bibitem[{{Schlegel} {et~al}\mbox{.}(1998){Schlegel}, {Finkbeiner}, \&
  {Davis}}]{Schlegel98}
{Schlegel} D.~J., {Finkbeiner} D.~P., {Davis} M., 1998, \apj, 500, 525

\bibitem[{{Scoville} {et~al}\mbox{.}(2007){Scoville}, {Aussel}, {Brusa},
  {Capak}, {Carollo}, {Elvis}, {Giavalisco}, {Guzzo}, {Hasinger}, {Impey},
  {Kneib}, {LeFevre}, {Lilly}, {Mobasher}, {Renzini}, {Rich}, {Sanders},
  {Schinnerer}, {Schminovich}, {Shopbell}, {Taniguchi}, \&
  {Tyson}}]{Scoville07}
{Scoville} N. {et~al.}, 2007, \apjs, 172, 1

\bibitem[{{Seljak} \& {Warren}(2004)}]{Seljak04}
{Seljak} U., {Warren} M.~S., 2004, \mnras, 355, 129

\bibitem[{{Semboloni} {et~al}\mbox{.}(2013){Semboloni}, {Hoekstra}, {Huang},
  {Cardone}, {Cropper}, {Joachimi}, {Kitching}, {Kuijken}, {Lombardi}, {Maoli},
  {Mellier}, {Miller}, {Rhodes}, {Scaramella}, {Schrabback}, \&
  {Velander}}]{Semboloni13}
{Semboloni} E. {et~al.}, 2013, \mnras, 432, 2385

\bibitem[{{Sheth} {et~al}\mbox{.}(2001){Sheth}, {Mo}, \& {Tormen}}]{Sheth01}
{Sheth} R.~K., {Mo} H.~J., {Tormen} G., 2001, \mnras, 323, 1

\bibitem[{{Sheth} \& {Tormen}(1999)}]{Sheth99}
{Sheth} R.~K., {Tormen} G., 1999, \mnras, 308, 119

\bibitem[{{Sif{\'o}n} {et~al}\mbox{.}(2014){Sif{\'o}n}, {Hoekstra}, {Cacciato},
  {Viola}, {K{\"o}hlinger}, {van der Burg}, {Sand}, \& {Graham}}]{Sifon14}
{Sif{\'o}n} C., {Hoekstra} H., {Cacciato} M., {Viola} M., {K{\"o}hlinger} F.,
  {van der Burg} R., {Sand} D., {Graham} M.~L., 2014, ArXiv:1406.5196

\bibitem[{{Simet} \& {Mandelbaum}(2014)}]{Simet14}
{Simet} M., {Mandelbaum} R., 2014, ArXiv:1406.4908

\bibitem[{{Sunyaev} \& {Zeldovich}(1972)}]{Sunyaev72}
{Sunyaev} R.~A., {Zeldovich} Y.~B., 1972, Comments on Astrophysics and Space
  Physics, 4, 173

\bibitem[{{Tinker} {et~al}\mbox{.}(2010){Tinker}, {Robertson}, {Kravtsov},
  {Klypin}, {Warren}, {Yepes}, \& {Gottl{\"o}ber}}]{Tinker10}
{Tinker} J.~L., {Robertson} B.~E., {Kravtsov} A.~V., {Klypin} A., {Warren}
  M.~S., {Yepes} G., {Gottl{\"o}ber} S., 2010, \apj, 724, 878

\bibitem[{{Umetsu} {et~al}\mbox{.}(2014){Umetsu}, {Medezinski}, {Nonino},
  {Merten}, {Postman}, {Meneghetti}, {Donahue}, {Czakon}, {Molino}, {Seitz},
  {Gruen}, {Lemze}, {Benitez}, {Biviano}, {Broadhurst}, {Ford}, {Grillo},
  {Koekemoer}, {Melchior}, {Mercurio}, {Moustakas}, \& {Zitrin}}]{Umetsu14}
{Umetsu} K. {et~al.}, 2014, ArXiv:1404.1375

\bibitem[{{van den Bosch} {et~al}\mbox{.}(2013){van den Bosch}, {More},
  {Cacciato}, {Mo}, \& {Yang}}]{vandenBosch13}
{van den Bosch} F.~C., {More} S., {Cacciato} M., {Mo} H., {Yang} X., 2013,
  \mnras, 430, 725

\bibitem[{{van Uitert} {et~al}\mbox{.}(2011){van Uitert}, {Hoekstra},
  {Velander}, {Gilbank}, {Gladders}, \& {Yee}}]{Uitert11}
{van Uitert} E., {Hoekstra} H., {Velander} M., {Gilbank} D.~G., {Gladders}
  M.~D., {Yee} H.~K.~C., 2011, \aap, 534, A14

\bibitem[{{Vikhlinin} {et~al}\mbox{.}(2009){Vikhlinin}, {Kravtsov}, {Burenin},
  {Ebeling}, {Forman}, {Hornstrup}, {Jones}, {Murray}, {Nagai}, {Quintana}, \&
  {Voevodkin}}]{Vikhlinin09}
{Vikhlinin} A. {et~al.}, 2009, \apj, 692, 1060

\bibitem[{{Viola} {et~al}\mbox{.}(2014){Viola}, {Kitching}, \&
  {Joachimi}}]{Viola14}
{Viola} M., {Kitching} T.~D., {Joachimi} B., 2014, \mnras, 439, 1909

\bibitem[{{von der Linden} {et~al}\mbox{.}(2014{\natexlab{a}}){von der Linden},
  {Allen}, {Applegate}, {Kelly}, {Allen}, {Ebeling}, {Burchat}, {Burke},
  {Donovan}, {Morris}, {Blandford}, {Erben}, \& {Mantz}}]{vonderLinden14}
{von der Linden} A. {et~al.}, 2014{\natexlab{a}}, \mnras, 439, 2

\bibitem[{{von der Linden} {et~al}\mbox{.}(2014{\natexlab{b}}){von der Linden},
  {Mantz}, {Allen}, {Applegate}, {Kelly}, {Morris}, {Wright}, {Allen},
  {Burchat}, {Burke}, {Donovan}, \& {Ebeling}}]{vonderLinden14b}
{von der Linden} A. {et~al.}, 2014{\natexlab{b}}, \mnras, 443, 1973

\end{thebibliography}

\appendix

\section{Effects of PSF anisotropy}
\label{sec:psfan}

To quantify the multiplicative bias in the shear measurement resulting
from noise in the images and limitations in the correction for the
size of the PSF and the weight function, we created simulated images
with a circular PSF. In real data, however, the PSF is generally
anisotropic, albeit to varying degree.  PSF anisotropy leads to
additive biases by introducing coherent alignments in the observed
shapes. In the case of cosmic shear this is a dominant source of
systematic, but in our case the signal is averaged azimuthally around
the clusters, which also averages out most of the PSF
anisotropy\footnote{We note that this is not always the case, as some
  imagers show strong radial patterns, which can bias the azimuthally
  averaged lensing signal if the cluster is located at the center of
  the field-of-view.}.

\begin{figure}
\begin{center}
\leavevmode \hbox{%
\includegraphics[width=8.5cm]{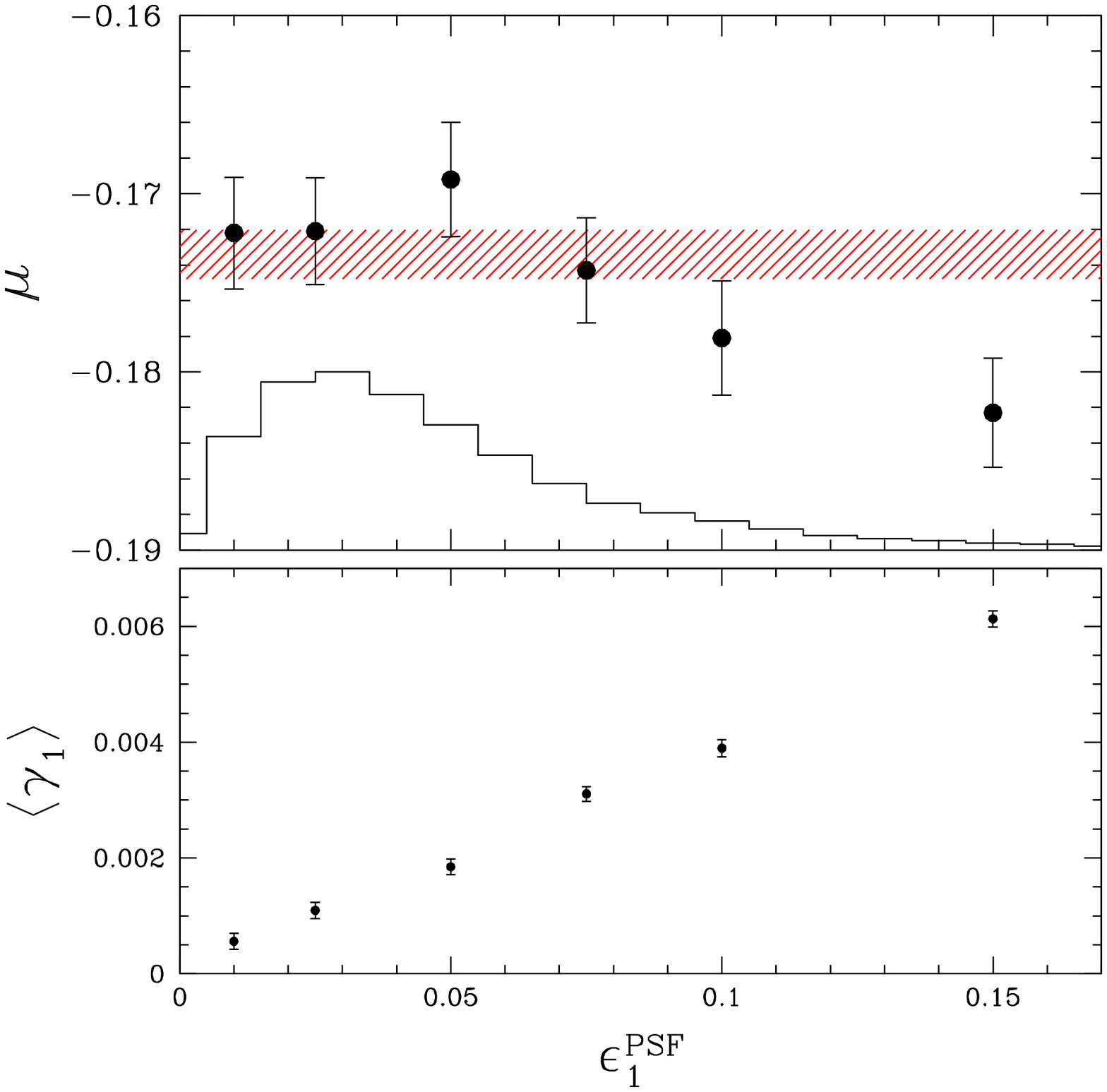}}
\caption{{\it Top panel:} Multiplicative bias as a function of the
ellipticity of the PSF for galaxies with $22<m_r<25$. The red hatched 
region indicates the results obtained for a circular PSF. The histogram 
indicates the distribution of PSF ellipticities observed in the CFHT data 
used in this paper. {\it Bottom panel:} The average shear $\gamma_1$ measured 
after PSF correction as a function of PSF ellipticity. The average should
vanish if the correction is perfect. These results suggest that the
residual is $\sim 4\%$ of the original PSF ellipticity.
\label{fig:psfan}}
\end{center}
\end{figure}

Nonetheless it is important to examine whether the results obtained
using circular PSFs can be applied to our CFHT data. We therefore
created a set of simulations in which the PSF is elliptical (along the
$x-$axis) using our fiducial ellipticity distribution
($\epsilon_0=0.25$). These images were analysed as usual, and the
resulting multiplicative bias as a function of PSF ellipticity is
presented in the top panel of Figure~\ref{fig:psfan}. Even for rather
ellongated PSFs ($\epsilon>0.1$) the increase in bias is at the
percent level at most.  For comparison, the histogram shows the
observed distribution of PSF ellipticity in the CCCP data. These
results suggest that we do not need to account for the PSF anisotropy
explicitly in order to quantify the multiplicative bias in our data.

We do not expect the correction to work perfectly, either due to
limitation of the correction method itself, or due to the fact that the
noise in the images biases the polarizabilities. The bottom panel
shows how well the correction for PSF anisotropy performs. We observe
a linear trend of the recovered average shear as a function of PSF
ellipticity. For the galaxies with $22<m_r<25$ this results in a
residual bias of $\sim 4\%$ of the original PSF ellipticity. Although the
bias is smaller for bright galaxies it does not vanish, suggesting
that a large part of the observed bias is due to a fundamental
limitation of the KSB method\footnote{The KSB algorithm assumes that
  the PSF is described as the convolution of a circular kernel
  with a compact anisotropic one.}.

\section{Contamination by stars}
\label{sec:star_contam}

\begin{figure}
\begin{center}
\leavevmode \hbox{%
\includegraphics[width=8.5cm]{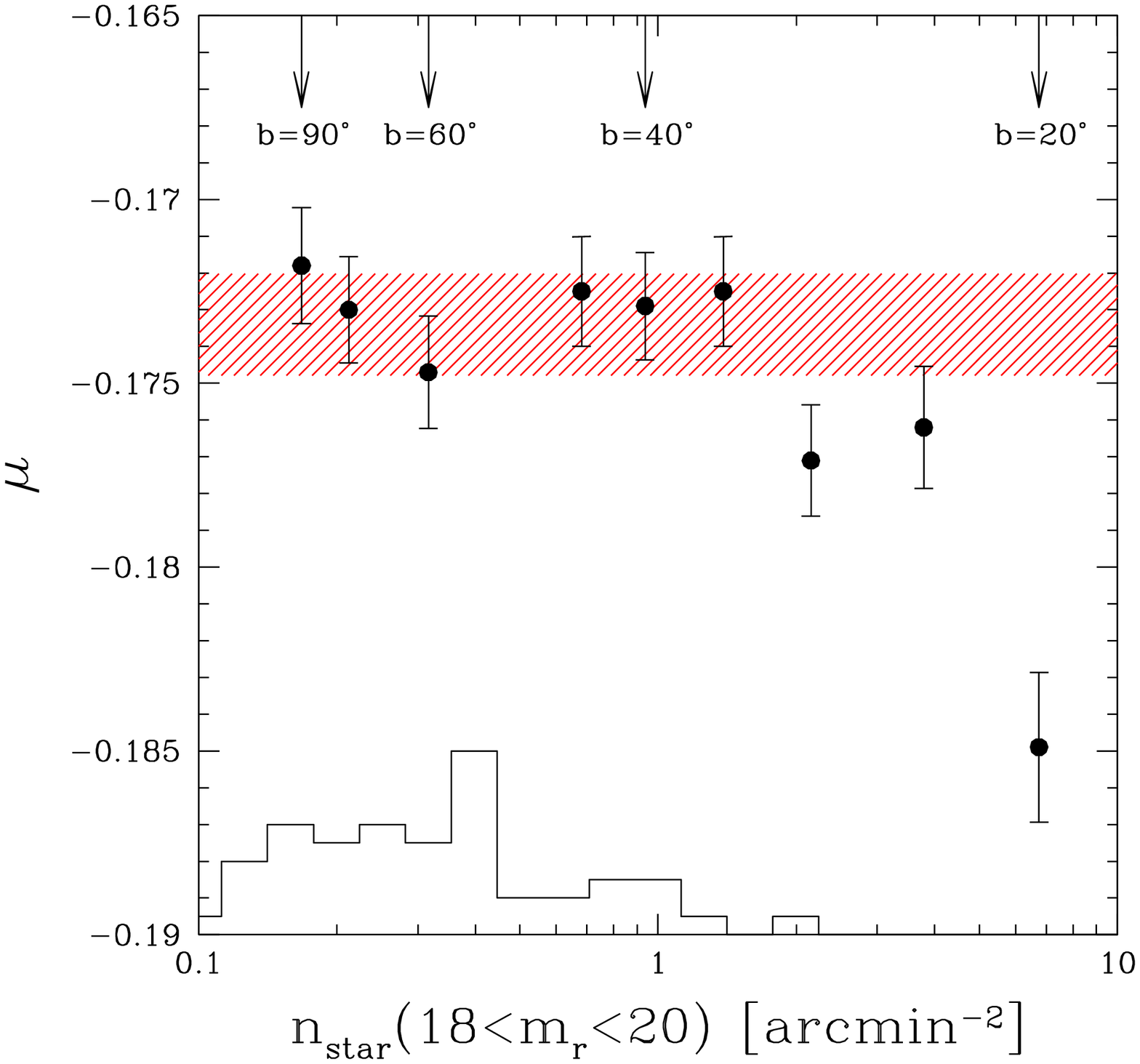}}
\caption{Multiplicative bias for galaxies with $22<m_r<25$ as a
  function of the density of stars with $18<m_r<20$, based on the
  Besan\c{c}on model of stellar population synthesis of the Galaxy
  \citep{Robin03}. The red hatched region indicates the results for the
  simulations without stars. The arrows indicate the corresponding
  Galactic latitudes (for the longitude of Abell~1835,
  $l=340^\circ$). The histogram shows the distribution of star
  densities of the CCCP data.\label{fig:starden}}
\end{center}
\end{figure}

The simulated data that were used to study the multiplicative bias
contained a small number of bright stars, which were only included to keep
track of the PSF properties. In real data, however, stars may blend
with galaxies or other stars. As a consequence they might be
misidentified and be included in the galaxy catalog, contributing to
multiplicative bias (as they are not sheared).

In this section we study how the multiplicative bias depends on star
density by including stars in the images with a realistic number
density and magnitude distribution. We use the Besan\c{c}on model of
stellar population synthesis of the Galaxy for the stars
\citep{Robin03}. As a reference we consider again Abell~1835, with
$(l,b)=(340,60)$, where we change the Galactic latitude $b$ to change
the star number density. We place the stars at random locations in the
images, with $\epsilon_0=0.25$, and proceed as before with the shape
analysis. 

The results for galaxies with $22<m_r<25$ are presented in
Figure~\ref{fig:starden}, which shows the multiplicative bias as a
function of the density of stars with $18<m_r<20$. For densities
$n_{\rm star}<1.5$ arcmin$^{-2}$ the bias is consistent with the
results without stars (indicated by the red hatched region). For the
adopted longitude of $l=340^\circ$ this corresponds to
$b>35^\circ$. For lower Galactic latitudes, the higher number density
of stars can lead to appreciable levels of bias. The histogram in
Figure~\ref{fig:starden} shows the distribution of star density in the
CCCP data, which suggests that we can safely ignore the contribution
from stars.

\section{Determining the empirical correction}
\label{app:correction}

In this appendix we describe the empirical correction used in the
analysis of the CCCP data. We assume that the bias is a function of
the signal-to-noise ratio \citep[e.g.][]{Melchior12,Kacprzak12} and
depends on how well a galaxy is resolved \citep[e.g.][]{Massey13,Miller13}.
That these parameters are important is also suggested by
Figures~\ref{fig:bias_mag} and~\ref{fig:bias_seeing}, which show a
larger bias for fainter galaxies and larger PSFs.

\begin{figure*}
\begin{center}
\leavevmode \hbox{%
\includegraphics[width=8.5cm]{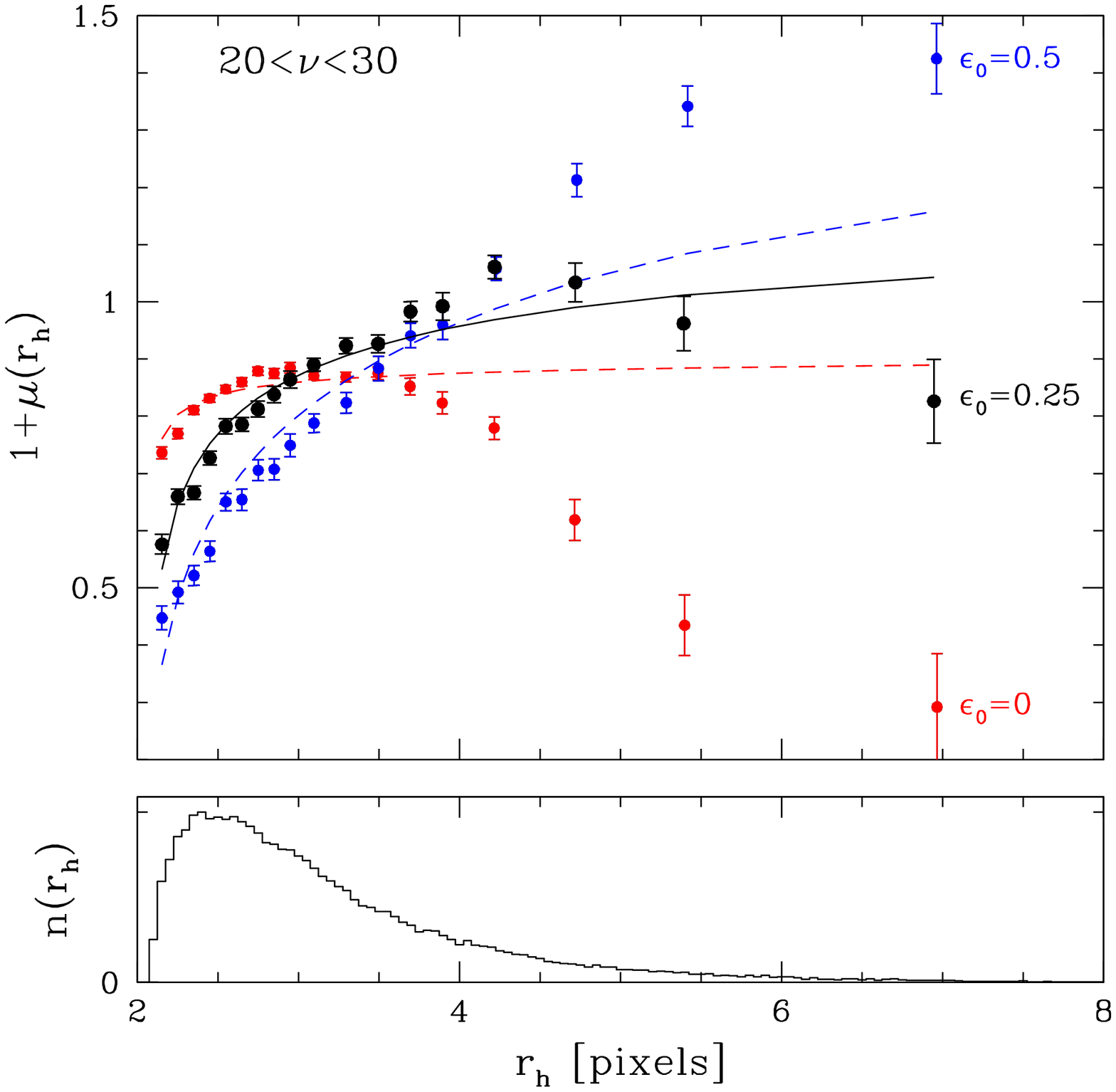}
\includegraphics[width=8.5cm]{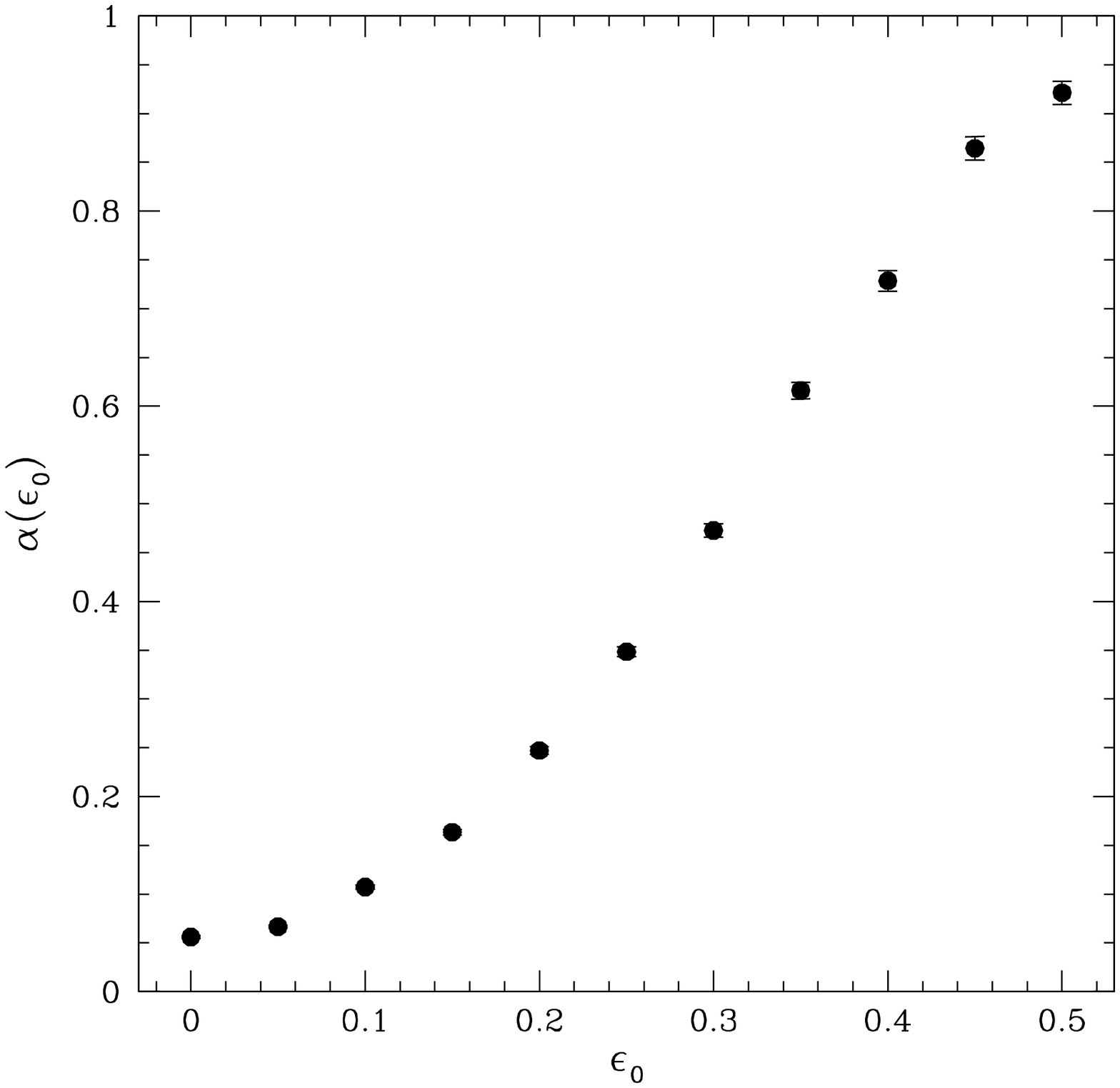}}
\caption{{\it Left panel:} bias $1+\mu$ as a function of half-light
  radius for galaxies with $20<\nu<30$. The black points indicate the
  results for our adopted value of $\epsilon_0=0.25$; the red (blue)
  points are for the extreme values of $\epsilon_0=0$
  ($\epsilon_0=0.5$). The lines indicate the fits using our best fit
  value for $\alpha(\epsilon_0)$. The bottom panel shows the
  distribution of half-light radii for these sources. {\it Right
    panel:} Resulting best-fit value for $\alpha(\epsilon_0)$ for our
  simulated CCCP data.  The values are listed in Table~\ref{tab:corpar}.}
\label{fig:alpha_par}
\end{center}
\end{figure*}

Our implementation of KSB provides an estimate for $\sigma_e$, the
uncertainty in the polarization \citep[see][for details]{Hoekstra00}.
The reciprocal of this quantity is a useful proxy for the
signal-to-noise ratio $\nu$, hence we use $\nu=1/\sigma_e$. To
quantify how well a galaxy is resolved we use

\begin{figure*}
\begin{center}
\leavevmode \hbox{%
\includegraphics[width=8.5cm]{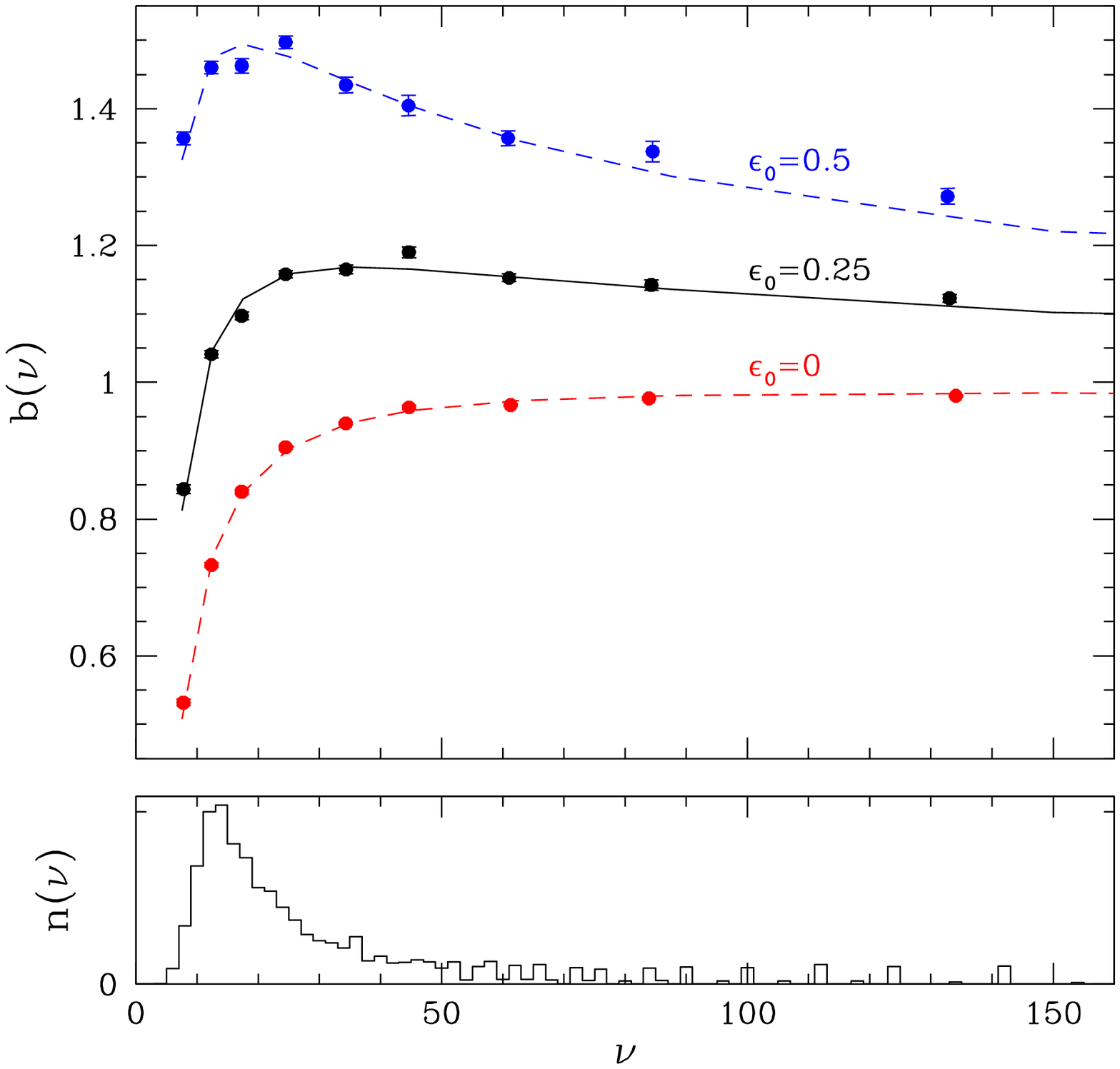}
\includegraphics[width=8.5cm]{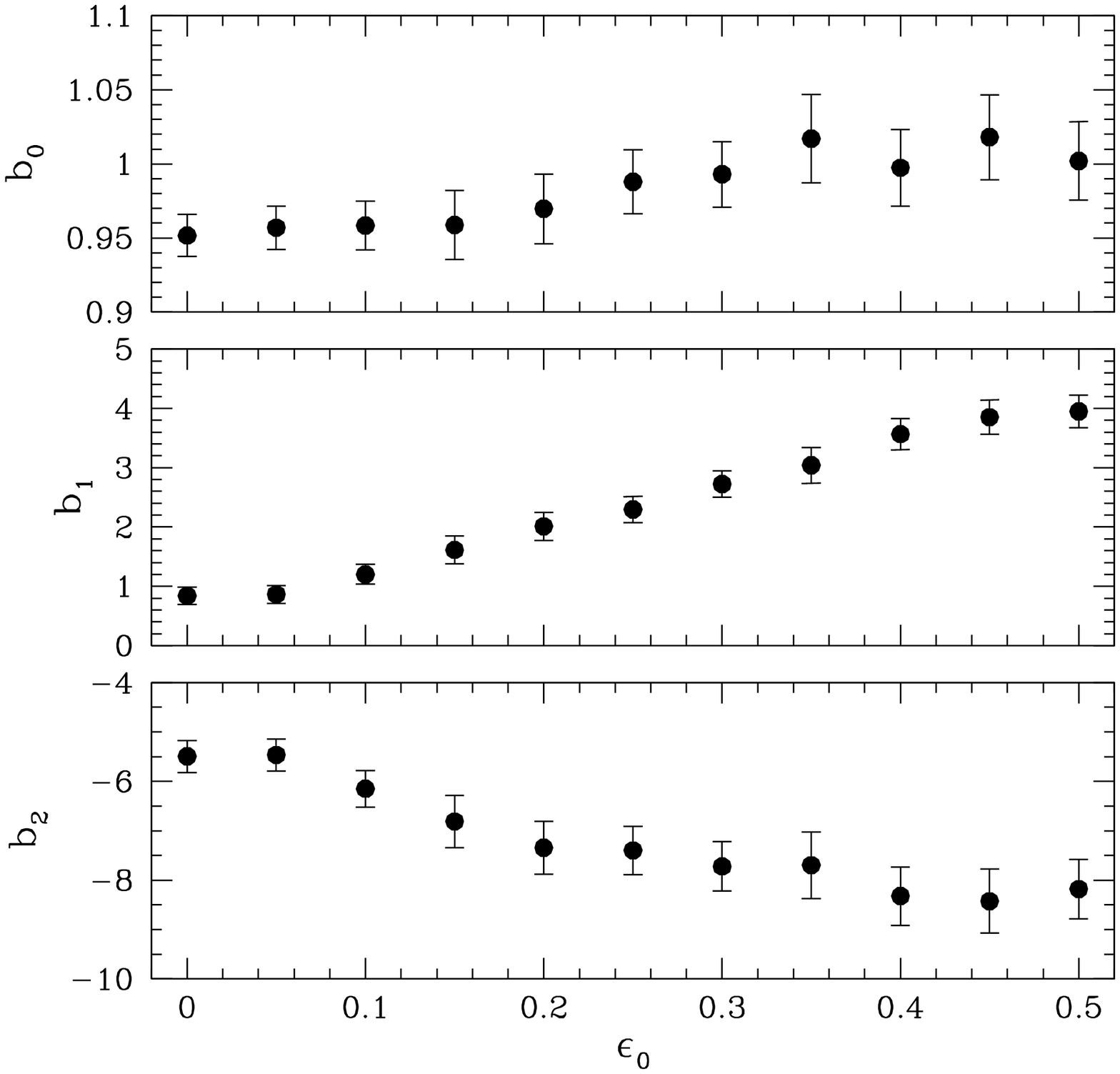}}
\caption{{\it Left panel:} $b(\nu)$ for the three ellipticity
  distributions, as well as the best fit models.  The bottom panel
  shows the distribution of signal-to-noise ratios for galaxies with
  $20<m_r<25$. {\it Right panel:} the model parameters $b_i$ as a
  function of $\epsilon_0$, which show a smooth dependence. The values
  are listed in Table~\ref{tab:corpar}. Note that $b_0$ is fairly
  close to unity.} \label{fig:bnu_par}
\end{center}
\end{figure*}

\begin{equation}
{\cal R}^2=\frac{r^2_{h,*}}{r^2_{h,{\rm gal}}-r^2_{h,*}},
\end{equation}

\noindent where $r_{h,*}$ denotes the half-light radius of the PSF and
$r_{h,{\rm gal}}$ that of the observed galaxy. The denominator
corresponds to the unconvolved size of a source if galaxies were
Gaussians. Importantly, these are quantities that can be measured for
individual sources.

\begin{table*}
\centering
\caption{Parameters for the empirical correction for multiplicative bias}\label{tab:corpar}
\begin{tabular}{lcccc}
\hline
\hline
$\epsilon_0$ & $\alpha$ & $b_0$ & $b_1$ & $b_2$ \\
\hline
0     &	$0.056\pm 0.0015$ & $0.952 \pm 0.0143$ & $0.84 \pm 0.15$ & $-5.50 \pm 0.32$ \\	
0.05  &	$0.067\pm 0.0017$ & $0.957 \pm 0.0145$ & $0.86 \pm 0.15$ & $-5.47 \pm 0.33$ \\
0.10  &	$0.107\pm 0.0022$ & $0.958 \pm 0.0164$ & $1.20 \pm 0.17$ & $-6.15 \pm 0.37$ \\ 
0.15  &	$0.164\pm 0.0029$ & $0.959 \pm 0.0233$ & $1.61 \pm 0.24$ & $-6.81 \pm 0.53$ \\ 
0.20  &	$0.247\pm 0.0038$ & $0.970 \pm 0.0235$ & $2.01 \pm 0.24$ & $-7.34 \pm 0.53$ \\
0.25  &	$0.348\pm 0.0051$ & $0.988 \pm 0.0217$ & $2.29 \pm 0.22$ & $-7.40 \pm 0.49$ \\ 
0.30  &	$0.473\pm 0.0067$ & $0.993 \pm 0.0220$ & $2.73 \pm 0.22$ & $-7.72 \pm 0.50$ \\
0.35  &	$0.616\pm 0.0087$ & $1.017 \pm 0.0298$ & $3.04 \pm 0.30$ & $-7.70 \pm 0.68$ \\
0.40  &	$0.729\pm 0.0104$ & $0.997 \pm 0.0260$ & $3.57 \pm 0.26$ & $-8.32 \pm 0.59$ \\
0.45  &	$0.864\pm 0.0120$ & $1.018 \pm 0.0289$ & $3.85 \pm 0.29$ & $-8.43 \pm 0.65$ \\
0.50  &	$0.921\pm 0.0121$ & $1.002 \pm 0.0265$ & $3.95 \pm 0.27$ & $-8.18 \pm 0.60$ \\
\hline
\hline
\end{tabular}
\end{table*}

We consider values of $\epsilon_0$ between 0 and 0.5 with steps of
0.05 and create 169 pairs of images, each 10,000 by 10,000 pixels,
with constant shears of $-0.06\le\gamma_i\le 0.06$ applied. As a
result, for each $\epsilon_0$ we analyse $\sim 10^7$ galaxies with
$20<m_r<25$ (note that the input catalog does contain fainter
galaxies). For a given $\epsilon_0$ we bin the measurements in
fine bins of $\nu$ and $r_h$ and determine the bias.

We considered various fitting functions, with the aim to find an
adequate correction that is also robust against the uncertainty in the
true value of $\epsilon_0$. We consider a conservative range of
$\epsilon_0\in[0.15,0.3]$, where we note that a stronger prior on the
input ellipticity distribution would allow the uncertainties to be
reduced, and the fitting functions to be optimized. 

Although further optimization is possible by introducing additional
parameters, we opted for a correction with 4 free parameters: one
to describe the size dependence of the bias for a given $\nu$, and
three to capture the dependence on $\nu$. The correction takes
the form

\begin{equation}
1+\mu(\nu,{\cal R})=\frac{b(\nu)}{1+\alpha(\epsilon_0) {\cal R}}.
\end{equation}

\noindent The first step is to determine $\alpha(\epsilon_0)$, which
is done by examining $\mu$ as a function of size for narrow bins in
$\nu$. The left panel of Figure~\ref{fig:alpha_par} shows the
measurements as a function of half-light radius for galaxies with
$20<\nu<30$ for three ellipticity distributions. The number of objects
as a function of observed half-light radii is shown in the bottom
panel; most sources are small (the value for the PSF is $r_{h,*}=2.056$
pixels).

For each ellipticity distribution we determine the best fit value for
$\alpha$, under the assumption it only depends on $\epsilon_0$. The
results are presented in the right panel of Figure~\ref{fig:alpha_par}
and listed in Table~\ref{tab:corpar}. The right panel in
Figure~\ref{fig:alpha_par} shows that $\alpha(\epsilon_0)$ increases
smoothly with increasing $\epsilon_0$. The lines in the left panel
indicate the predicted bias, with the amplitude a free parameter
(which is used to determine $b(\nu)$, see below). Our parametrization
for the size dependence does fairly well for the bulk of the sources,
but it does not capture the results for $r_h>4$ pixels.  Better
agreement is obtained if we include an additional term $\propto{\cal
  R}^2$, but we found that this did not improve the robustness of the
correction. Similarly, we found that we could have included a
dependence on $1/\nu$, again with limited effect. Note there is
covariance between some of the parameters. For instance, for small
values for $\nu$ we do expect $b(\nu)$ and $\alpha(\epsilon_0)$ to be
correlated because those galaxies have on average larger values for
${\cal R}$.

Closer inspection of the objects with large observed sizes revealed
that most of these are faint objects for which the input sizes were
much smaller or blends with other galaxies. As the latter might be
particularly relevant for the study of galaxy clusters, we decided to
only use galaxies with $r_h<5$ pixels for the actual lensing
analysis. The ensemble averaged lensing signal as a function of source
size presented in Figure~\ref{fig:re_bgsize} indicates that the
results for large sources are indeed biased low. Note that we do not
apply this cut when fitting for $\alpha(\epsilon_0)$, because we found
that the correction performed a bit better when we considered the full
range in sizes.

The next step is to quantify how $\mu$ depends on the signal-to-noise
ratio. The results for $\epsilon_0=0,0.25$ and 0.5 are presented in
the left panel of Figure~\ref{fig:bnu_par}. For low $\epsilon_0$ the
bias increases monotonically, asymptoting to a value $\sim 0.95$ for
large $\nu$, whereas for $\epsilon_0=0.5$, $b(\nu)$ increases first
before declining. To capture the variation in $b(\nu)$ we adopt

\begin{equation}
b(\nu)=b_0(\epsilon_0)+\frac{b_1(\epsilon_0)}{\sqrt{\nu}}+
\frac{b_2((\epsilon_0)}{\nu}.
\end{equation}

\noindent We note that fixing $b_0 \sim 0.95$ also gave reasonable
results. Furthermore $b_1$ and $b_2$ are highly (anti-)correlated, and
it thus might be possible to reduce the number of free parameters in
principle. Interestingly the value for $b_0$, which corresponds to the
bias for bright, large sources, is fairly close to unity. We list
the best fit parameters in Table~\ref{tab:corpar} and the right
panel in Figure~\ref{fig:bnu_par} shows that the parameters vary smoothly
with $\epsilon_0$. 

In \S\ref{sec:correction} we discuss how well the correction performs
as a function of ellipticity distribution and seeing. We also examined
$\mu_{\rm cor}$ as a function of apparent magnitude and found that
residuals are $<1\%$ for galaxies fainter than $m_r=22$, whereas
$\mu_{\rm cor}\sim 0.02$ for galaxies with $20<m_r<22$. This
overcorrection is probably caused by the fact that our parametrization
tends to underestimate the bias for large objects.

\begin{figure}
\centering
\leavevmode \hbox{%
\includegraphics[width=8.5cm]{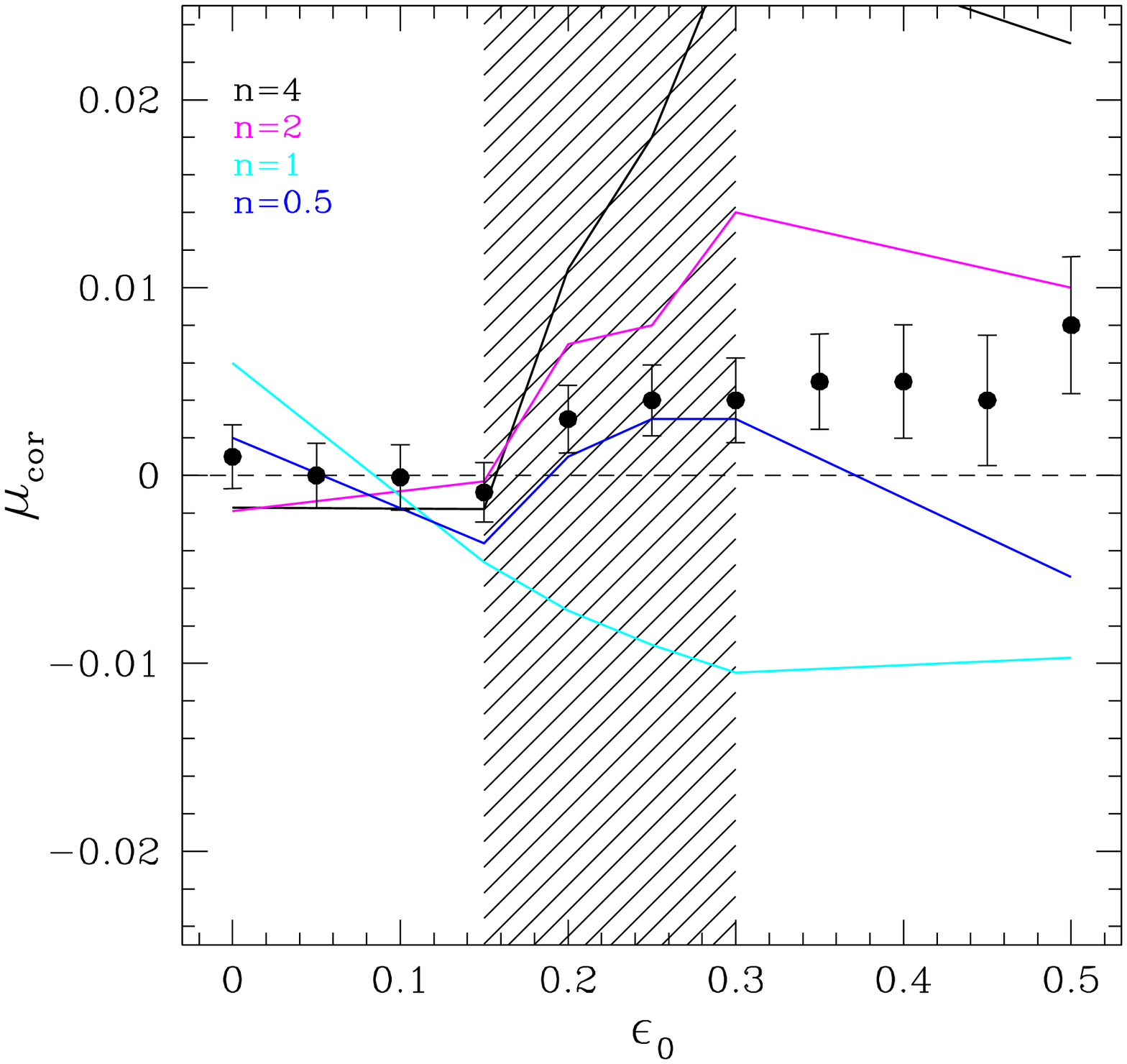}}
\caption{{\it Bottom panel:} Residual multiplicative bias for galaxies
  with different values for the Sersic index as a function of
  $\epsilon_0$.  The simulated galaxies have the same distribution in
  half-light radii as the regular simulations (indicated by the black
  points with errorbars) and $22<m_r<25$. As in the real data we
    only include galaxies with an observed size $r_h<5$ pixels. The
    full range spans about 0.03, but the uncertainty for the ensemble
    of sources is much less as the distribution of Sersic profiles has
    been obtained from HST observations.} \label{fig:profile_cor}
\end{figure}

The bias depends on Sersic index (see Fig.~\ref{fig:bias_profile}),
and therefore the performance of the empirical correction may differ.
Figure~\ref{fig:profile_cor} shows the residual bias as a function of
ellipticity distribution for four different Sersic indices. The black
points with errorbars indicate the results for the ensemble of
galaxies, with half-light radii and Sersic indices drawn from the
observed distribution of values. The lines indicate the results for
the simulations where the Sersic indices were fixed to the indicated
values. The range in bias is similar to what was observed in
Figure~\ref{fig:bias_profile}, with a positive bias in the expected
range of $\epsilon_0$ for $n=4$ and a negative bias for $n=1$. Given
the relatively weak dependence of Sersic index and the fact that the
distribution of Sersic values is well determined from HST observations
\citep{Rix04}, we conclude that our correction for the ensemble is
robust. Hence, we ignore the dependence of the bias on Sersic index.

\section{Application to simulated RCS2 data}
\label{sec:rcs2}

\begin{figure}
\centering
\leavevmode \hbox{%
\includegraphics[width=8.5cm]{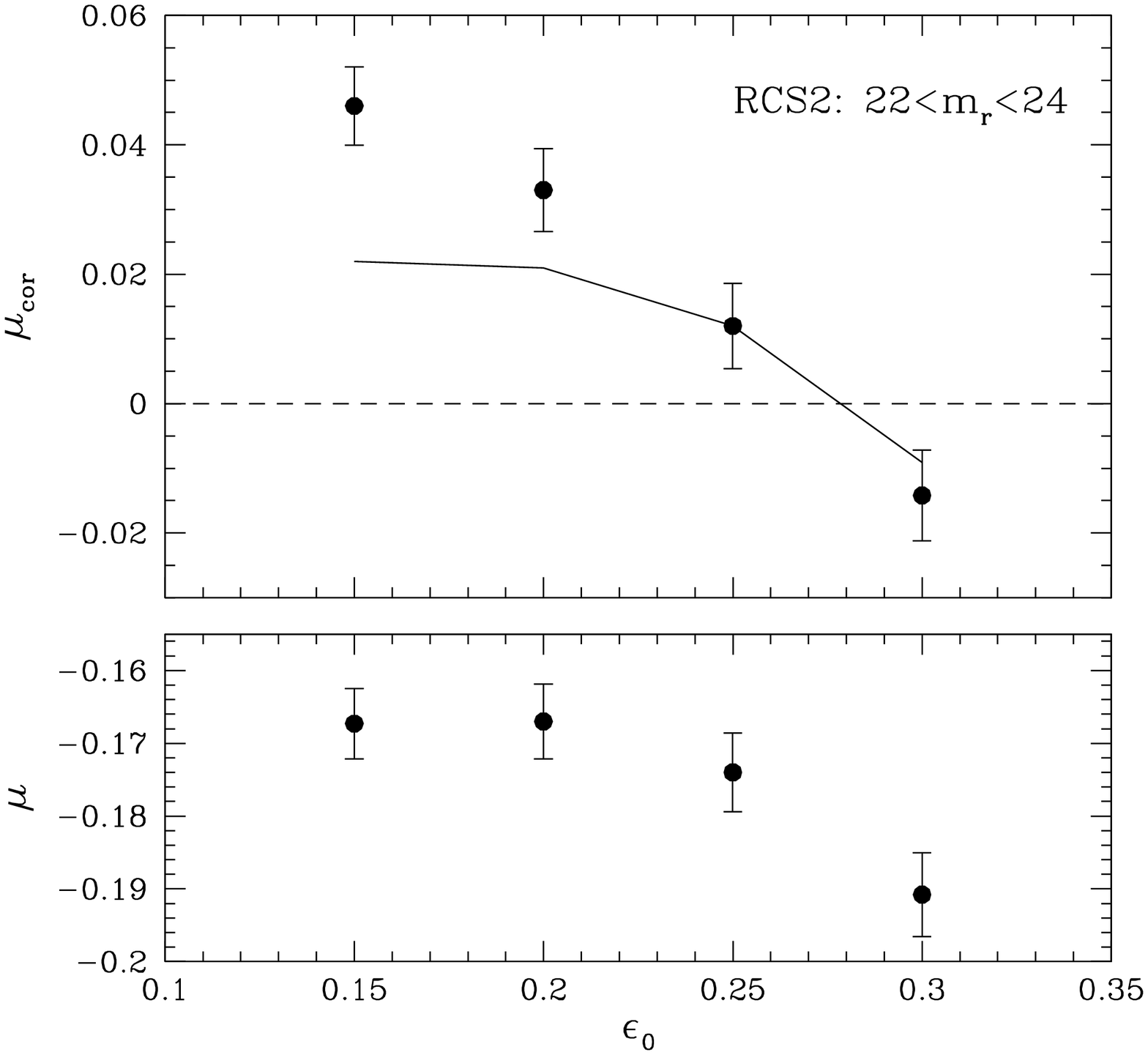}}
\caption{{\it Bottom panel:} Multiplicative bias for galaxies with
  $20<m_r<24$ as a function of $\epsilon_0$ for a simulation of RCS2
  data, which are shallower than the CCCP data. {\it Top panel:}
  Residual bias after we apply our correction, using parameters
  optimized for CCCP data, to the simulated RCS2 data. The solid line
  indicates $\mu_{\rm cor}$ if we use $\epsilon_0=0.25$ to correct the
  simulations for other ellipticity distributions, suggesting our
  approach is adequate for these data as well.}  \label{fig:rcs2_cor}
\end{figure}

Our KSB implementation has also been used by \cite{Uitert11} to
measure the lensing signal around galaxies using data from the second
Red-sequence Cluster Survey (RCS2). It is therefore interesting to
examine the impact of our findings on those results. Compared to our
CCCP data (with a total integration time of 3600s per position) the
RCS2 data are much shallower, consisting of a single 480s exposure in
the $r'$-band. We create a separate set of simulations where the noise
level matches that of the RCS2 data. The resulting bias for galaxies
with $22<m_r<24$ (the range used by \cite{Uitert11} for the RCS2
analysis) as a function of $\epsilon_0$ is presented in the lower
panel of Figure~\ref{fig:rcs2_cor}. The actual bias is smaller because
\cite{Uitert11} used the STEP1 implementation, also used in the CCCP
analysis of H12.  We also note that \cite{Uitert11} used
{\tt SExtractor} to detect objects, and used the resulting half-light
radius for the weight function in the subsequent shape analysis.

Here we are interested whether our correction scheme can be used to
data with different noise properties.  We therefore apply the
correction to the simulated RCS2 results and present the residual bias
in the top panel of Figure~\ref{fig:rcs2_cor}.  The black line
indicates $\mu_{\rm cor}$ if we use our reference value of
$\epsilon_0=0.25$ to correct the measurements for the various
ellipticity distributions. The results suggest that the parameters
that were optimized to correct CCCP can be used for the shallower RCS2
data as well.

\section{Obscuration by cluster members}
\label{app:obscure}

\begin{figure}
\centering
\leavevmode \hbox{%
\includegraphics[width=8.5cm]{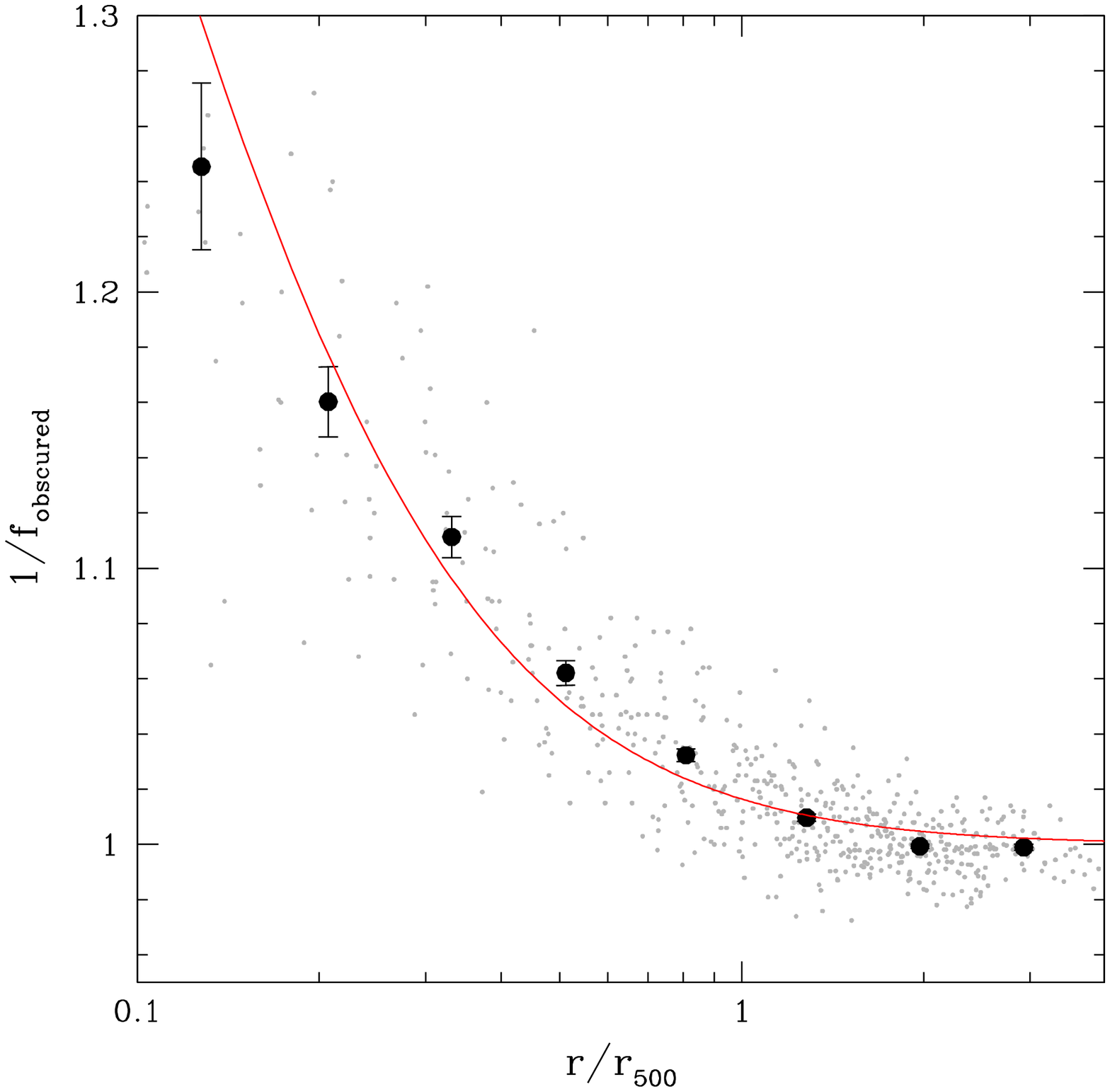}}
\caption{Plot of the correction of the counts of galaxies with
$22<m_r<25$ for the obscuration by cluster galaxies as a function
of radius in units of $r_{500}$. The black points indicate the 
average from a representative subset of clusters, whereas the
light grey points show the individual measurements. The red
line is the best fit model as described in the text.} \label{fig:obscure}
\end{figure}

As described in \S\ref{sec:contam}, we account for the dillution of
the lensing signal due to cluster galaxies in the source galaxy sample
using the excess galaxy counts as a function of cluster-centric
radius. Figure~\ref{fig:contam_mag} shows that the number density of
bright cluster galaxies is substantial at small radii, which may
affect the detection of sources. As shown in \cite{Simet14}, this is
an important source of bias for the measurement of the magnification
signal, but also is relevant here as it leads to biases in the
correction for contamination by cluster galaxies.

To quantify the impact of the obscuration by cluster galaxies on the
source galaxy counts, we use our image simulations: we add the cluster
observations to the simulated image and perform the object detection
and analysis on the images with and without the cluster
added\footnote{We ignore magnification which increases the fluxes of
  sources and thus reduces the effects of obscuration somewhat.  Note
  that the number density of sources is not affected significantly by
  magnification because the power law slope of the number counts ${\rm
    d}\log N_{\rm gal}/{\rm d}M\sim 0.38-0.4$, as discussed in
  \S\ref{sec:redshift}.}. We identify the objects detected in both
catalogs and measure their number density as a function of
cluster-centric radius. As expected, at small radii we observe a
decrease in the recovered number density.

Figure~\ref{fig:obscure} shows the corresponding correction for the
source counts as a function of radius in units of $r_{500}$. The black
points correspond to the average from a representative sample of
clusters that we used in this study. The individual measurements are
indicated by the lightgrey points. For radii larger than $r_{500}$ the
observed excess counts are biased low by a few percent. Hence, our
aperture mass estimates for $M_{500}$ are unaffected. On the smallest
scales considered for the NFW fits, the observed counts are biased low
by $\sim 10\%$. Even in this case the impact is small, as this is a
$10\%$ correction to a correction that itself is $30\%$; the resulting
change in the best fit mass is a $\sim 1-2\%$ increase. We find that
the correction can be described by
\begin{equation}
\frac{1}{f_{\rm obscured}}=1+\frac{0.021}{0.14+(r/r_{500})^2},
\end{equation}
which is the red line shown in Figure~\ref{fig:obscure}. We use 
this model to correct the measurements of the contamination by
cluster members. 

\section{Mass estimates using Duffy et al. (2008) $c(M)$-relation}

\label{app:duffy}
\begin{table*}
\begin{center}
  \caption{Weak lensing mass estimates for the CCCP sample using the
    mass-concentraton from \citet{Duffy08}\label{tab:duffy}}
\begin{tabular}{lccccccccccc}
\hline
(1) & (2) & (3) & (4) & (5) & (6) & (7) & (8) & (9) & (10) \\ 
name   & $M^{\rm proj}_{0.5}$ & $M^{\rm proj}_{1.0}$ & $r_{2500}$ & $M^{\rm ap}_{2500}$ & $r_{500}$ & $M^{\rm ap}_{500}$ & $M_{\rm vir}^{\rm NFW}$ & $M_{2500}^{\rm NFW}$ & $M_{500}^{\rm NFW}$ \\   
\hline
Abell~68 & $5.0\pm0.4$ & $8.7\pm1.2$ & $560$ & $3.2^{+0.3}_{-0.3}$ &$1391$ & $9.9^{+1.5}_{-1.6}$ &$14.2^{+3.1}_{-3.1}$ &$2.8^{+0.6}_{-0.6}$ & $7.8^{+1.7}_{-1.7}$\\
Abell~209 & $3.7\pm0.5$ & $7.8\pm1.5$ & $508$ & $2.3^{+0.4}_{-0.4}$ &$1249$ & $6.8^{+1.6}_{-1.5}$ &$9.3^{+2.7}_{-2.5}$ &$1.9^{+0.5}_{-0.5}$ & $5.1^{+1.5}_{-1.4}$\\
Abell~267 & $4.6\pm0.5$ & $7.8\pm1.5$ & $546$ & $2.9^{+0.4}_{-0.4}$ &$1222$ & $6.6^{+1.6}_{-1.6}$ &$8.6^{+2.8}_{-2.6}$ &$1.8^{+0.6}_{-0.5}$ & $4.8^{+1.5}_{-1.4}$\\
Abell~370 & $8.1\pm0.6$ & $16.9\pm1.7$ & $661$ & $6.1^{+0.6}_{-0.6}$ &$1659$ & $19.3^{+2.5}_{-2.5}$ &$34.6^{+6.6}_{-6.3}$ &$6.2^{+1.2}_{-1.1}$ & $18.7^{+3.6}_{-3.4}$\\
Abell~383 & $2.8\pm0.6$ & $6.4\pm1.4$ & $445$ & $1.5^{+0.5}_{-0.6}$ &$1232$ & $6.4^{+2.2}_{-2.4}$ &$6.2^{+2.9}_{-2.6}$ &$1.3^{+0.6}_{-0.6}$ & $3.5^{+1.6}_{-1.5}$\\
Abell~963 & $4.5\pm0.5$ & $7.9\pm1.4$ & $555$ & $3.0^{+0.5}_{-0.5}$ &$1274$ & $7.2^{+1.7}_{-1.7}$ &$13.6^{+3.5}_{-3.4}$ &$2.7^{+0.7}_{-0.7}$ & $7.5^{+1.9}_{-1.8}$\\
Abell~1689 & $8.6\pm0.5$ & $15.0\pm1.4$ & $734$ & $6.7^{+0.6}_{-0.6}$ &$1616$ & $14.4^{+2.4}_{-2.2}$ &$35.0^{+6.0}_{-5.7}$ &$6.4^{+1.1}_{-1.0}$ & $18.5^{+3.2}_{-3.0}$\\
Abell~1763 & $5.4\pm0.5$ & $10.5\pm1.4$ & $615$ & $4.1^{+0.6}_{-0.5}$ &$1529$ & $12.7^{+3.3}_{-2.9}$ &$19.0^{+4.1}_{-3.9}$ &$3.6^{+0.8}_{-0.8}$ & $10.3^{+2.2}_{-2.1}$\\
Abell~2218 & $5.7\pm0.5$ & $9.3\pm1.4$ & $646$ & $4.6^{+0.7}_{-0.6}$ &$1403$ & $9.4^{+2.2}_{-2.2}$ &$18.3^{+4.4}_{-4.2}$ &$3.5^{+0.8}_{-0.8}$ & $9.9^{+2.4}_{-2.3}$\\
Abell~2219 & $4.5\pm0.6$ & $10.1\pm1.4$ & $562$ & $3.2^{+0.6}_{-0.6}$ &$1418$ & $10.2^{+1.9}_{-1.7}$ &$12.5^{+2.9}_{-2.9}$ &$2.5^{+0.6}_{-0.6}$ & $6.9^{+1.6}_{-1.6}$\\
Abell~2390 & $6.1\pm0.5$ & $11.6\pm1.3$ & $648$ & $4.9^{+0.6}_{-0.6}$ &$1407$ & $10.0^{+1.7}_{-1.5}$ &$26.4^{+4.5}_{-4.3}$ &$4.9^{+0.8}_{-0.8}$ & $14.2^{+2.4}_{-2.3}$\\
MS~0015.9+1609 & $8.5\pm0.6$ & $20.8\pm2.0$ & $644$ & $6.9^{+0.7}_{-0.7}$ &$1654$ & $23.4^{+3.1}_{-3.0}$ &$33.4^{+8.5}_{-8.0}$ &$5.9^{+1.5}_{-1.4}$ & $18.4^{+4.7}_{-4.4}$\\
MS~0906.5+1110 & $4.5\pm0.5$ & $9.6\pm1.3$ & $574$ & $3.2^{+0.5}_{-0.5}$ &$1457$ & $10.4^{+1.7}_{-1.8}$ &$14.2^{+3.2}_{-3.2}$ &$2.8^{+0.6}_{-0.6}$ & $7.7^{+1.8}_{-1.8}$\\
MS~1224.7+2007 & $2.0\pm0.6$ & $2.6\pm1.6$ & $359$ & $0.9^{+0.4}_{-0.4}$ &$794$ & $2.0^{+0.8}_{-0.7}$ &$5.2^{+2.6}_{-2.3}$ &$1.1^{+0.6}_{-0.5}$ & $3.0^{+1.5}_{-1.3}$\\
MS~1231.3+1542 & $1.2\pm0.4$ & $0.5\pm1.2$ & $359$ & $0.8^{+0.2}_{-0.2}$ &$584$ & $0.7^{+0.4}_{-0.4}$ &$1.9^{+1.3}_{-1.2}$ &$0.5^{+0.3}_{-0.3}$ & $1.1^{+0.8}_{-0.7}$\\
MS~1358.4+6245 & $4.8\pm0.7$ & $9.4\pm1.7$ & $542$ & $3.2^{+0.7}_{-0.7}$ &$1316$ & $9.1^{+2.2}_{-2.2}$ &$15.0^{+3.8}_{-3.6}$ &$2.9^{+0.7}_{-0.7}$ & $8.3^{+2.1}_{-2.0}$\\
MS~1455.0+2232 & $4.3\pm0.4$ & $8.1\pm1.3$ & $538$ & $2.9^{+0.4}_{-0.4}$ &$1191$ & $6.2^{+1.3}_{-1.2}$ &$14.6^{+2.8}_{-2.8}$ &$2.9^{+0.6}_{-0.6}$ & $8.0^{+1.6}_{-1.6}$\\
MS~1512.4+3647 & $1.5\pm0.6$ & $4.5\pm1.6$ & $274$ & $0.4^{+0.3}_{-0.3}$ &$829$ & $2.4^{+1.8}_{-1.4}$ &$4.1^{+2.0}_{-1.7}$ &$0.9^{+0.4}_{-0.4}$ & $2.4^{+1.2}_{-1.0}$\\
MS~1621.5+2640 & $5.4\pm0.6$ & $12.3\pm1.6$ & $565$ & $4.0^{+0.7}_{-0.7}$ &$1301$ & $9.9^{+2.0}_{-1.8}$ &$21.8^{+5.0}_{-4.8}$ &$4.0^{+0.9}_{-0.9}$ & $12.0^{+2.7}_{-2.6}$\\
CL~0024.0+1652 & $6.7\pm0.6$ & $12.5\pm1.8$ & $597$ & $4.6^{+0.6}_{-0.6}$ &$1357$ & $10.7^{+1.9}_{-1.9}$ & $27.7^{+6.8}_{-6.4}$ & $5.1^{+1.2}_{-1.2}$ & $15.1^{+3.7}_{-3.5}$\\
\hline
Abell~115N & $1.4\pm0.5$ & $5.3\pm1.3$ & $283$ & $0.4^{+0.3}_{-0.4}$ &$1087$ & $4.4^{+1.3}_{-1.9}$ &$6.6^{+2.2}_{-2.2}$ &$1.4^{+0.5}_{-0.5}$ & $3.7^{+1.2}_{-1.2}$\\
Abell~115S & $2.6\pm0.5$ & $6.0\pm1.3$ & $399$ & $1.1^{+0.5}_{-0.5}$ &$1116$ & $4.8^{+1.3}_{-1.2}$ &$7.9^{+2.4}_{-2.4}$ &$1.6^{+0.5}_{-0.5}$ & $4.4^{+1.3}_{-1.3}$\\
Abell~222 & $2.9\pm0.4$ & $6.9\pm1.2$ & $433$ & $1.4^{+0.5}_{-0.7}$ &$1165$ & $5.6^{+1.2}_{-1.1}$ &$7.1^{+2.4}_{-2.3}$ &$1.5^{+0.5}_{-0.5}$ & $4.0^{+1.4}_{-1.3}$\\
Abell~223N & $3.0\pm0.5$ & $7.6\pm1.2$ & $448$ & $1.6^{+0.5}_{-0.7}$ &$1226$ & $6.5^{+1.3}_{-1.2}$ &$10.0^{+3.0}_{-2.8}$ &$2.0^{+0.6}_{-0.6}$ & $5.5^{+1.7}_{-1.6}$\\
Abell~223S & $3.0\pm0.4$ & $8.4\pm1.2$ & $451$ & $1.6^{+0.5}_{-0.6}$ &$1355$ & $8.7^{+1.6}_{-1.5}$ &$8.8^{+2.9}_{-2.7}$ &$1.8^{+0.6}_{-0.6}$ & $4.9^{+1.6}_{-1.5}$\\
Abell~520 & $3.6\pm0.5$ & $7.4\pm1.4$ & $513$ & $2.3^{+0.5}_{-0.5}$ &$1201$ & $6.0^{+1.5}_{-1.3}$ &$16.9^{+3.5}_{-3.4}$ &$3.3^{+0.7}_{-0.7}$ & $9.2^{+1.9}_{-1.8}$\\
Abell~521 & $3.1\pm0.5$ & $9.0\pm1.4$ & $401$ & $1.2^{+1.0}_{-0.5}$ &$1321$ & $8.5^{+1.7}_{-1.6}$ &$12.2^{+3.5}_{-3.4}$ &$2.4^{+0.7}_{-0.7}$ & $6.7^{+2.0}_{-1.9}$\\
Abell~586 & $2.5\pm0.6$ & $6.1\pm1.4$ & $430$ & $1.3^{+0.5}_{-0.4}$ &$1203$ & $5.9^{+2.4}_{-2.2}$ &$5.0^{+2.4}_{-2.2}$ &$1.1^{+0.5}_{-0.5}$ & $2.8^{+1.3}_{-1.2}$\\
Abell~611 & $3.7\pm0.5$ & $9.0\pm1.3$ & $489$ & $2.2^{+0.6}_{-0.5}$ &$1226$ & $7.1^{+1.2}_{-1.2}$ &$10.1^{+3.4}_{-3.1}$ &$2.0^{+0.7}_{-0.6}$ & $5.6^{+1.9}_{-1.7}$\\
Abell~697 & $4.6\pm0.5$ & $10.5\pm1.4$ & $551$ & $3.2^{+0.5}_{-0.5}$ &$1417$ & $10.8^{+1.7}_{-2.0}$ &$15.7^{+3.8}_{-3.6}$ &$3.1^{+0.7}_{-0.7}$ & $8.6^{+2.1}_{-2.0}$\\
Abell~851 & $5.4\pm0.5$ & $12.2\pm1.4$ & $540$ & $3.4^{+0.4}_{-0.4}$ &$1348$ & $10.7^{+1.8}_{-1.7}$ &$24.1^{+6.2}_{-5.7}$ &$4.4^{+1.1}_{-1.1}$ & $13.2^{+3.4}_{-3.2}$\\
Abell~959 & $5.0\pm0.5$ & $10.9\pm1.4$ & $580$ & $3.7^{+0.6}_{-0.6}$ &$1333$ & $9.0^{+1.5}_{-1.5}$ &$22.3^{+4.4}_{-4.4}$ &$4.2^{+0.8}_{-0.8}$ & $12.1^{+2.4}_{-2.4}$\\
Abell~1234 & $2.5\pm0.5$ & $4.4\pm1.4$ & $439$ & $1.4^{+0.3}_{-0.3}$ &$982$ & $3.2^{+1.3}_{-1.1}$ &$8.2^{+2.6}_{-2.4}$ &$1.7^{+0.5}_{-0.5}$ & $4.5^{+1.4}_{-1.3}$\\
Abell~1246 & $2.7\pm0.5$ & $5.7\pm1.1$ & $426$ & $1.3^{+0.4}_{-0.5}$ &$1082$ & $4.4^{+0.9}_{-0.9}$ &$9.7^{+2.7}_{-2.5}$ &$2.0^{+0.5}_{-0.5}$ & $5.4^{+1.5}_{-1.4}$\\
Abell~1758 & $5.5\pm0.5$ & $12.2\pm1.4$ & $633$ & $4.8^{+0.6}_{-0.6}$ &$1491$ & $12.6^{+1.9}_{-1.9}$ &$21.6^{+3.6}_{-3.6}$ &$4.1^{+0.7}_{-0.7}$ & $11.7^{+2.0}_{-2.0}$\\
Abell~1835 & $5.3\pm0.4$ & $10.7\pm1.2$ & $603$ & $4.0^{+0.4}_{-0.4}$ &$1387$ & $9.8^{+1.5}_{-1.5}$ &$22.4^{+4.2}_{-4.2}$ &$4.2^{+0.8}_{-0.8}$ & $12.1^{+2.3}_{-2.3}$\\
Abell~1914 & $3.7\pm0.5$ & $7.9\pm1.2$ & $522$ & $2.4^{+0.4}_{-0.4}$ &$1287$ & $7.2^{+1.4}_{-1.3}$ &$15.1^{+2.8}_{-2.7}$ &$3.0^{+0.5}_{-0.5}$ & $8.2^{+1.5}_{-1.5}$\\
Abell~1942 & $3.8\pm0.5$ & $7.6\pm1.2$ & $519$ & $2.5^{+0.4}_{-0.4}$ &$1204$ & $6.2^{+1.3}_{-1.2}$ &$15.1^{+3.2}_{-3.1}$ &$3.0^{+0.6}_{-0.6}$ & $8.2^{+1.8}_{-1.7}$\\
Abell~2104 & $4.2\pm0.5$ & $10.3\pm1.2$ & $583$ & $3.3^{+0.6}_{-0.6}$ &$1422$ & $9.5^{+1.9}_{-1.6}$ &$17.8^{+4.0}_{-3.8}$ &$3.4^{+0.8}_{-0.7}$ & $9.6^{+2.2}_{-2.1}$\\
Abell~2111 & $3.9\pm0.5$ & $6.6\pm1.3$ & $517$ & $2.5^{+0.4}_{-0.4}$ &$1156$ & $5.5^{+1.6}_{-1.5}$ &$10.1^{+2.5}_{-2.5}$ &$2.1^{+0.5}_{-0.5}$ & $5.6^{+1.4}_{-1.4}$\\
Abell~2163 & $4.4\pm0.5$ & $9.5\pm1.4$ & $562$ & $3.1^{+0.6}_{-0.5}$ &$1456$ & $10.8^{+2.2}_{-2.1}$ &$19.6^{+4.4}_{-4.2}$ &$3.7^{+0.8}_{-0.8}$ & $10.6^{+2.4}_{-2.3}$\\
Abell~2204 & $4.8\pm0.5$ & $11.0\pm1.0$ & $619$ & $3.9^{+0.5}_{-0.5}$ &$1490$ & $10.9^{+1.4}_{-1.3}$ &$22.4^{+3.8}_{-3.6}$ &$4.2^{+0.7}_{-0.7}$ & $12.0^{+2.0}_{-1.9}$\\
Abell~2259 & $2.4\pm0.5$ & $5.6\pm1.3$ & $417$ & $1.2^{+0.4}_{-0.4}$ &$1106$ & $4.5^{+1.2}_{-1.3}$ &$8.8^{+3.1}_{-2.7}$ &$1.8^{+0.6}_{-0.6}$ & $4.8^{+1.7}_{-1.5}$\\
Abell~2261 & $6.0\pm0.4$ & $14.2\pm1.3$ & $666$ & $5.3^{+0.5}_{-0.5}$ &$1654$ & $16.1^{+1.6}_{-1.6}$ &$28.1^{+4.9}_{-4.7}$ &$5.2^{+0.9}_{-0.9}$ & $15.1^{+2.6}_{-2.5}$\\
Abell~2537 & $5.4\pm0.6$ & $10.1\pm1.4$ & $585$ & $3.9^{+0.6}_{-0.6}$ &$1302$ & $8.5^{+1.6}_{-1.5}$ &$23.6^{+4.6}_{-4.6}$ &$4.4^{+0.9}_{-0.9}$ & $12.9^{+2.5}_{-2.5}$\\
MS~0440.5+0204 & $2.9\pm0.5$ & $2.8\pm1.3$ & $457$ & $1.6^{+0.4}_{-0.4}$ &$890$ & $2.4^{+0.7}_{-0.7}$ &$3.5^{+2.0}_{-1.8}$ &$0.8^{+0.4}_{-0.4}$ & $2.0^{+1.1}_{-1.0}$\\
MS~0451.6-0305 & $4.4\pm0.7$ & $8.4\pm2.1$ & $453$ & $2.4^{+0.6}_{-0.6}$ &$1071$ & $6.3^{+1.7}_{-1.9}$ &$19.4^{+6.7}_{-6.2}$ &$3.6^{+1.2}_{-1.1}$ & $10.8^{+3.7}_{-3.5}$\\
MS~1008.1-1224 & $4.1\pm0.4$ & $8.2\pm1.4$ & $507$ & $2.5^{+0.4}_{-0.3}$ &$1168$ & $6.2^{+1.3}_{-1.2}$ &$18.2^{+4.0}_{-3.8}$ &$3.5^{+0.8}_{-0.7}$ & $10.0^{+2.2}_{-2.1}$\\
RX~J1347.5-1145 & $5.2\pm0.8$ & $10.1\pm1.9$ & $530$ & $3.4^{+0.9}_{-0.8}$ &$1301$ & $10.1^{+3.2}_{-3.0}$ &$22.0^{+6.0}_{-5.7}$ &$4.1^{+1.1}_{-1.0}$ & $12.2^{+3.3}_{-3.1}$\\
RX~J1524.6+0957 & $2.1\pm0.9$ & $6.5\pm2.1$ & $245$ & $0.4^{+0.4}_{-1003.0}$ &$961$ & $4.4^{+2.2}_{-1.9}$ &$7.1^{+4.8}_{-4.2}$ &$1.4^{+1.0}_{-0.9}$ & $4.1^{+2.8}_{-2.4}$\\
MACS~J0717.5+3745 & $6.4\pm0.9$ & $19.3\pm2.3$ & $586$ & $5.2^{+1.1}_{-3.1}$ &$1470$ & $16.4^{+3.4}_{-3.0}$ &$44.8^{+10.8}_{-10.3}$ &$7.6^{+1.8}_{-1.8}$ & $24.4^{+5.9}_{-5.6}$\\
MACS~J0913.7+4056 & $3.1\pm0.5$ & $5.3\pm1.5$ & $380$ & $1.2^{+0.5}_{-0.5}$ &$935$ & $3.7^{+1.0}_{-0.9}$ &$6.9^{+3.7}_{-3.2}$ &$1.4^{+0.8}_{-0.7}$ & $4.0^{+2.1}_{-1.8}$\\
CIZA~J1938+54 & $5.3\pm0.5$ & $11.3\pm1.3$ & $588$ & $3.8^{+0.4}_{-0.4}$ &$1557$ & $14.0^{+2.2}_{-2.2}$ &$19.4^{+5.0}_{-4.7}$ &$3.7^{+1.0}_{-0.9}$ & $10.6^{+2.7}_{-2.6}$\\
3C295 & $4.6\pm0.6$ & $8.1\pm1.9$ & $488$ & $2.7^{+0.5}_{-0.5}$ &$1090$ & $6.0^{+1.5}_{-1.6}$ &$13.4^{+4.6}_{-4.2}$ &$2.6^{+0.9}_{-0.8}$ & $7.5^{+2.6}_{-2.4}$\\
\hline
\hline
\end{tabular}
\bigskip
\begin{minipage}{\linewidth} {\footnotesize Column 1: cluster name;
    Columns~2 \& 3: projected mass within an aperture of
    $0.5h_{70}^{-1}$ and $1h_{70}^{-1}$Mpc, resp.; Columns~4 \& 6:
    $r_{\Delta}$ (in units of $h_{70}^{-1}$kpc) determined using
    aperture masses; Columns~5 \& 7: deprojected aperture masses
    within $r_\Delta$; Columns~8-10: masses from best fit NFW
    model. All masses are listed in units of
    $[10^{14}h_{70}^{-1}$\msun]}
\end{minipage}
\end{center}
\end{table*}

Previous cluster weak lensing studies, including H12, presented mass
measurements using the relation between mass and concentration from
\cite{Duffy08}, which is based on numerical simulatons assuming a
{\it WMAP5} cosmology \citep{Komatsu09}. The first results from {\it
  Planck} presented by \cite{PlanckXVI} suggest higher values for both
the normalization of the matter power spectrum $\sigma_8$ and the mean
density $\Omega_m$. For this reason we adopted the relation from
\cite{Dutton14}, which yields a concentration that is $\sim 20\%$
higher than \cite{Duffy08} for a given mass.

This affects the masses inferred from parametric NFW fits and the
deprojected aperture masses, but not necessarily in the same
sense. For instance, if we switch to the \cite{Duffy08} relation, the
estimate for $M_{2500}$ decreases by $\sim 7\%$, no matter whether we
consider the aperture mass or the best fit NFW model. On the other
hand, the estimate for $M_{500}$ increases by $5\%$ for the NFW model
fit, whereas the deprojected aperture mass decreases by $\sim 3\%$
(also see \S4.3 in H12). Hence, to allow for a straightforward
comparison with previous mass measurements, we present in
Table~\ref{tab:duffy} the results if we use the mass-concentration
relation from \cite{Duffy08}.

\end{document}